\documentclass[prd,aps,superscriptaddress,twocolumn,driverfallback=dvips]{revtex4}
\usepackage{feynmp}
\usepackage{amsmath}
\usepackage{amssymb}
\usepackage{epsfig}
\usepackage{graphicx}
\usepackage{subfigure}
\usepackage{hyperref}
\usepackage{bbm}
\usepackage{epsfig}
\usepackage{color}
\usepackage{float}
\usepackage{mhchem}
\usepackage{epstopdf}

\newcommand{\Rmnum}[1]{\expandafter\@slowromancap\romannumeral #1@}
\allowdisplaybreaks[3]

\begin{document}

\title{Superfluid quantum criticality and the thermal evolution of neutron stars}

\author{Hao-Fu Zhu}
\affiliation{Department of Astronomy, University of Science and
Technology of China, Hefei, Anhui 230026, China}\affiliation{School
of Astronomy and Space Science, University of Science and Technology
of China, Hefei, Anhui 230026, China}
\author{Guo-Zhu Liu}
\altaffiliation{Corresponding author: gzliu@ustc.edu.cn}
\affiliation{Department of Modern Physics, University of Science and
Technology of China, Hefei, Anhui 230026, China}
\author{Jing-Rong Wang}
\affiliation{Anhui Province Key Laboratory of Condensed Matter
Physics at Extreme Conditions, High Magnetic Field Laboratory of the
Chinese Academy of Sciences, Hefei, Anhui 230031, China}
\author{Xufen Wu}
\affiliation{Department of Astronomy, University of Science and
Technology of China, Hefei, Anhui 230026, China}\affiliation{School
of Astronomy and Space Science, University of Science and Technology
of China, Hefei, Anhui 230026, China}

\begin{abstract}
The neutron star starts to cool down shortly after its birth by
emitting neutrinos. As it becomes cold enough, the Cooper pairs of
neutrons are formed, triggering a superfluid transition. Previous
studies on neutron superfluidity focused on finite-temperature
transitions, with little attention paid to the potentially important
quantum critical phenomena associated with superfluidity. Here, we
provide the first theoretical analysis of superfluid quantum
criticality, concentrating on its impact on neutron star cooling.
Extensive calculations found that superfluidity occurs within a
finite range of neutron star density $\rho$. The density serves as a
nonthermal parameter for a superfluid quantum phase transition. In a
broad quantum critical region, gapless neutrons are strongly coupled
to the quantum critical fluctuations of the superfluid order parameter.
We handle this coupling using both perturbation theory and
renormalization group methods and find that it leads to non-Fermi
liquid behavior, which yields a logarithmic $T\ln(1/T)$ correction
to the neutron specific heat $c_{\mathrm{n}}\propto T$ and also
dramatically alters the neutrino emissivity. Quantum critical
phenomena emerge much earlier than the onset of superfluidity and
persist throughout almost the entire lifetime of a neutron star. At
low temperatures, these phenomena coexist with superfluidity in the
neutron star interior but occupy different layers. We incorporate
superfluid quantum criticality into the theoretical description of
neutron star cooling and show that it substantially prolongs the
thermal relaxation time. By varying the strength of superfluid
fluctuations and other quantities, we obtain an excellent fit to the
observed cooling data of a number of neutron stars. Our results
indicate an intriguing correlation between superfluid quantum
criticality and the thermal evolution of neutron stars.
\end{abstract}

\maketitle


\section{Introduction \label{Sec:introduction}}

The thermal evolution of neutron stars (NSs) \cite{Lattimer04,
Lattimer16} is a pivotal area of study as it provides profound
insights into their internal structure and composition. After its
birth from a supernova explosion, the NS loses a significant portion
of its energy within a short time, with its internal temperature
falling rapidly from $\sim$$10^{11}~\mathrm{K}$ \cite{Yakovlev05}.
Afterwards, the NS cooling is primarily driven by the emission of
neutrinos, until, about $10^{5}$ to $10^{6}$ yr later, the
neutrino emission is overshadowed by the surface thermal radiation
of photons \cite{Yakovlev99, Yakovlev01, Yakovlev04, Page06a,
Potekhin15, Tsuruta23}.

Neutrinos are emitted from the interiors of NSs via several
different nucleon (neutrons/protons)-associated processes
\cite{Yakovlev99, Yakovlev01, Yakovlev04, Page06a, Potekhin15,
Tsuruta23}, including the direct Urca (DU) process, the modified
Urca (MU) process, the nucleon-nucleon bremsstrahlung (NNB) process,
and so on. The DU process is not the dominant cooling scenario for
two reasons. First, it makes NSs lose energies at an extremely high
speed, which is at odds with observations. In the inner core region,
whose composition is in fierce debate, the DU process
\cite{Lattimer91} might take place with some exotic particles like
hyperons and quarks \cite{Prakash92, Prakash98}. However, the
cooling rates induced by such DU processes are still excessively
large. Second, the DU process is forbidden unless the proton density
$\rho^{~}_{p}$ exceeds $(11$-$15)\% \rho_b$ \cite{Lattimer91} with
$\rho_b$ being the baryon density. The critical proton fraction
$Y^{~}_{p}=\rho^{~}_{p}/\rho_b$ for triggering the DU process is
complicated by the dependence of the electron/muon fraction ratio on
the equation of state (EOS) \cite{Dohi19, Lopes24}, and is generally
only reached at high densities within NSs. This indicates that the
DU process occurs only when the NS mass $\mathrm{M}$ exceeds some
threshold $\mathrm{M}_{\mathrm{DU}}$, whose value is uncertain and
depends heavily on the EOS. Actually, the fast cooling scenario
associated with DU processes occurs only under special situations,
such as in massive NSs \cite{Marino24}.

In comparison, MU processes are usually regarded as the standard
scenario of neutrino emission since they occur in all NSs and cool
them down at a moderate speed. However, MU processes cannot explain
the rapid cooling of certain NSs. A notable example is the young
($\approx340$ yr old) center compact object (CCO) within the
supernova remnant Cassiopeia A (Cas A), first observed by the
\emph{Chandra X-ray Observatory} in 1999 \cite{Hughes00}. The NS in
Cas A is a typical weakly magnetized thermally emitting isolated NS
(TINS). Its surface temperature had been observed \cite{Ho09, Ho10}
to decrease from $2.12\times 10^{6}$ to $2.04\times
10^{6}~\mathrm{K}$ during 2000-2009. More recent analysis adjusted
the cooling rate from 4$\%$ to 2$\%$ over a decade \cite{Posselt18,
Wijngaarden19, Ho21, Shternin23}. This revised cooling rate is much
higher than that predicted by MU processes, but is far lower than
that induced by DU processes. Moreover, such a rapid cooling occurs
at a relatively late stage ($\sim 300$ yr), whereas DU processes
operate shortly after birth. It appears that the cooling rate of the
Cas A NS is governed by a novel mechanism different from both DU and
MU processes.

It is universally accepted that the formation of neutron
superfluidity is central to the thermal evolution of NSs
\cite{Dean03, Sedrakian19, Pagereview}. As the NS cools down to a
sufficiently low temperature $T$ in its isothermal interior, the
attractive force between neutrons can bind gapless neutrons into
Cooper pairs, which triggers an instability of the degenerate
neutron liquid and drives a superfluid transition. The pairing gap
is quite small slightly below the transition temperature
$T_{\mathrm{cn}}$, and hence thermal fluctuations \cite{Inotani20}
lead to the constant breaking and recombination of Cooper pairs.
Flowers \emph{et al.} \cite{Flowers76} first studied the effects of
pairing breaking and formation (PBF) and revealed that PBF results
in an enhanced neutrino emission due to the weak interaction between
the neutral current of Bogoliubov quasiparticles and the neutrino
current. Later, the PBF scenario was used \cite{Voskresensky87,
Yakovlev99b} to understand the cooling history of NSs. In
particular, Page \emph{et al.} \cite{Page11} and Shternin \emph{et
al.} \cite{Shternin11} proposed a minimal cooling paradigm
\cite{Page04, Page06, Page09} based on the PBF mechanism along with
some additional assumptions to explain the rapid cooling of the Cas
A NS. However, other studies \cite{Potekhin18, Leinson22} indicated
that the neutrino emission of the PBF scenario is not efficient
enough to account for the observed cooling rate.

Besides the Cas A NS, there exist several other NSs that exhibit
peculiar cooling histories. For example, two rotation-powered
pulsars (PSRs), denoted by PSR J0205+6449 \cite{Kothes13} and PSR
B2334+61 \cite{Yar-Uyaniker04}, are observed to be younger than
$10^{4}$ yr, but their surface temperatures are exceptionally low,
i.e., $\sim$$5\times 10^{5}~\mathrm{K}$ \cite{Potekhin20}. As a
comparison, some very old NSs like X-ray emitting isolated NSs
(XINSs) \cite{Haberl07}, whose ages are roughly $\sim$$10^{6}$ yr,
are estimated to be as warm as $\sim$$10^{6}~\mathrm{K}$
\cite{Potekhin20}. At present, the microscopic origins of the
cooling trajectories of these NSs remain poorly understood. In view
of such a situation, it is interesting to explore new scenarios that
might dramatically affect the NS cooling history.

In this paper, we present a theoretical analysis of the thermal
evolution of NSs based on a careful examination of the effects
caused by superfluid quantum criticality. Quantum criticality is one
of the cornerstones of current condensed-matter physics
\cite{Sachdev00, Vojta03, Chubukov03AFMQCP, Coleman05, Loehneysen07,
Sachdevbook, Keimer15}. Unfortunately, it has attracted little
attention in other branches of physics. While the superfluid
transition in NSs has been studied for several decades \cite{Dean03,
Sedrakian19, Pagereview}, the striking phenomena resulting from
superfluid quantum criticality and their observational effects have
not been previously considered in the NS community. We show
that the superfluid quantum criticality leads to the breakdown of
the Fermi liquid (FL) description of the degenerate neutron gas and
produces a non-Fermi liquid (NFL) behavior. Once this NFL behavior
is incorporated, the cooling trajectories obtained in our numerical
simulations are in accordance with the observational data of the NS
in Cas A and some other NSs mentioned above.

It is necessary to first sketch the general picture of quantum
criticality \cite{Sachdev00, Vojta03, Chubukov03AFMQCP, Coleman05,
Loehneysen07, Sachdevbook} before applying this concept to study NS
cooling. Consider a system cooled down to $T=0$. Upon tuning a
nonthermal parameter $\delta$, which might be pressure or particle
density, this system undergoes a continuous phase transition at some
critical value $\delta=\delta_{c}$. This transition is classified as
a quantum phase transition since it is driven by the
quantum-mechanical effects instead of thermal fluctuations. The
critical value $\delta=\delta_{c}$ defines a quantum critical point
(QCP). Without loss of generality, we assume that a certain symmetry
is broken for $\delta>\delta_{c}$ but preserved for $\delta \leq
\delta_{c}$. The order parameter $\Phi$ for this transition has a
vanishing expectation value, i.e., $\langle\Phi\rangle=0$, in the
symmetric phase ($\delta \leq \delta_{c}$), but acquires a finite
expectation value, i.e., $\langle\Phi \rangle\neq 0$, in the
symmetry-broken phase ($\delta>\delta_{c}$). Even though
$\langle\Phi \rangle=0$ at the QCP, the quantum critical
fluctuations of the order parameter are very strong and can lead to
unusual quantum critical phenomena under proper conditions. At
finite temperatures, the zero-$T$ QCP is broadened into a $V$-shaped
quantum critical region on the $T$-$\delta$ plane, where thermal and
quantum fluctuations are both important. Quantum critical phenomena
can emerge in the whole quantum critical region. In the last
decades, quantum criticality has been extensively investigated in
condensed-matter physics \cite{Hertz, Millis, Sachdev00, Vojta03,
Chubukov03AFMQCP, Coleman05, Loehneysen07, Sachdevbook, Keimer15,
Chubukov04FMQCP, Chubukov04FM, Rech06, Lee07, Metlitski10,
Metlitski102, Liu12, Grover14, Maciejko16, Zerf16, Wang17, Liu18,
Pan18, Liu19, Wang20, Sachdev23, CaoYuan20, Polshyn19, Jaoui22,
Hartnoll22, Zaanen19, Phillips22}. Numerous experimental findings
suggest that the quantum criticality may be responsible for many
salient features of a large number of strongly correlated
condensed-matter systems, including high-$T_{c}$ cuprate
superconductors \cite{Sachdev00, Vojta03, Sachdevbook, Keimer15},
heavy fermion compounds \cite{Coleman05, Loehneysen07}, Dirac/Weyl
semimetals \cite{Grover14, Maciejko16, Zerf16, Wang17, Liu18, Pan18,
Liu19, Wang20}, magic-angle twisted bilayer graphene
\cite{CaoYuan20, Polshyn19, Jaoui22}, as well as some Planckian
strange metals \cite{Sachdev23, Hartnoll22, Zaanen19, Phillips22}.

\begin{widetext}

\begin{figure*}[htbp]
\centering
\includegraphics[width=5.4in]{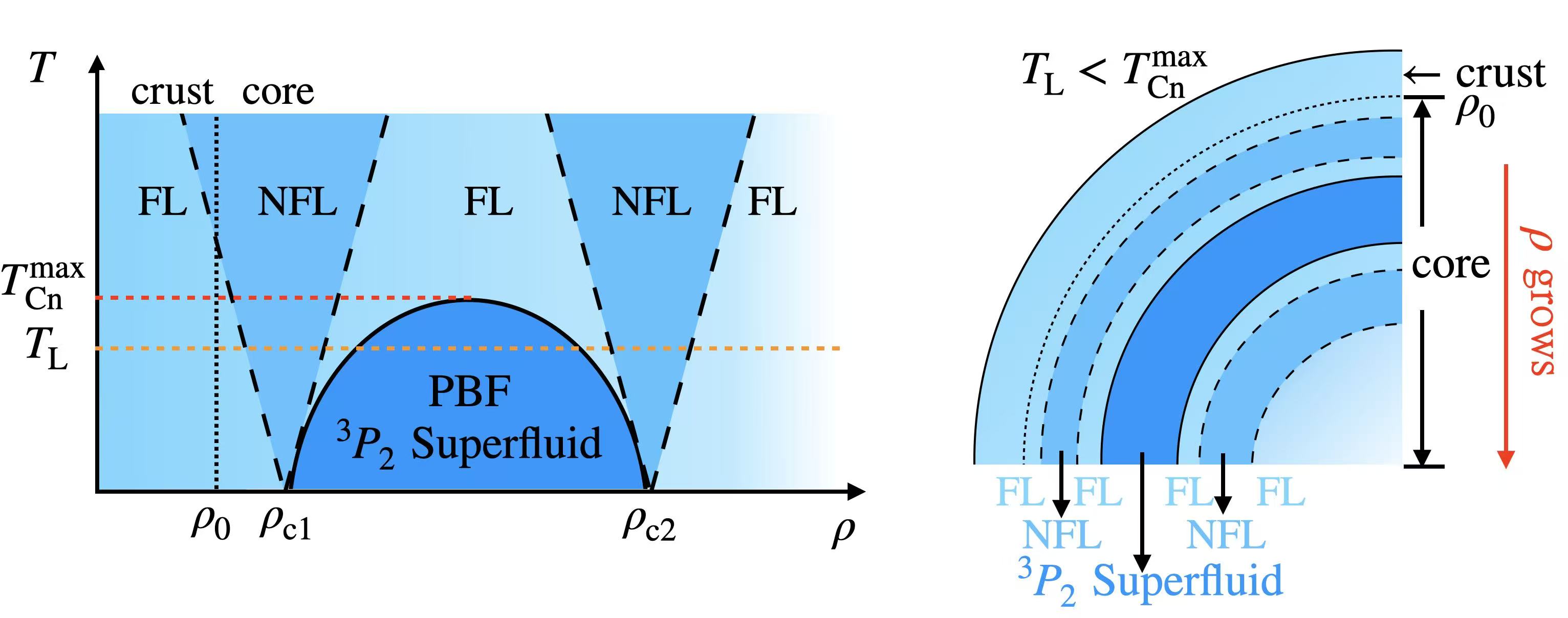}
\caption{The left panel depicts a schematic global phase diagram on
the $T$-$\rho$ plane. The zero-temperature QCPs are broadened at
finite temperatures into two quantum critical regions exhibiting NFL
behavior. The superfluid state is gapped, whereas FL and NFL states are
gapless. The PBF process occurs in the neutron $^3P_2$ superfluid phase,
at temperatures not much lower than
$T^{\mathrm{max}}_{\mathrm{cn}}$. Landau quasiparticles are
well defined in the FL state, but destroyed in the NFL state. The
right panel shows the coexistence of different layers in NS
interior. The thickness of each layer depends on the temperature or
the age of the NS. In particular, the thickness of the NFL layer is roughly
$\sim k_{\mathrm{B}}T$. {For simplicity, the positions of the
neutron $^1S_0$ superfluid and proton $^1S_0$ superconductor have
been omitted. The density $\rho_0 \approx 2.0 \times 10^{14}
~\mathrm{g/cm^3}$ marks the boundary between crust and core.}}
\label{fig:phasediagram}
\end{figure*}

\end{widetext}

We anticipate that quantum criticality plays a vital role in NSs as
well.  This can be understood as follows. Extensive calculations
using Bardeen-Cooper-Schrieffer (BCS) mean-field theory
\cite{Dean03, Sedrakian19, Pagereview} revealed that $^1S_0$-wave
superfluid and $^3P_2$-wave superfluid could occur within finite
ranges of NS density $\rho$. In the case of $^3P_2$-wave pairing,
the gap $\Delta_{\mathrm{n}}$ or the transition temperature
$T_{\mathrm{cn}}$ attains its maximal value at a certain density
$\rho_{m}$, decreases as $\rho$ deviates from $\rho_{m}$, and
vanishes once $\rho$ becomes smaller than $\rho_{c1}$ or larger than
$\rho_{c2}$. A schematic illustration is shown in the left panel of
Fig.~\ref{fig:phasediagram}. Although the accurate values of
$\rho_{c1}$ and $\rho_{c2}$ are unknown, a dome-shaped
({Gaussian-like}) phase boundary on the $\Delta_{\mathrm{n}}$-$\rho$
plane or the $T_{\mathrm{cn}}$-$\rho$ plane is found in most
BCS-level calculations \cite{Dean03, Sedrakian19, Pagereview,
Page04, Ho15} and in some phenomenological analyses of the cooling
history of NSs \cite{Pagereview}. At any density between $\rho_{c1}$
and $\rho_{c2}$, the superfluid transition temperature
$T_{\mathrm{cn}}$ takes a specific nonzero value. It is interesting
to view such transitions from a different perspective. One can
alternatively fix $T$ at $T=0$ and raise $\rho$ from zero to large
values, which amounts to moving inwards toward the NS interior from
the surface. In this process, two superfluid transitions occur at
$\rho_{c1}$ and $\rho_{c2}$. One could think of $\rho$ as a
nonthermal tuning parameter for a superfluid quantum phase
transition, and regard $\rho_{c1}$ and $\rho_{c2}$ as two zero-$T$
QCPs. The NS temperature is never lowered down to absolutely zero,
and thus QCPs are not directly observable. However, remarkable
superfluid quantum critical phenomena can emerge at finite
temperatures near $\rho_{c1}$ and $\rho_{c2}$.

Superfluid quantum criticality in NSs exhibits distinctive
characteristics not manifested in condensed-matter physics.
Condensed-matter systems are usually uniform, and thus they become
quantum critical as a whole at $\delta = \delta_{c}$. For $\delta
\neq \delta_{c}$, the system is in either the ordered or disordered
phase. For an NS interior, however, the NS density $\rho$ depends
strongly on the radius of any position $\mathbf{r}$. According to
the Tolman-Oppenheimer-Volkof equation \cite{Tolman, OV}, the
balance between the gravitational force and the degenerate pressure
of the neutron liquid requires that $\rho(\mathbf{r})$ should grow
as $|\mathbf{r}|$ decreases. Thus, the quantum ordered phase
(superfluid), quantum disordered phase (normal liquid), and quantum
critical region can coexist in one NS. They occupy different layers,
as illustrated in the right panel of Fig.~\ref{fig:phasediagram}.
The width of each layer is $T$ dependent. In comparison, such a
coexistence rarely occurs in condensed-matter systems. After the
internal thermal relaxation stage is ended, the NS interior is
thought to be isothermal with a finite $T$. {At a given temperature
above $T^{\mathrm{max}}_{\mathrm{cn}}$, PBF processes are absent,
but quantum critical phenomena emerge already.} At a lower
temperature $T^{~}_{\mathrm{L}}$ below
$T^{\mathrm{max}}_{\mathrm{cn}}$, PBF layers and quantum critical
layers are both present, but separated by normal FL layers. We
observe from Fig.~\ref{fig:phasediagram} that quantum critical
phenomena last for a much longer time scale than the PBF scenario.
Their influence on the NS cooling rate deserves a careful
investigation.

In the condensed-matter community, it is widely accepted that
quantum criticality is fundamentally different from classical
criticality \cite{Vojta03, Chubukov03AFMQCP, Coleman05,
Loehneysen07, Sachdevbook}. For classical phase transitions driven
by the variation of temperature, thermal fluctuations operate in
space, but not in time. Hence, classical criticality can be well
described by pure $\Phi^{4}$ theory in the framework of the
Ginzburg-Landau-Wilson (GLW) paradigm \cite{Vojta03, Sachdevbook}.
However, the GLW paradigm breaks down for quantum criticality, since
quantum fluctuations are significant in both space and time. It is
now established \cite{Vojta03, Chubukov03AFMQCP, Coleman05,
Loehneysen07, Sachdevbook} that the key characteristic of quantum
criticality is the presence of a strong Yukawa-type (a terminology
borrowed from nuclear physics) coupling between the gapless
fermionic excitations and the quantum fluctuation of the associated
order parameter. It has been revealed that this coupling can induce
a variety of unusual quantum critical phenomena, such as NFL
behavior \cite{Chubukov03AFMQCP, Rech06, Metlitski10, Metlitski102,
Wang17, Pan18, Wang20, Sachdev23}, strange metallicity
\cite{Sachdev23, Hartnoll22, Zaanen19, Phillips22}, and emergent
low-energy symmetry \cite{Lee07, Grover14, Maciejko16, Zerf16,
Liu19}.

We perform a field-theoretic study of superfluid quantum
criticality. This criticality is characterized by the coupling of
gapless neutrons excited on the Fermi surface to superfluid quantum
fluctuations. Similar to its many condensed-matter counterparts,
such a coupling is sharply peaked at zero-momentum scattering
processes, which allows us to derive an effective low-energy field
theory to describe quantum criticality. We first compute the neutron
self-energy $\Sigma(i\omega)$ by employing the perturbation theory
based on a $1/N$ expansion scheme, where $N=2$. We show that the
neutron damping rate $\Gamma(\omega)$, which is linked to the
imaginary part of the retarded self-energy
$\mathrm{Im}\Sigma^{~}_{R}(\omega)$, exhibits a linear $\omega$
dependence. This is an NFL behavior. To verify the reliability of
such a result, we handle the same effective field theory using the
renormalization group (RG) theory. After solving the RG flow
equations of all model parameters, we find the same
linear-in-$\omega$ NFL behavior of $\Gamma(\omega)$. We demonstrate
that this NFL behavior generates a logarithmic $T\ln(1/T)$
correction to the original linear specific heat
$c_{\mathrm{n}}(T)\propto T$ of the neutron FL. The effective
neutron mass is also significantly renormalized and acquires a
logarithmic $T$ dependence.

\begin{widetext}

\begin{figure}[htbp]
\centering
\includegraphics[width=4.5in]{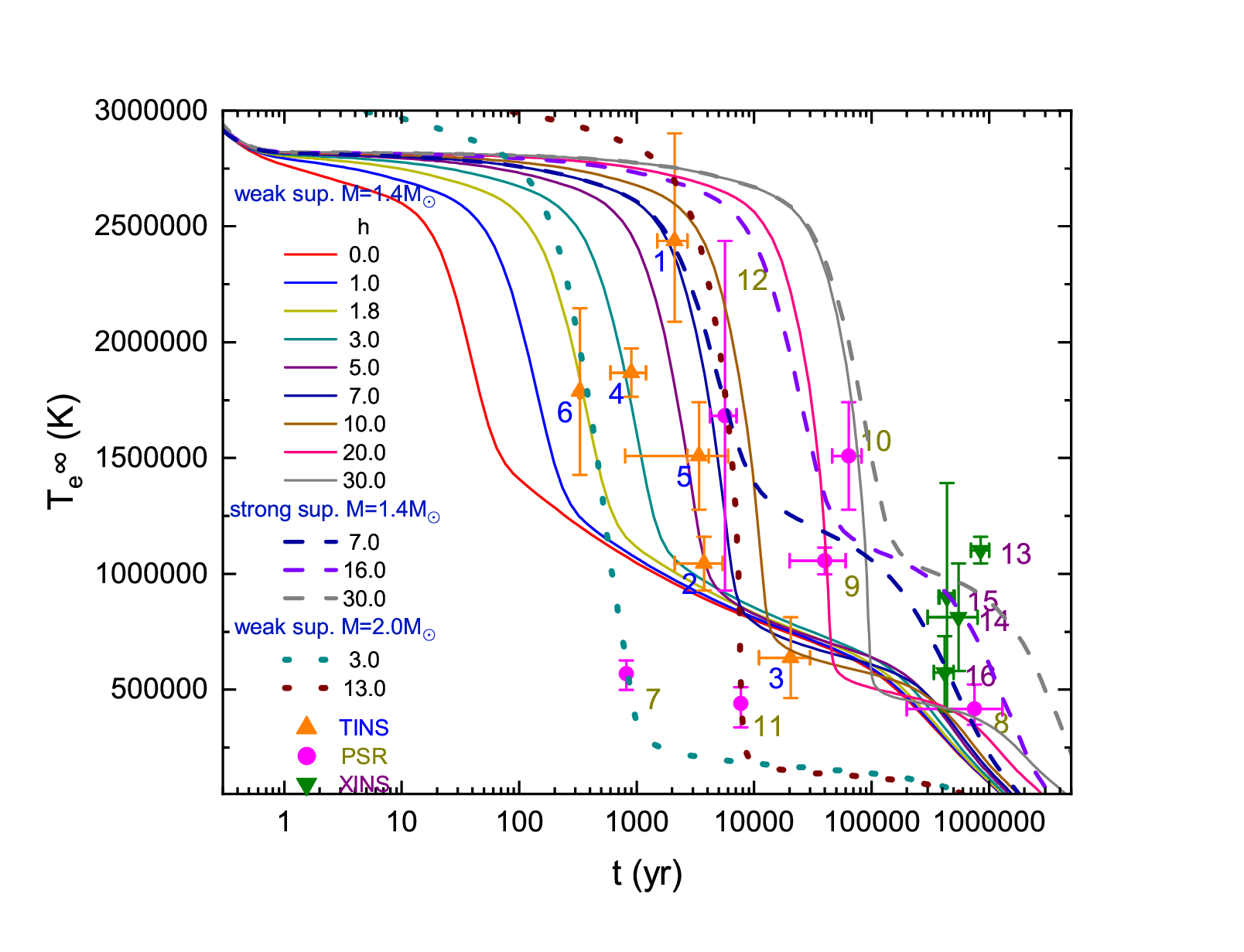}
\caption{Redshifted effective temperature,
$T^{\infty}_{\mathrm{e}}$~(K), plotted against the age
$\mathrm{t}$~(yr) for various values of $h$, which is the strength
parameter of the coupling between neutrons and superfluid quantum
fluctuations. Solid lines are the cooling curves for an isolated NS
that has a mass $\mathrm{M}=1.4\mathrm{M}_{\odot}$ and exhibits weak
neutron $^{3}P_2~(m_J=0)$-wave superfluidity {(the superfluid model
``a'' in \cite{Page04, Baldo98} with
$T^{\mathrm{max}}_{\mathrm{cn}}\sim 10^{9}~\mathrm{K}$) }inside its
core. Dashed lines are the cooling curves for an isolated NS that
has a mass $\mathrm{M}=1.4\mathrm{M}_{\odot}$ but exhibits strong
neutron $^{3}P_2~(m_J=0)$-wave superfluidity (the superfluid model
``c'' in \cite{Page04, Baldo98} with
$T^{\mathrm{max}}_{\mathrm{cn}}\sim 10^{10}~\mathrm{K}$). Dotted
lines are the cooling curves for an isolated massive NS that has a
mass $\mathrm{M}=2.0 \mathrm{M}_{\odot}$ and exhibits weak neutron
$^{3}P_2~(m_J=0)$-wave superfluidity. The observational data are
taken from Ref.~\cite{Potekhin20}. The error bars represent a
$1\sigma$ confidence interval. Three classes of different NSs,
namely TINS, PSR, and XINS, are considered. TINS (including CCO): 1)
1E 0102.2-7219; 2) CXOU J085201.4-461753 (in Vela Jr.); 3) 2XMM
J104608.7-594306 (in Homunculus); 4) XMMU J172054.5-372652; 5) CXOU
J181852.0-150213; 6) CXOU J232327.8+584842 (in Cas A). PSR
(including a high-B pulsar): 7) PSR J0205+6449 (in 3C 58); 8) PSR
J0357+3205, also known as ``Morla''; 9) PSR J0538+2817; 10) PSR
B1951+32 (in CTB 80); 11) PSR B2334+61; 12) PSR J1119-6127 (High-B).
XINS: 13) RX J0720.4-3125; 14) RX J1308.6+2127; 15) RX J1605.3+3249;
16) RX J1856.5-3754. A more in-depth analysis can be found in
Sec.~\ref{sec:numericalresults}.} \label{14feweak}
\end{figure}

\end{widetext}

Then we further show that the NFL quantum critical behavior enhances
not only the specific heat of neutrons but also the neutrino
emissivities of DU, MU, and NNB processes. Obviously, these two
effects are competitive. The enhancement of specific heat
decelerates the cooling, whereas the enhancement of neutrino
emissivity accelerates the cooling. The ultimate fate of the cooling
trajectory is determined by the complicated interplay of these two
opposite trends. Our simulations indicate that the thermal
relaxation of the crust is dramatically slowed down compared with
the minimal cooling paradigm \cite{Page04, Page06, Page09} and the
fast cooling paradigm \cite{Page92, Pethick92}.

We perform extensive simulations of the NS cooling curves after
incorporating the influence of the NFL behavior and depict the
results in Fig.~\ref{14feweak}. More detailed discussions on the
features of these cooling curves will be presented in
Sec.~\ref{sec:numericalresults}. For each NS, with No.~13 being the
only exception, there exists at least one cooling curve that matches
perfectly the observed cooling data within the $1\sigma$ confidence
interval error range when the NS mass, the maximum $^3P_2$-wave
neutron superfluid $T_{\mathrm{cn}}^{\mathrm{max}}$, and the
coupling constant $h$ take suitable values. The agreement between
our theoretical results and NS observations suggests that the
previously known cooling scenarios and the superfluid quantum
criticality can be integrated into a new cooling paradigm that
provides a more comprehensive understanding of the thermal evolution
of NSs.

The rest of the paper is organized as follows. In
Sec.~\ref{sec:modelqcp}, we derive an effective model of quantum
criticality. In Sec.~\ref{sec:nflbehavior}, we compute the neutron
damping rate and other quantities, and discuss the crossover between
NFL and FL behaviors. In Sec.~\ref{sec:nscooling}, we demonstrate
that the NFL behavior leads to logarithmic corrections to the heat
capacity and the total neutrino emissivity. In
Sec.~\ref{sec:numericalresults}, we compare our theoretical results
of NS cooling curves to astrophysical observations. In
Sec.~\ref{sec:summary}, we summarize the results and discuss some
future research projects. In Appendices \ref{sec:appa} and
\ref{sec:appb}, we present detailed calculations using perturbation
theory and RG theory, respectively. In Appendix \ref{sec:appc}, we
briefly analyze the property of thermal conductivity. In Appendix
\ref{sec:appd}, we discuss the relation between the $h$ parameter
and nuclear potential.

\section{Effective model of superfluid quantum criticality \label{sec:modelqcp}}

Shortly after the BCS theory of superconductivity \cite{BCS} was
developed, Bohr, Mottelson, and Pines \cite{Bohr58} proposed that
some peculiar features of finite nuclei can be explained by the
formation of Cooper pairs of nucleons. Migdal \cite{Migdal60} first
hypothesized the existence of neutron superfluid in a star composed
primarily of neutrons. In 1969, Baym \emph{et al.} \cite{Baym69}
proposed to attribute the glitch, which refers to the sudden change
of rotational period, observed in some NSs to the relative motion
between the normal fluid and superfluid as well as the pinning of
vortices. Since then, many theoretical efforts \cite{Dean03,
Lombardoreview, Pagereview, Sedrakian19, Baldo98} have been devoted
to computing the superfluid gap $\Delta_{\mathrm{n}}$ and the
superfluid transition temperature $T_{\mathrm{cn}}$. These studies
have unveiled \cite{Dean03, Lombardoreview, Pagereview, Sedrakian19,
Baldo98} the presence of $^{1}S_{0}$-wave superfluidity in the
lower-density regions of the crust and $^{3}P_{2}$-wave
superfluidity in the higher-density regions of the core. For each of
these two distinct pairing states, the transition temperature
$T_{\mathrm{cn}}$ exhibits a Gaussian-like density dependence, as
depicted in Fig.~\ref{fig:phasediagram}. Each pairing phase is
characterized by two critical densities, defining two separate QCPs.
Quantum criticality is expected to occur at all four QCPs. We can
select any one of these QCPs as a representative case to examine the
intriguing properties of quantum criticality. The findings are
applicable to the other QCPs.

Let us consider the left QCP ($\rho_{c1}$) shown in
Fig.~\ref{fig:phasediagram} as an example. According to the GLW
paradigm, the superfluid transition can be described by the
following field theory:
\begin{eqnarray}
S_{\Phi} = \int d\tau d^3 \mathbf{r}\left[\Phi^{\ast}
\left(\alpha-\partial^{2}_\tau-\nabla^2\right)\Phi +
\beta|\Phi^{\ast}\Phi|^{2}\right], \label{eq:Landautheory}
\end{eqnarray}
where $\Phi$ is superfluid order parameter. The coefficient $\beta$
of the quartic term is positive, whereas the mass $\alpha$ could be
positive, negative, or zero. When the density $\rho$ lies between
$\rho_{c1}$ and $\rho_{c2}$ such that $\alpha<0$, the ground state
of the system is located at $\langle \Phi \rangle =
\pm\sqrt{-\alpha/2\beta}$, indicating the presence of
$^{3}P_{2}$-wave neutron superfluidity. If the density $\rho <
\rho_{c1}$ with $\alpha>0$, the ground state is at $\langle \Phi
\rangle = 0$, corresponding to a normal FL ground state. At the QCP
of $\rho_{c1}$ where $\alpha=0$, the ground state is also at
$\langle\Phi\rangle = 0$. However, although $\langle\Phi\rangle = 0$
in both cases of $\alpha=0$ and $\alpha>0$, the QCP and the FL state
exhibit distinct properties. The distinction is most evident in the
features of the correlation length $\varsigma$, which characterizes
the spatial correlation between fluctuations of the order parameter.
At the QCP, the correlation length diverges according to the scaling
relation $\varsigma \propto |\alpha|^{-1/2}$
\cite{Vojta03,Sachdevbook}, which significantly enhances the quantum
fluctuations of the superfluid order parameter. In stark contrast,
when the system is away from the QCP, these quantum fluctuations are
substantially suppressed. This is because the correlation length
$\varsigma$ remains finite whenever $\alpha$ is finite
\cite{Vojta03,Sachdevbook}. In the language of condensed-matter
physics, the superfluid quantum fluctuations are gapless at the QCP
where $\alpha=0$, but they become gapped if the system is not at the
QCP, i.e., when $\alpha \neq 0$.

It is worth noting that the QCP $\rho_{c1}$ is located at the border
of the superfluid phase. At this QCP, Cooper pairs of neutrons are
actually already formed, but the neutron pairing gap
$\Delta_{\mathrm{n}}$ is extremely small. This has two consequences.
First, the Cooper pairs are fragile and can be readily broken by
quantum fluctuations. Second, it costs little energy to recombine
the nearly gapless neutrons into new Cooper pairs. In regions where
the density is close to $\rho_{c1}$, there are numerous small
droplets of Cooper pairs. Within each droplet, the Cooper pairs may
exhibit phase coherence. However, Cooper pairs in different droplets
are uncorrelated. Importantly, there is no long-range order at the
QCP, consistent with the characteristic $\langle \Phi \rangle = 0$.

The quantum fluctuation of the superfluid order parameter around its
expectation value $\langle \Phi \rangle$ can be described by a
collective bosonic mode, similar to how the phonons describe lattice
vibrations in a crystal. Conventionally, a scalar field
$\phi(\tau,\mathbf{r})$ is introduced to represent this bosonic
mode, defined as $\phi = \Phi - \langle \Phi \rangle$. To capture
the continuous breaking and reformation of Cooper-pair droplets, it
is customary to introduce a Yukawa-type coupling between the boson
field $\phi(\tau,\mathbf{r})$ and the neutron field
$\psi(\tau,\mathbf{r})$. At the QCP, the boson mode is gapless
because its effective mass $\alpha=0$, and the neutrons excited on
the Fermi surface are also gapless due to the vanishing of the
superfluid gap, i.e., $\Delta_{\mathrm{n}}(\rho_{c1})=0$. The
coupling between gapless bosons and gapless neutrons is strongly
peaked at zero boson momentum. According to the extensive research
experience accumulated in condensed-matter physics
\cite{Chubukov03AFMQCP, Rech06, Metlitski10, Metlitski102, Wang17,
Pan18, Wang20, Sachdev23}, such extreme forward scattering can
result in the breakdown of FL theory and the emergence of NFL
behavior. This NFL behavior would lead to singular corrections to
both the specific heat and the neutrino emissivity, thereby
affecting the NS cooling rate. In the FL and superfluid states, the
bosonic mode $\phi$ becomes gapped, which substantially weakens its
coupling to neutrons and prevents the NFL behavior. Moreover,
neutrons are also gapped in the superfluid state, further precluding
the presence of NFL behavior. Therefore, NFL behavior occurs only in
the close vicinity of superfluid QCP.

At finite temperatures, thermal fluctuations become important and
cooperate with superfluid quantum fluctuations. Their cooperation
broadens the single zero-$T$ QCP into a finite quantum critical
region on the $T$-$\rho$ plane depicted in
Fig.~\ref{fig:phasediagram}. NFL behavior exists in this region,
where the neutron's thermal energy $\sim k_{\mathrm{B}}T$ exceeds
the energy scale set by the boson mass $\alpha$. Thus, the width of
the quantum critical region is proportional to the thermal energy,
i.e., $\propto k_{\mathrm{B}}T$.

We emphasize that the mechanism of quantum criticality is
essentially different from the PBF scenario \cite{Flowers76,
Voskresensky87, Yakovlev99b}. NFL-type quantum critical phenomena
occur in the quantum critical region around the superfluid QCP,
whereas PBF processes originate from the thermal fluctuations and
exist only in the gapped superfluid phase. As a consequence of this
difference, quantum criticality affects the cooling process of an NS
for a much longer time than PBF processes, which occur only when the
NS becomes sufficiently cold. For a young NS, its interior is
composed of alternating FL and NFL layers. For an older NS, the FL,
NFL, and PBF layers coexist in the interior, as intuitively
illustrated in the right panel of Fig.~\ref{fig:phasediagram}. The
overall cooling history of NSs should be determined by the
cooperation of all the layers.


There are two routes to obtain the effective field theory of a
quantum criticality \cite{Vojta03, Sachdevbook}. One could begin
with a four-fermion type pairing interaction characterized by a
potential function $V(\mathbf{r})$, such as $V(\mathbf{r})
\psi^{\ast}_{\uparrow}\psi^{\ast}_{\downarrow}
\psi_{\downarrow}\psi_{\uparrow}$. One can introduce an auxiliary
bosonic field $\phi$ and then perform a standard
Hubbard-Stratonovich transformation \cite{Stratonovich58, Hubbard59}
to this interaction to convert it into a fermion-boson coupling,
following the procedure illustrated in Appendix \ref{sec:appd}.
Alternatively, one can write down a suitable quantum field theory on
generic grounds (e.g., unitarity, symmetry, etc.) to describe the
kinetics and dynamics of all the low-energy degrees of freedom. Such
a generic field theory usually contains many terms. Fortunately, the
RG theory can be applied to find out the terms that remain important
as the lowest energy limit is taken. Both routes are widely adopted
in condensed-matter physics. The first approach is difficult to
implement in practice because of the formal complexity of
$V(\mathbf{r})$. It appears more convenient to take the second
route.

Among the particles appearing in the interior of an NS, the
low-energy degrees of freedom related to superfluidity are the
neutrons (on Fermi surface), the mesons (origin of nuclear force),
and the collective boson (order-parameter quantum fluctuation).
Protons and electrons are bystanders of the superfluid
transition. The neutrons and the collective boson are gapless in the
quantum critical region, but the mesons are massive, with the
lightest pion having a mass of $\sim140~\mathrm{MeV}$. A
well-established notion of quantum field theory and condensed-matter
physics is that a particle plays a minor role at energies lower than
its mass/gap. According to this notion, the neutrons and the
collective boson are the dominant degrees of freedom for superfluid
quantum criticality, and the mesons play only a secondary role. At
present, we consider only the neutrons and the collective boson. The
effects of mesons will be discussed later.

The gapless neutrons and the gapless collective boson are equally
important at low energies and thus should be treated on an equal
footing. {We continue to regard the left QCP ($\rho_{c1}$) shown in
Fig.~\ref{fig:phasediagram} as an example.} Based on the elementary
rules of quantum field theory, the superfluid quantum criticality
can be modeled by the following action:
\begin{eqnarray}
S = S_{\psi}+S_{\phi}+S_{\phi^4}+S_{\psi\phi}.
\label{eq:qcpaction}
\end{eqnarray}
The free action for neutrons is
\begin{eqnarray}
S_{\psi} = \int\frac{d\omega}{2\pi} \frac{d^{3}\mathbf{k}}{(2\pi)^3}
\bar{\psi}(\omega,\mathbf{k})\left[-i\omega\gamma^{0} +
H_{\psi}(\mathbf{k})\right]\psi(\omega,\mathbf{k}),
\label{eq:actionfreeneutron}
\end{eqnarray}
which contains a Hamiltonian
\begin{eqnarray}
H_{\psi}(\mathbf{k}) = c_{f}\mathbf{k}\cdot\mathbf{\gamma}
-\mu\gamma^{0} + M_{n}.
\end{eqnarray}
Here, $\mu$ is the chemical potential, $M_{n}$ is the neutron mass, $c_{f}$
is the neutron velocity parameter, and $\gamma^{0,1,2,3}$ are Dirac
matrices. To describe thermal effects, we use $i\omega$ to denote
the Matsubara frequency (energy). The spinor field is
$\psi=(\psi_{+}, \psi_{-})^T=(\psi_{+\uparrow}, \psi_{+\downarrow},
\psi_{-\uparrow}, \psi_{-\downarrow})^T$ for
particle$(+)$, antiparticle$(-)$, spin-up$(\uparrow)$,
and spin-down$(\downarrow)$ degrees of freedom. Its conjugate
is $\bar{\psi} = \psi^{\dag}\gamma^0$. The free action of the
collective boson is
\begin{eqnarray}
S_{\phi} = \int\frac{d\Omega}{2\pi} \frac{d^3
\mathbf{q}}{(2\pi)^3}\phi^{\ast}[\Omega^2 + c^{2}_{b}\mathbf{q}^{2} +
c^{4}_{b}m^{2}_{b}]\phi, \label{eq:actionfreeboson1}
\end{eqnarray}
where $c_{b}$ is the boson velocity and $m_{b}$ is the boson mass.
The self-coupling of the boson takes the form
\begin{eqnarray}
S_{\phi^4} &=& \frac{\lambda}{4}\int\prod^4_{i=1}
\int\frac{d\Omega_i}{2\pi}\frac{d^3\mathbf{q}_{i}}{(2\pi)^3}
\delta(\Omega_1+\Omega_3-\Omega_2-\Omega_4) \nonumber \\
&& \times \delta^3(\mathbf{q}_{1} + \mathbf{q}_{3} - \mathbf{q}_{2}
- \mathbf{q}_{4})|\phi^{\ast}\phi|^2,
\label{eq:phifouraction}
\end{eqnarray}
where $\lambda$ is its coupling constant. The boson mass parameter
is $r=c^{4}_{b}m^{2}_{b}$, with $r=0$ defining the superfluid QCP.
For concreteness, we assume $r>0$. The results are also valid for
$r<0$. The Yukawa-type fermion-boson coupling reads
\begin{eqnarray}
S_{\psi\phi} &=& h\int\prod^2_{i=1}\int\frac{d\omega_i}{2\pi}
\frac{d^3\mathbf{k}_{i}}{(2\pi)^3}\frac{d\Omega}{2\pi}
\frac{d^3\mathbf{q}}{(2\pi)^3}\delta(\omega_1+\omega_2 -
\Omega)\nonumber \\
&& \times \delta^3(\mathbf{k}_{1}+\mathbf{k}_{2}-\mathbf{q})
\left[\phi^{\ast}\psi^{T}\left(i\gamma^{2}\gamma^{0}\Upsilon\right)
\psi + \mathrm{H.c.}\right],\nonumber \\
\end{eqnarray}
where $h$ is the coupling parameter. There are various possible
structures of the matrix $\Upsilon$. For instance, the gap matrix
$\Upsilon = \gamma^{5} = i\gamma^{0}\gamma^{1}\gamma^{2}\gamma^{3}$
embodies the overall antisymmetry of the Cooper pair, and it also
corresponds to the even-parity, spin-singlet pairing, a state in
which the fermions with the same chirality form a Cooper pair.

The above action is apparently not the most generic form. In
principle, one might consider some additional self-coupling terms,
such as $(\bar{\psi}\psi)^{2}$ and $(\phi^{\ast}\phi)^{6}$.
Moreover, the interactions between neutrons and various mesons are
not included in the action. We will explain later why these
additional terms can be safely neglected.

\begin{figure}[htbp]
\centering
\includegraphics[width=2.6in]{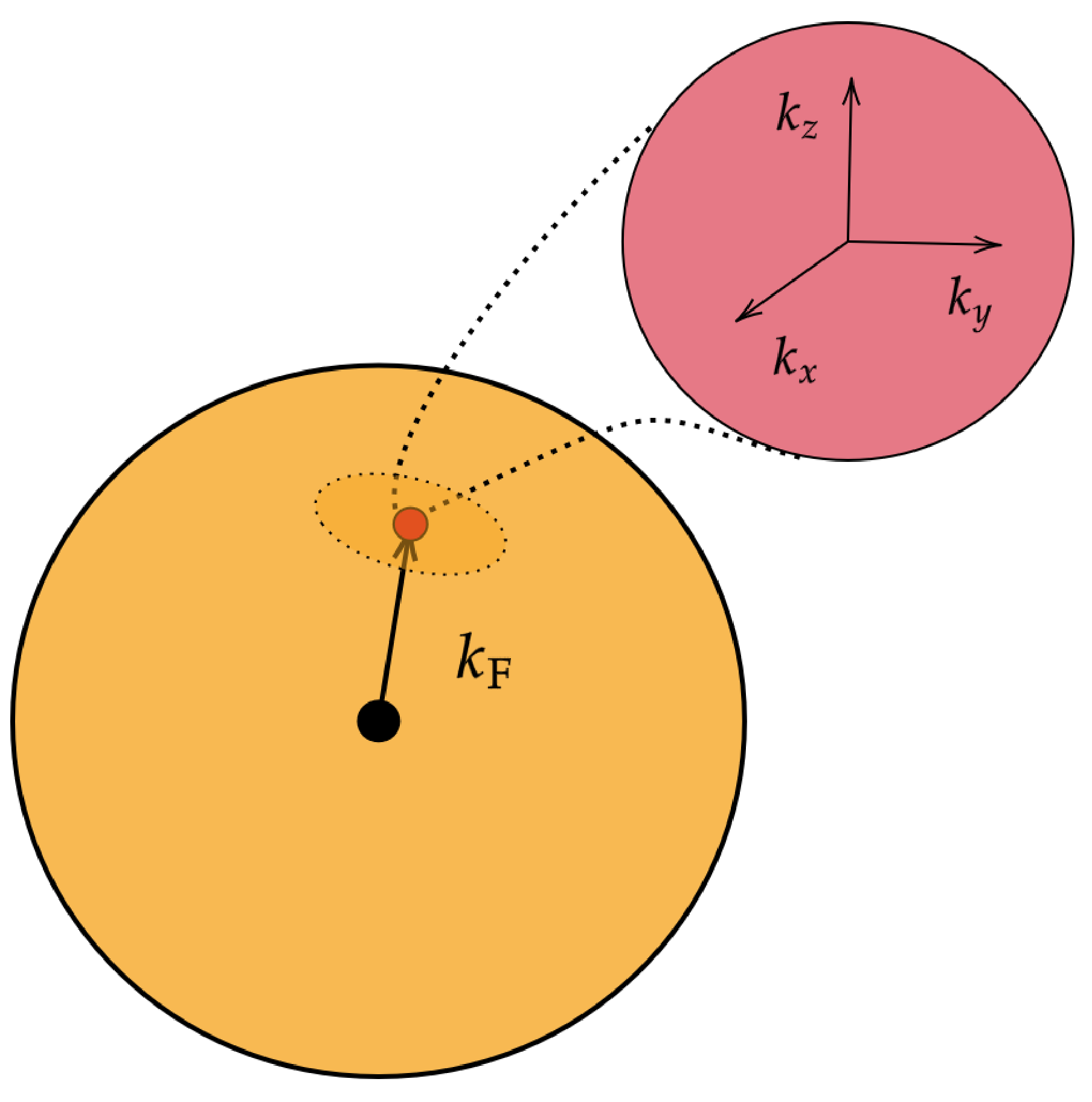}
\caption{Local coordinate frame at one given point on the Fermi
surface. Scattering happens only between the neutrons inside a small
patch around each point.} \label{fig:Fermisurface}
\end{figure}

The free neutron and boson propagators are
\begin{eqnarray}
G_{\mathrm{n}0}(\omega,\mathbf{k}) &=& \frac{1}{-i\omega \gamma^0 +
c_{f}\mathbf{k}\cdot\mathbf{\gamma}-\mu\gamma^0+M_{n}} \\
D_{0}(\Omega,\mathbf{q}) &=& \frac{1}{\Omega^2+c_{b}^{2}
\mathbf{q}^2}.
\end{eqnarray}
The chemical potential $\mu = \sqrt{c^2_{f}k^{2}_{\mathrm{F}} +
c^{4}_{f}M_{n}^2}$, where $k_{\mathrm{F}}$ is the Fermi momentum of
neutrons, enters into the neutron propagator.
{In principle, the chemical potential
should be temperature dependent. In the low-$T$ region, the neutron
chemical potential $\mu(T)$ can be expressed as $\mu(T)\approx \mu
\left[1 - O\left[\left(k_{\mathrm{B}}T/\mu\right)^2\right]\right]$
via the Sommerfeld expansion \cite{Ashcroftbook}. The zero-$T$ value
$\mu \approx 1000~\mathrm{MeV}$ is much greater than the
characteristic temperature ($\approx 0.01-0.1~\mathrm{MeV}$) within
NSs. This implies that
$O\left[\left(k_{\mathrm{B}}T/\mu\right)^2\right]$ is extremely
small and can be neglected. It is therefore justified to approximate
$\mu(T)$ by a constant $\mu$.}

In the presence of $\mu$, performing field-theoretical calculations
becomes challenging, since the conventional Feynman integral trick
is not directly applicable. To facilitate calculations, we now adopt
a suitable approximation to the model of quantum criticality. It is
important to note that the free boson propagator behaves like
$D_{0}(\Omega,\mathbf{q}) \sim 1/\mathbf{q}^2$ in the static limit
$\Omega \rightarrow 0$. Apparently, the fermion-boson interaction is
primarily dominated by scattering processes with $|\mathbf{q}|=0$ at
the QCP. This is closely analogous to the extreme forward-scattering
realized in many (e.g., nematic, antiferromagnetic) quantum critical
systems \cite{Hertz, Millis, Sachdev00, Vojta03, Chubukov03AFMQCP,
Coleman05, Loehneysen07, Sachdevbook, Keimer15, Chubukov04FMQCP,
Chubukov04FM, Rech06, Lee07, Metlitski10, Metlitski102, Liu12,
Grover14, Maciejko16, Zerf16, Wang17, Liu18, Pan18, Liu19, Wang20,
Sachdev23}. Moreover, it shares similarities with the U(1)
electron-gauge-boson coupling in two-dimensional metals with a
finite Fermi surface \cite{Holstein, Lee89, Lee92, Polchinski,
Nayak94, Altshuler94, Lee09, Lee18}, which serves as an effective
theory of high-$T_{c}$ superconductors \cite{Lee89, Lee92,
Polchinski, Nayak94, Altshuler94}. The methods developed in the
extensive studies of these systems can be applied to handle the
action given by Eq.~(\ref{eq:qcpaction}). As demonstrated in
Refs.~\cite{Polchinski, Nayak94, Esterlis21}, this kind of
interaction scatters fermions from a given point on the Fermi
surface to a neighboring point. To treat such interactions, it is
convenient to establish a local coordinate frame at a specific point
on the Fermi surface, as illustrated in Fig.~\ref{fig:Fermisurface}.
It is sufficient to consider fermions within a small patch around
this point. Fermions in different patches are nearly independent, as
scattering processes involving large transferred momenta are
significantly suppressed.

In the noninteracting limit, the neutron spectrum is
\begin{eqnarray}
\varepsilon(\mathbf{k})=
\begin{cases}
+\sqrt{c^2_{f}\mathbf{k}^2 +
c^4_{f}M_{n}^2} - \mu~~~\mathrm{particles}, \\
-\sqrt{c^2_{f}\mathbf{k}^2+c^4_{f}M_{n}^2} +
\mu~~~\mathrm{particle~holes},\\
-\sqrt{c^2_{f}\mathbf{k}^2+c^4_{f}M_{n}^2} -
\mu~~~\mathrm{antiparticles}, \\
+\sqrt{c^2_{f}\mathbf{k}^2 + c^4_{f}M_{n}^2} +
\mu~~~\mathrm{antiparticle~holes}.\nonumber
\end{cases}
\end{eqnarray}
For a neutron near the Fermi surface, its momentum in the
$z$ direction can be written, according to
Fig.~\ref{fig:Fermisurface}, as $k^{~}_{\mathrm{F}}+k_{z}$, where
$k_{z}$ is extremely small. The other two components $k_{x}$ and
$k_{y}$ are also small (since the patch is small), i.e., $k_{x,y}\ll
k^{~}_{\mathrm{F}}$, but usually larger than $k_{z}$. The particle
spectrum can be approximately handled as follows
\begin{eqnarray}
\varepsilon_{+}(\mathbf{k}) &=& \sqrt{c^2_{f}k^2_x +
c^{2}_{f}k^{2}_{y}+c^{2}_{f}(k^{~}_{\mathrm{F}} + k_{z})^{2} +
c^4_{f}M_{n}^{2}} \nonumber \\
&& -\sqrt{c^2_{f}k^{2}_{\mathrm{F}}+c^{4}_{f}M_{n}^{2}} \nonumber \\
&\approx& \sqrt{c^2_{f}k^{2}_{x}+c^2_{f}k^{2}_{y} +
2c^2_{f}k^{~}_{\mathrm{F}}k_{z}+c^2_{f}k^{2}_{\mathrm{F}}
+c^4_{f}M_{n}^{2}} \nonumber \\
&& -\sqrt{c^2_{f}k^{2}_{\mathrm{F}}+c^{4}_{f}M_{n}^{2}} \nonumber \\
&=& \sqrt{c^2_{f}k^{2}_{\mathrm{F}}+c^{4}_{f}M_{n}^{2}} \nonumber \\
&& \times \left(\sqrt{1+\frac{c^2_{f}k^{2}_{x} + c^2_{f}k^{2}_{y} +
2c^{2}_{f}k^{~}_{\mathrm{F}}k_{z}}{c^2_{f}k^{2}_{\mathrm{F}}
+ c^4_{f}M_{n}^{2}}}-1\right) \nonumber \\
&\approx& \frac{c^2_{f}}{2\mu}(k^{2}_{x}+k^{2}_{y}) +
\frac{c^2_{f}k^{~}_{\mathrm{F}}}{\mu}k_{z},
\end{eqnarray}
where the inequality $c^2_{f}k^{2}_{x}+c^{2}_{f}k^{2}_{y} +
2c^{2}_{f}k^{~}_{\mathrm{F}}k_{z} \ll \mu^{2}$ is used. The energy
spectra for the remaining three cases can be handled similarly. Then
the neutron dispersion has a new expression
\begin{eqnarray}
\varepsilon(\mathbf{k})=
\begin{cases}
+\frac{c^{2}_{f}}{2\mu}(k^{2}_{x}+k^{2}_{y}) +
\frac{c^{2}_{f}k^{~}_{\mathrm{Fn}}}{\mu}k_{z}
~~~~~~~~\mathrm{particles}, \\
-\frac{c^{2}_{f}}{2\mu}(k^{2}_{x}+k^{2}_{y}) -
\frac{c^{2}_{f}k^{~}_{\mathrm{F}}}{\mu}k_{z}
~~~~~~~~~\mathrm{particle~holes},\\
-\frac{c^{2}_{f}}{2\mu}(k^{2}_{x}+k^{2}_{y}) -
\frac{c^{2}_{f}k^{~}_{\mathrm{F}}}{\mu}k_{z} - 2\mu
~~\mathrm{antiparticles}, \\
+\frac{c^{2}_{f}}{2\mu}(k^{2}_{x}+k^{2}_{y}) +
\frac{c^{2}_{f}k^{~}_{\mathrm{F}}}{\mu}k_{z} +
2\mu~~\mathrm{antiparticle~holes}. \nonumber
\end{cases}
\label{eq:dispersion}
\end{eqnarray}

Due to the Pauli's exclusion principle, the antiparticle states can
only be excited beyond an exceedingly high energy threshold. This
allows us to ignore their influence on the particle states. We thus
arrive at the following form of the free neutron action
\begin{eqnarray}
S_{\psi_{+}} &=& \sum_{\sigma=\uparrow,\downarrow}\int
\frac{d\omega}{2\pi}\frac{d^3\mathbf{k}}{(2\pi)^3}
\psi^\ast_{+\sigma}[-i\omega+H_{\psi_+}(\mathbf{k})]
\psi_{+\sigma} \nonumber \\
&=& \int\frac{d\omega}{2\pi}\frac{d^3\mathbf{k}}{(2\pi)^3}
\psi^\ast_{+}[-i\omega+H_{\psi_+}(\mathbf{k})]\psi_{+},
\label{eq:neutronfreeaction}
\end{eqnarray}
where
\begin{eqnarray}
H_{\psi_+}(\mathbf{k})=\frac{c^2_{f}}{2\mu}\left(k^{2}_{x} +
k^{2}_{y}\right) + \frac{c^2_{f}k^{~}_{\mathrm{F}}}{\mu}k_{z}.
\label{eq:hamiltonianhplus}
\end{eqnarray}
The new effective free fermion propagator has the form
\begin{eqnarray}
G_{+}(\omega, \mathbf{k}) = \frac{1}{-i\omega +
\frac{c^2_{f}}{2\mu}\left(k^{2}_{x} + k^{2}_{y}\right) +
\frac{c^2_{f}k^{~}_{\mathrm{F}}}{\mu}k_z}.
\label{eq:freeneutrongplus}
\end{eqnarray}
The action of the Yukawa coupling is also changed. In NSs, the
superfluid gap has three possible structures \cite{Hoffberg70,
Tamagaki70}: $^1S_0$-wave gap, $^3P_2~(m_J=\pm2)$-wave gap, and
$^3P_2~(m_J=0)$-wave gap. The corresponding order parameters
\cite{Hoffberg70, Tamagaki70} are
\begin{eqnarray}
\Delta_{S}&=&\begin{bmatrix}0 & 1 \\ 1 & 0\end{bmatrix},\\
\Delta_{P,\pm 2}&=&\begin{bmatrix}-\frac{1}{\sqrt{2}}
\left(\hat{q}_{x}+i\hat{q}_{y}\right) & 0 \\ 0 &\frac{1}{\sqrt{2}}
\left(\hat{q}_{x}-i\hat{q}_{y}\right)\end{bmatrix},\\
\Delta_{P,0} &=& \begin{bmatrix}\frac{1}{\sqrt{2}}
\left(\hat{q}_{x}-i\hat{q}_{y}\right) & \sqrt{2}\hat{q}_{z} \\
\sqrt{2}\hat{q}_{z} & -\frac{1}{\sqrt{2}} \left(\hat{q}_{x} +
i\hat{q}_{y}\right)\end{bmatrix}.
\end{eqnarray}
The Yukawa-type fermion-boson couplings $S_{\psi_+\phi}$ are defined
as follows
\begin{eqnarray}
S_{\psi_+\phi}^{S} &=& h\prod^{2}_{i=1}\int\frac{d\omega_i}{2\pi}
\frac{d^{3} \mathbf{k}_{i}}{(2\pi)^3}\frac{d\Omega}{2\pi}
\frac{d^{3}\mathbf{q}}{(2\pi)^3} \nonumber \\
&& \times \delta\left(\omega_1+\omega_2-\Omega\right)
\delta^3\left(\mathbf{k_1}+\mathbf{k_2}-\mathbf{q}\right)\nonumber \\
&& \times \big[\phi^\ast\psi_{+\downarrow}\psi_{+\uparrow} -
\phi^\ast\psi_{+\uparrow}\psi_{+\downarrow} \nonumber \\
&& +\phi\psi^\ast_{+\uparrow}\psi^{\ast}_{+\downarrow}-
\phi\psi^\ast_{+\downarrow}\psi^{\ast}_{+\uparrow}\big],
\label{eq:yukawaactionswave} \\
S_{\psi_+\phi}^{P,\pm 2} &=&
h\prod^2_{i=1}\int \frac{d\omega_i}{2\pi}
\frac{d^3\mathbf{k}_{i}}{(2\pi)^3} \frac{d\Omega}{2\pi}
\frac{d^3\mathbf{q}}{(2\pi)^3} \nonumber \\
&& \times \delta\left(\omega_{1}+\omega_{2}-\Omega\right)
\delta^3\left(\mathbf{k_1}+\mathbf{k_2}-\mathbf{q}\right)\nonumber
\\
&& \times\big[\left(-\hat{q}_{x}+i\hat{q}_{y}\right)\phi^{\ast}
\psi_{+\uparrow}\psi_{+\uparrow} \nonumber \\
&& + \left(\hat{q}_{x}+i\hat{q}_y\right)\phi^{\ast}
\psi_{+\downarrow}\psi_{+\downarrow} \nonumber \\
&& +\left(-\hat{q}_{x}-i\hat{q}_{y}\right)\phi
\psi^{\ast}_{+\uparrow}\psi^{\ast}_{+\uparrow} \nonumber \\
&& +\left(\hat{q}_{x}-i\hat{q}_y\right)\phi
\psi^{\ast}_{+\downarrow}\psi^{\ast}_{+\downarrow}\big],
\label{eq:yukawaactionp2wave} \\
S_{\psi_+\phi}^{P,0} &=& h\prod^2_{i=1}\int\frac{d\omega_{i}}{2\pi}
\frac{d^3\mathbf{k}_{i}}{(2\pi)^{3}}\frac{d\Omega}{2\pi}
\frac{d^3\mathbf{q}}{(2\pi)^3} \nonumber \\
&& \times \delta\left(\omega_{1}+\omega_{2}-\Omega\right)\delta^{3}
\left(\mathbf{k_1}+\mathbf{k_2}-\mathbf{q}\right)\nonumber \\
&& \times \big[\left(\hat{q}_{x}+i\hat{q}_{y}\right)\phi^{\ast}
\psi_{+\uparrow}\psi_{+\uparrow} \nonumber \\
&& +\left(-\hat{q}_{x}+i\hat{q}_{y}\right)\phi^{\ast}
\psi_{+\downarrow}\psi_{+\downarrow} \nonumber\\
&& +2\hat{q}_z\phi^{\ast} \left(\psi_{+\uparrow}\psi_{+\downarrow}
+\psi_{+\downarrow}\psi_{+\uparrow}\right) \nonumber \\
&& +\left(\hat{q}_x-i\hat{q}_y\right)\phi\psi^{\ast}_{+\uparrow}
\psi^{\ast}_{+\uparrow} \nonumber \\
&& +\left(-\hat{q}_x-i\hat{q}_y\right)\phi \psi^{\ast}_{+\downarrow}
\psi^{\ast}_{+\downarrow} \nonumber \\
&& +2\hat{q}_z\phi\left(\psi^{\ast}_{+\uparrow}
\psi^{\ast}_{+\downarrow}
+\psi^{\ast}_{+\downarrow}\psi^{\ast}_{+\uparrow}\right)\big].
\label{eq:yukawaactionp0wave}
\end{eqnarray}
Here, we have defined
\begin{eqnarray}
\hat{\mathbf{q}} &=& \frac{\mathbf{q}}{|\mathbf{q}|} = (\hat{q}_{x},
\hat{q}_{y}, \hat{q}_{z}), \nonumber \\
\hat{q}_{x} &=& \sin\theta \sin\phi, \nonumber \\
\hat{q}_{y} &=& \sin\theta \cos\phi, \nonumber \\
\hat{q}_{z} &=& \cos\theta.\nonumber
\end{eqnarray}
in spherical coordinates. Now, the effective action of superfluid
quantum criticality becomes $S = S_{\psi_{+}} + S_{\phi} +
S_{\phi^{4}} + S_{\psi_+\phi}$, where $S_{\psi_{+}}$, $S_{\phi}$,
and $S_{\phi^{4}}$ are given by Eq.~(\ref{eq:neutronfreeaction}),
Eq.~(\ref{eq:actionfreeboson1}), and Eq.~(\ref{eq:phifouraction}),
respectively, and $S_{\psi_+\phi}$ is given by one of
Eqs.~(\ref{eq:yukawaactionswave}-\ref{eq:yukawaactionp0wave}).

\section{Non-Fermi liquid behavior \label{sec:nflbehavior}}

The gravitational collapse of NSs is prevented by the degenerate
pressure of neutron matter. In previous works, the neutron matter is
treated as an ordinary FL. In the noninteracting limit, the
neutrons constitute an ideal quantum gas, which has a sharp Fermi
surface separating the empty and occupied states. The specific heat
(i.e., the heat capacity per unit volume) exhibits the following
$T$ dependence
\begin{eqnarray}
c_{\mathrm{n}} = \frac{M_{n}k^{~}_{\mathrm{F}}
k^{2}_{\mathrm{B}}T}{3\hbar^{3}}, \label{eq:flcn}
\end{eqnarray}
where $M_{n}$ is the bare neutron mass, $k_{\mathrm{B}}$ is the
Boltzmann constant, and $\hbar$ is the reduced Planck constant. When
the nuclear force, mediated by the exchange of massive mesons, is
considered, the ideal neutron gas is turned into a neutron FL. The
FL still has a well-defined Fermi surface and its specific heat
continues to display a linear $T$ dependence. The only difference is
that the bare mass $M_{n}$ is replaced by an effective mass
$M^{\ast}_{n}$, which is a function of the baryon density $\rho_b$
and depends sensitively on the EOS. At the saturation nuclear
density, $M^{\ast}_{n} = 0.6-0.8 M_{n}$ \cite{Page04} due to the
renormalization from neutron-meson and/or neutron-neutron
interactions in the FL state. In our subsequent calculations of
cooling curves, we will use Akmal-Pandharipande-Ravenhall (APR) EOS
\cite{Akaml98} along with the corresponding values of
$M^{\ast}_{n}$.

The FL behavior may be fundamentally altered by the superfluid
quantum criticality. It is well established in condensed-matter
physics that the quantum critical fluctuations of some (e.g.,
ferromagnetic, antiferromagnetic) order parameter can destroy the FL
behavior and yield a singular correction to the linear-$T$ specific
heat. We now examine whether similar singular corrections are
induced by the superfluid quantum fluctuations.

\subsection{Perturbative calculation \label{sec:perturbative}}

To study the fate of FL behavior, we need to first compute the
neutron self-energy $\Sigma(i\omega,\mathbf{p})$ and then use it to
calculate the quasiparticle residue $Z_{f}$ as well as other
quantities. The critical neutron-boson coupling could be quite
strong, and the coupling parameter $h$ may be large. To put
calculations under control, we do not take $h$ as the expansion
parameter. Instead, we perform series expansion in powers of
$1/N$, where $N$ is an effective neutron flavor. Consider the
$^{3}P_2~(m_J=0)$-wave pairing as an illustrative example. Its
neutron-boson coupling term has two parts:
$(\hat{q}_{x}+i\hat{q}_{y})\phi^{\ast}\psi_{+\uparrow}
\psi_{+\uparrow}+(\hat{q}_{x}-i\hat{q}_{y})\phi\psi^{\ast}_{+\uparrow}
\psi^{\ast}_{+\uparrow} + 2\hat{q}_{z}(\phi^{\ast}\psi_{+\uparrow}
\psi_{+\downarrow} +\phi\psi^{\ast}_{+\downarrow}
\psi^{\ast}_{+\uparrow})$ and $(-\hat{q}_{x}+i\hat{q}_{y})
\phi^{\ast}\psi_{+\downarrow}\psi_{+\downarrow} +
(-\hat{q}_{x}-i\hat{q}_{y})\phi \psi^{\ast}_{+\downarrow}
\psi^{\ast}_{+\downarrow} + 2\hat{q}_{z}(\phi^{\ast}
\psi_{+\downarrow}\psi_{+\uparrow} + \phi\psi^{\ast}_{+\uparrow}
\psi^{\ast}_{+\downarrow})$. During the loop-diagram computations,
each part contributes the same factor of $(1 + 3\cos^2\theta)$, and thus
the two parts lead to an identical contribution. We distinguish
these two parts by labeling them as spin-up and spin-down,
respectively. Then the effective neutron flavor is $N=2$. The same
analysis is applicable to the other two types of pairing gap.
For the convenience of applying the $1/N$ expansion technique, we
can rescale $c^{2}_{b}\rightarrow N c^{2}_{b}$ and $r\rightarrow
Nr$. Now the action for superfluid order parameter has been revised
\begin{eqnarray}
S_{\phi} = \int\frac{d\Omega}{2\pi} \frac{d^{3}
\mathbf{q}}{(2\pi)^3}\phi^{\ast}\left[\Omega^2 + Nc^{2}_{b} \mathbf{q}^2
+Nr\right]\phi. \label{eq:actionfreeboson}
\end{eqnarray}

According to the research experience of quantum criticality, the
low-energy dynamics of the critical boson mode (i.e., order
parameter fluctuation) is controlled by the boson self-energy, also
known as the polarization function, rather than the kinetic term. At
the one-loop level, the polarization function is defined as
\begin{eqnarray}
\Pi(i\Omega,\mathbf{q}) &=& Nh^{2}\int\frac{R(\theta)d\omega d^{3}
\mathbf{k}}{(2\pi)^{4}}G_{+}(\omega,\mathbf{k}) \nonumber \\
&& \times G_{+}(\omega+\Omega,\mathbf{k}+\mathbf{q}),
\end{eqnarray}
where a function $R(\theta)$ is introduced to accommodate the angle
dependence of three pairing gaps:
\begin{eqnarray}
R(\theta)=
\begin{cases}
1 & ^{1}S_0, \\
\sin^2\theta & ^{3}P_{2,\pm 2},\\
(1+3\cos^2\theta) & ^{3}P_{2,0}.\\
\end{cases}
\label{eq:Rtheta}
\end{eqnarray}
After performing straightforward computations (see Appendix
\ref{sec:appa} for details), we find that the polarization,
for small $\Omega$, has the form
\begin{eqnarray}
\Pi(i\Omega,\mathbf{q}) \approx N\gamma
\frac{|\Omega|}{\sqrt{q^2_x+q^2_y}},
\end{eqnarray}
where $\gamma=\frac{h^{2}\mu^{2}\Lambda}{2k^{~}_{\mathrm{F}}
c^{4}_{f}\pi^2}$ with $\Lambda$ being a UV cutoff. For very large
$\Lambda$, the polarization $\Pi(i\Omega,\mathbf{q})$ has the same
expression for three different pairing gaps.

Inserting $\Pi(i\Omega,\mathbf{q})$ into the Dyson equation
$\tilde{D}^{-1} = D_{0}^{-1} + \Pi$ leads to the renormalized boson
propagator
\begin{eqnarray}
\tilde{D}(\Omega,\mathbf{q};r) &=& \frac{1}{D_{0}^{-1}
(\Omega,\mathbf{q})+\Pi(i\Omega,\mathbf{q})+r} \nonumber \\
&\approx& \frac{1}{N}\frac{1}{c^{2}_{b}\mathbf{q}_{\perp}^2 + \gamma
\frac{|\Omega|}{|\mathbf{q}_{\perp}|}+r},
\label{eq:renormalizeddphi}
\end{eqnarray}
where $|\mathbf{q}_{\perp}|=\sqrt{q^{2}_{x}+q^{2}_{y}}$. Here, to
make our analysis generic, a finite $r$ is introduced. We have
omitted the boson kinetic term $\Omega^{2}$, since $\Omega^{2}\ll
\gamma|\Omega|/|\mathbf{q}_{\perp}|$ at low energies. Additionally,
the $q_z$ component has been disregarded as it is irrelevant
compared with the $|\mathbf{q}_{\perp}|$ dependence, especially
under the subsequent rescaling.

If $r=0$, $\tilde{D}(\Omega,\mathbf{q};r)$ is sharply peaked at
$\mathbf{q}_{\perp} = 0$ in the static limit $|\Omega| \rightarrow
0$. We show that this characteristic plays the key role in the
formation of NFL behavior. Away from the QCP, i.e., when $r \neq 0$,
the scattering processes with $\mathbf{q}_{\perp} = 0$ are strongly
suppressed and only produce ordinary FL behavior.

Notice that a factor $1/N$ appears in
$\tilde{D}(\Omega,\mathbf{q};r)$. The Feynman diagrams that contain
more boson lines are suppressed by powers of $1/N$. If $N$ is large,
the diagrams having less boson lines make more significant
contributions \cite{Altshuler94, Metlitski10, Metlitski102, Wang17,
Esterlis21}. This indicates that the leading contribution to the
neutron self-energy comes from the one-loop diagram. At the one-loop
level, the neutron self-energy is given by
\begin{eqnarray}
\Sigma(i\omega,\mathbf{k};r) &=& -h^2\int\frac{R(\theta)d\Omega
d^{3}q}{(2\pi)^{4}}\tilde{D}(\Omega,\mathbf{q};r) \nonumber \\
&& \times G_{+}(\omega-\Omega,\mathbf{k}-\mathbf{q}).
\label{eq:neutronselfenergy}
\end{eqnarray}
Now, to simplify calculations, we redefine $r$
in Eq.~(\ref{eq:renormalizeddphi}) as $r/|\mathbf{q}_{\perp}|$.
Based on the detailed derivation in Appendix
\ref{sec:appa}, we find that, for small $\omega$, this
self-energy is given by
\begin{eqnarray}
\Sigma(i\omega;r) &\approx& -\frac{i\mu h^{2}}{6N c^{2}_{b}
k^{~}_{\mathrm{F}}c^{2}_{f}\pi^{2}} \left(\int^{2\pi}_{0}
\frac{R(\theta)d\theta}{2\pi}\right)\nonumber \\
&& \times \mathrm{sign}(\omega)|\omega|\ln\left(\frac{c^{2}_{b}
\Lambda^3}{r+\gamma|\omega|}\right), \label{eq:sigmaomegar}
\end{eqnarray}
where the integration over the angle $\theta$ is
\begin{eqnarray}
\int^{2\pi}_0\frac{R(\theta) d\theta}{2\pi}=
\begin{cases}
1 & ^{1}S_0, \\
\frac{1}{2} & ^{3}P_{2,\pm 2},\\
\frac{5}{2} & ^{3}P_{2,0}.\\
\end{cases}
\label{eq:}
\end{eqnarray}
Obviously, this self-energy has the same $\omega$ dependence in the
three cases. For simplicity, we drop the coefficients and
concentrate on the $\omega$ dependence.


If the system is tuned to the QCP with $r=0$, the corresponding
self-energy becomes
\begin{eqnarray}
\Sigma(i\omega;r=0) \approx -i\omega\ln\left(\frac{{c^{2}_{b}
\Lambda^3}}{\gamma \sqrt{\omega^2}}\right).\label{eq:sigmaomegar0}
\end{eqnarray}
Now perform the analytical continuation $i\omega\rightarrow \omega +
i\epsilon$, where $\epsilon$ is an infinitesimal factor. Then
we obtain the retarded self-energy
\begin{eqnarray}
\Sigma_{R}(\omega;r=0) &\approx& -\omega\ln\left(\frac{{c^2_{b}
\Lambda^3}}{\gamma\sqrt{-\omega^2}}\right) \nonumber \\
&=& -\omega \ln\left(\frac{{c^2_{b}\Lambda^3}}{\gamma
\sqrt{\omega^{2} e^{i\pi}}}\right) \nonumber \\
&=& -\omega\left[\ln\left(\frac{{c^2_{b}\Lambda^3}}{\gamma |\omega|}
\right)+\ln\left(e^{-i\frac{\pi}{2}}\right)\right]\nonumber \\
&=& -\omega \ln\left(\frac{{c^2_{b}\Lambda^3}}{\gamma|\omega|}\right)
+ i\frac{\pi}{2}\omega.
\end{eqnarray}
The real and imaginary parts of this complex function can be readily
determined. Obviously, the imaginary part
$\mathrm{Im}\Sigma_{R}(\omega;r=0)$ is a linear function of
$\omega$.

Next we discuss the physical implication of the above results. It is
useful to first make a generic analysis of the criterion of FL
theory. If one inserts $\Sigma_{R}(\omega;r=0)$ into
Eq.~(\ref{eq:freeneutrongplus}) via the Dyson equation
$\tilde{G}^{-1}_{+R} = G^{-1}_{+R} + \Sigma_{R}$, one would get a
retarded renormalized neutron propagator
\begin{eqnarray}
\tilde{G}_{+R}(\omega) = \frac{1}{-\omega +
\mathrm{Re}\Sigma_{R}(\omega)+i\mathrm{Im}\Sigma_{R}(\omega) +
\ldots }, \label{eq:freeneutrongplusretarded}
\end{eqnarray}
where the momentum terms are not shown. Define an energy-dependent
function $\Gamma(\omega)$ as
\begin{eqnarray}
\Gamma(\omega) = \mathrm{Im}\Sigma_{R}(\omega).
\label{eq:dampingrate}
\end{eqnarray}
Then $\tilde{G}_{+R}(\omega)$ is rewritten as
\begin{eqnarray}
\tilde{G}_{+R}(\omega) = \frac{1}{-\omega +
\mathrm{Re}\Sigma_{R}(\omega) + i\Gamma(\omega) + \ldots }.
\label{eq:freeneutrondampingrate}
\end{eqnarray}
After Fourier transformation, this propagator exhibits the following
time dependence:
\begin{eqnarray}
\tilde{G}_{+R}(t) \propto e^{-i\left(\omega -
\mathrm{Re}\Sigma_{R}\right)t} e^{-\Gamma t}.
\end{eqnarray}
The single neutron state decays as the time $t$ grows if $\Gamma
\neq 0$. The function $\Gamma(\omega)$ is thus called the damping
rate or the decay rate of neutron quasiparticles. The quasiparticle
lifetime is proportional to $\Gamma^{-1}(\omega)$. The Pauli
exclusion principle guarantees that $\Gamma(\omega)$ goes to zero as
$\omega \rightarrow 0$, because the space of the final states into
which a neutron is scattered must vanish on the Fermi surface. The
$\omega$-dependence of $\Gamma(\omega)$ is model dependent. For
instance, previous calculations \cite{Giulianibook} have confirmed
that the screened short-range Coulomb interaction leads to
$\Gamma(\omega)\sim \omega^{2}$, whereas the electron-phonon
interaction gives rise to $\Gamma(\omega)\sim \omega^{3}$ in
three-dimensional normal metals. If a two-dimensional metal is tuned
to a nematic or ferromagnetic QCP \cite{Chubukov04FMQCP, Rech06,
Metlitski10, Esterlis21}, the quantum critical fluctuations of the
nematic or ferromagnetic order parameter yield $\Gamma(\omega)\sim
\omega^{2/3}$. In general, one can assume that the damping rate
exhibits a power-law behavior
\begin{eqnarray}
\Gamma(\omega)\sim \omega^{\eta},
\end{eqnarray}
where $\eta$ is a positive constant. Based on the Kramers-Kronig
relation \cite{Giulianibook}, the real part of the retarded
self-energy is computed as
\begin{equation}
\mathrm{Re}\Sigma_{R}(\omega) = \frac{1}{\pi}
\int_{-\infty}^{+\infty} d\omega^{\prime} \frac{\mathrm{Im}
\Sigma_{R}(\omega^{\prime})}{\omega^{\prime}-\omega},
\label{eqn:KKRelation}
\end{equation}
which then can be used to define the quasiparticle residue
$Z_{f}(\omega)$ as follows
\begin{eqnarray}
Z_{f}(\omega) = \frac{1}{1-\frac{\partial \mathrm{Re}
\Sigma_{R}(\omega)}{\partial \omega}}.\label{eq:zfdefinition}
\end{eqnarray}
In quantum many-body theory \cite{Giulianibook, Varma}, $Z_{f}$
plays a unique role: it measures the overlap between an interacting
fermion liquid and a noninteracting fermion gas. When $Z_{f}$ takes
a finite value, the system can be regarded as either an FL or an
ideal gas containing long-lived Landau quasiparticles. In contrast,
if $Z_{f} \rightarrow 0$ in the $\omega \rightarrow 0$ limit, the FL
theory breaks down and the system has no Landau quasiparticles. It
can be verified that $Z_{f}\neq 0$ if the exponent $\eta>1$ and that
$Z_{f}\rightarrow 0$ if $\eta\leq 1$. According to this criterion, the
Coulomb and electron-phonon interactions result in FL behavior,
whereas the nematic and ferromagnetic QCPs exhibit NFL behavior.

Let us take $^{3}P_2~(m_J=0)$ wave as an example. The neutron
damping rate is
\begin{eqnarray}
\Gamma(\omega) = \mathrm{Im}\Sigma_{R}(\omega)= \frac{5\mu
h^2}{24N c^{2}_{b}k^{~}_{\mathrm{F}}c^2_{f}\pi}\omega,
\label{eq:gammaomegaperturbative}
\end{eqnarray}
and the corresponding residue is
\begin{eqnarray}
Z_{f}^{-1} &=& 1-\frac{\partial\mathrm{Re}
\Sigma_{R}(\omega)}{\partial \omega} \nonumber \\
&=& 1+\frac{5\mu h^{2}}{12N c^{2}_{b}k^{~}_{\mathrm{F}}
c^2_{f}\pi^{2}}\frac{\partial}{\partial \omega}\left[\omega
\ln\left(\frac{{c^{2}_{b} \Lambda^{3}}}{\gamma
|\omega|}\right)\right] \nonumber \\
&\sim& \frac{5\mu h^{2}}{12N c^{2}_{b}k^{~}_{\mathrm{F}}
c^2_{f}\pi^{2}} \ln\left(\frac{{c^{2}_{b}
\Lambda^{3}}}{\gamma|\omega|}\right).
\end{eqnarray}
Apparently, $Z_{f}$ vanishes as $\omega \rightarrow 0$. This
conclusion also holds for the other two gap symmetries. Therefore,
the dense neutron system tuned to the vicinity of the superfluid QCP
should be identified as an NFL. The linear damping rate is
particularly noticeable as it is usually believed to be the minimal
violation of FL theory. Such a behavior is also known as marginal
FL behavior \cite{Varma}. Marginal FL behavior plays a pertinent role
in the understanding of the abnormal properties of normal-state of
high-$T_{c}$ copper-oxide superconductors \cite{Varma}.

If the system departs from the QCP and enters the disordered
(nonsuperfluid) phase, the boson has a finite mass with $r \neq 0$.
The self-energy $\Sigma(i\omega;r)$ given by
Eq.~(\ref{eq:sigmaomegar}) can be equivalently written as
\begin{eqnarray}
\Sigma(i\omega;r) \sim \mathrm{sign}(\omega)|\omega|
\left[-\ln\left(\frac{{c^{2}_{b} \Lambda^3}}{r}\right) +
\ln\left(1+\frac{\gamma|\omega|}{r} \right)\right].
\label{eq:sigmaomegar2}
\end{eqnarray}
It is interesting to analyze two limiting cases. First consider the
low-energy regime with $\gamma|\omega| \ll r$, where this function
has the form
\begin{eqnarray}
\Sigma(i\omega;r) \sim \mathrm{sign}(\omega)|\omega|
\left[-\ln\left(\frac{{c^{2}_{b}\Lambda^3}}{r}\right) +
\frac{\gamma|\omega|}{r}\right]. \label{eq:}
\end{eqnarray}
Making analytical continuation $i\omega \rightarrow \omega +
i\epsilon$ yields the retarded self-energy:
\begin{eqnarray}
\Sigma_{\mathrm{R}}(\omega;r) \sim -\omega\ln \left(\frac{{c^{2}_{b}
\Lambda^3}}{r}\right)-i\mathrm{sign}(\omega)
\frac{\gamma}{r}\omega^{2}. \label{eq:}
\end{eqnarray}
The damping rate $\Gamma(\omega;r) \sim \omega^{2}$, which is a
typical FL behavior. Then consider the opposite limit with
$\gamma|\omega| \gg r$. In such a limit, the self-energy of
Eq.~(\ref{eq:sigmaomegar}) becomes Eq.~(\ref{eq:sigmaomegar0}) and
NFL behavior exists at energies much higher than $r/\gamma$.

The above analysis indicates that NFL behavior occurs at all
energies at the superfluid QCP with $r=0$ and at energies higher
than the energy scale set by the ratio $r/\gamma$ for $r \neq 0$.
Such a departure from FL behavior will have a dramatic impact on the
bulk properties of the NSs. If $r$ is finite but negative, the
system enters the superfluid phase, which is fundamentally distinct
from the FL/NFL phase.

The above results are applicable to $T=0$. At finite temperatures,
the neutron has two energies: the single-particle energy $\omega$
and the thermal energy $\sim k_{\mathrm{B}}T$. For any finite $r$,
the NFL behavior is ruined as the total energy scale is lower than
$r/\gamma$. At a sufficiently high $T$, however, the thermal energy
$k_{\mathrm{B}}T$ is large enough to revive the NFL behavior,
regardless of the magnitude of $\omega$. This is the reason why the
zero-$T$ QCP is broadened into a $V$-shaped quantum critical region
on the $T$-$\rho$ plane. When the system is far away from the QCP
with the ratio $r/\gamma \gg k_{\mathrm{B}}T$, the NFL behavior is
entirely destroyed and gives its position to ordinary FL behavior.

\subsection{RG analysis \label{sec:rganalysis}}

Wilson's RG theory \cite{Wilson71} is one of the most powerful tools
for studying interacting many-particle systems. It has achieved
great success in the theoretical description of classical critical
phenomena \cite{Wilson71}, and also plays an essential role in the
exploration of quantum criticality \cite{Vojta03, Sachdevbook}.
Below, we provide an RG study of the effective field theory of
superfluid criticality. The RG results enable us to determine
the interaction corrections to all the model parameters. The damping
rate $\Gamma(\omega)$ and the residue $Z_{f}$ can also be computed
from RG results.

The NS interior contains several sorts of particles, which could
interact with each other in many possible ways. Needless to say,
studying all particles and all interactions at once is a formidable
task. Fortunately, such a task can be greatly simplified by noting that
most observable quantities are primarily governed by the long-time
and large-distance properties. Given the existence of time-energy
and distance-momentum correspondences, one could make an effort to find
the particles and their interactions that determine the
low-energy (which also represents the small-momentum) physics. RG
theory provides an ideal approach for such a manipulation
\cite{Shankar94}. The essence of RG theory is to integrate out the
degrees of freedom defined at high energies within the framework of
functional integrals. This operation would give rise to a relatively
simple effective model that adequately captures the low-energy
behaviors. The influence of high-energy degrees of freedom on
low-energy physics is embodied in a set of coupled RG flow
equations fulfilled by all the model parameters. The
interaction-induced many-body effects can be extracted from the RG
solutions.

We now define the Fermi velocity of the neutrons as
$v^{~}_{\mathrm{F}} = c^2_{f}k^{~}_{\mathrm{F}}/\mu$ and make the
rescaling transformations $c_b \phi \rightarrow \phi$,
$h/c_b\rightarrow h$, and $\lambda/c^{4}_{b}\rightarrow \lambda$. At
the superfluid QCP, the partition function of the system is
\begin{eqnarray}
Z=\int D\phi D\phi^\ast D\psi_{+} D\psi^\ast_{+}e^{-S}.
\end{eqnarray}
The total action can be separated into the free part and the
interaction part as follows
\begin{eqnarray}
S &=& S_{0} + S_{I}.
\end{eqnarray}
The free part $S_{0} = S_{\psi} + S_{\phi}$ has the form
\begin{eqnarray}
&& \int^{\Lambda}_0\frac{d\omega}{2\pi}\frac{d^{3}
\mathbf{k}}{(2\pi)^3}\psi^{\ast}_{+}\big[-i\omega +
\frac{v^{~}_\mathrm{F}}{2 k^{~}_{\mathrm{F}}}\left(k^{2}_{x} +
k^{2}_{y}\right)+v^{~}_\mathrm{F}k_{z}\big]\psi_{+}
\nonumber \\
&& +N\int^{\Lambda}_0\frac{d\Omega}{2\pi} \frac{d^{3}
\mathbf{q}}{(2\pi)^{3}}\phi^{\ast}\left[\mathbf{q}_{\perp}^{2} +
\gamma\frac{|\Omega|}{|\mathbf{q}_{\perp}|}\right]\phi.
\label{eq:rgfreeaction}
\end{eqnarray}
Here, the one-loop polarization is included in the free boson action
$S_{\phi}$ as it becomes more important than the kinetic term at
very low energies. The interaction term $S_{I} = S_{\phi^4} +
S_{\psi_{+}\phi}$, given in Sec.~\ref{sec:modelqcp}, remains
unchanged. The total action has four model parameters, including
$v^{~}_{\mathrm{F}}$, $\gamma$, $\lambda$, and $h$. They take
constant values at the highest energy scale $\Lambda$, beyond which
the effective theory of quantum criticality is no longer applicable,
but acquire an explicit dependence on the varying energy scale due
to the quantum corrections resulting from fermion-boson coupling.
Superfluid critical phenomena rely heavily on the scale dependence
of these parameters.

{A detailed RG analysis of the above
action is provided in Appendix \ref{sec:appb}. Here, we directly
present the flow equations of all the model parameters. We take the
case of $^{3}P_2~(m_J=0)$-wave pairing as an example to analyze the
results. It is straightforward to extend the analysis to the other
two cases. The RG flow equations of $Z_{f}$, $v^{~}_{\mathrm{F}}$,
$h$ $\gamma$ and $\lambda$ are given by}
\begin{eqnarray}
\frac{dZ_{f}}{dl} &=& -\frac{5h^2}{4N v^{~}_{\mathrm{F}}\pi^2} Z_{f}, \label{Zf1}\\
\frac{dv^{~}_{\mathrm{F}}}{dl} &=& -\frac{5h^2}{4N\pi^2},\label{vf1} \\
\frac{dh}{dl} &=&  -\frac{5}{4Nv^{~}_{\mathrm{F}}\pi^2}h^3,\label{hf1} \\
\frac{d\gamma}{dl} &=& \gamma, \label{gamma1}\\
\frac{d\lambda}{dl} &=& -2\lambda. \label{lambda1}
\end{eqnarray}
From Eq.(\ref{hf1}), we can obtain
\begin{eqnarray}
\frac{dh^2}{dl} = -\frac{5}{2Nv^{~}_{\mathrm{F}}\pi^2}h^4.
\label{hf2}
\end{eqnarray}
By combining Eqs (\ref{Zf1}), (\ref{vf1}), and (\ref{hf2}), we can
derive the analytical solutions:
\begin{eqnarray}
Z_{f}(l) &=& -\frac{c_3}{c_1l-\frac{4N}{5}\pi^2c_2},
\label{solution11}\\
v^{~}_{\mathrm{F}}(l) &=& \frac{\frac{4N}{5}\pi^2}{c_1
l-\frac{4}{5N}\pi^2c_2},
\label{solution22} \\
h^2(l) &=& \frac{\frac{16N^2}{25}\pi^4 c_1}{(c_1 l-\frac{4N}{5}
\pi^{2}c_{2})^2}, \label{solution33} \label{h0}
\end{eqnarray}
and we can directly obtain
\begin{eqnarray}
\gamma(l) &=& \gamma e^{l},\\
\lambda(l) &=& \lambda e^{-3l}.
\label{h0}
\end{eqnarray}
from Eqs.~(\ref{gamma1}) and (\ref{lambda1}). To determine the
unknown constants $c_1$, $c_2$ and $c_3$, we combine the initial
conditions $Z_{f}(l=0)=1$, $v^{~}_{\mathrm{F}}(l=0) =
v^{~}_{\mathrm{F}}$ and $h^2(l=0)=h^{2}$ with
Eqs.(\ref{solution11})-(\ref{solution33}), which yield
$c_1=h^{2}/v^{2}_{\mathrm{F}}$, $c_{2}=-1/v^{~}_{\mathrm{F}}$, and
$c_3=-\frac{4N}{5}\pi^{2}/v^{~}_{\mathrm{F}}$. Collecting all the
above results, we eventually obtain
\begin{eqnarray}
Z_{f}(l) &=& \frac{\frac{4N}{5}\pi^2 v^{~}_{\mathrm{F}}}{h^{2}l +
\frac{4N}{5}\pi^{2}v^{~}_{\mathrm{F}}}. \label{solution111}\\
v^{~}_{\mathrm{F}}(l) &=& \frac{\frac{4N}{5}\pi^{2}
v^{2}_{\mathrm{F}}}{h^{2}l + \frac{4N}{5} \pi^2
v^{~}_{\mathrm{F}}}, \label{solution222} \\
h^{2}(l) &=& \frac{\frac{16N^2}{25}\pi^4 h^{2} v^{2}_{\mathrm{F}
}}{(h^{2}l + \frac{4N}{5} \pi^{2}v^{~}_{\mathrm{F}})^2}.
\label{solution333}
\end{eqnarray}
{In the nonrelativistic limit, the
renormalized Fermi velocity $v^{~}_{\mathrm{F}}(l)$ can be
approximated as
\begin{eqnarray}
v^{~}_{\mathrm{F}}(l) = \frac{c^2_{f} k^{~}_{\mathrm{F}}}{
\sqrt{c^2_{f}k^{2}_{\mathrm{F}}+c^4_{f}M^{\ast2}_{n}(l)}} \approx
\frac{k^{~}_{\mathrm{F}}}{M^{\ast}_{n}(l)}.\label{nonrelavf}
\end{eqnarray}
The corrections to the Fermi velocity $v^{~}_{\mathrm{F}}$ from NFL
behavior are all incorporated into the effective neutron mass
$M^{\ast}_{n}$, whereas the Fermi momentum $k^{~}_{\mathrm{F}}$
remains unchanged. By substituting Eq.~(\ref{solution222}) into
Eq.~(\ref{nonrelavf}), we obtain the NFL-corrected effective neutron
mass $M^{\ast}_{n}(l)$, which depends on $l$ as follows
\begin{eqnarray}
M^{\ast}_{n}(l) = \frac{k^{~}_{\mathrm{F}}(h^{2}l+\frac{4N}{5}\pi^{2}
v_{\mathrm{F}})}{\frac{4N}{5}\pi^{2}v^{2}_{\mathrm{F}}}.
\label{solutionmmm}
\end{eqnarray}}

It is evident that the NFL-corrected mass $M^{\ast}_{n}(l)$ and the
effective mass $M^{\ast}_{n}$ are distinct. Within the framework of
quantum hadrodynamics, the effective mass $M^{\ast}_{n}$ primarily
originates from the neutron-meson interaction \cite{Walecka74,
Zhu24}. This interaction induces fermion-antifermion pairing
\cite{Garani22}, described by nonzero $\langle \bar{\psi}\psi
\rangle$, modifying the bare neutron mass $M_{n}$ to the effective
neutron mass $M^{\ast}_{n}$. In contrast, the NFL-corrected mass
$M^{\ast}_{n}(l)$ arises from the coupling of neutrons to the boson
mode that represents the superfluid quantum fluctuation. In the
limit $l=0$, $M^{\ast}_{n}(l)$ is reduced to $M^{\ast}_{n}$.

One can see from Eq.~(\ref{solution333}) that the fermion-boson
coupling parameter $h(l) \sim l^{-1}$ in the long-wavelength limit
$l\rightarrow \infty$. Thus, $h$ approaches zero in the
low-energy regime. However, the vanishing of $h$ does not mean that
the superfluid quantum fluctuations play a negligible role at low
energies. The reason is that the importance of the fermion-boson
coupling is indeed determined by the ratio of the potential energy
over the kinetic energy, rather than the potential energy alone.
Notice that the kinetic energy is a decreasing function of the
energy, since the velocity $v^{~}_{\mathrm{F}}(l) \sim l^{-1}$ for
large $l$. To assess the impact of the fermion-boson coupling, we need
to examine the asymptotic behavior of the residue $Z_{f}$. As shown
in Eq.~(\ref{solution111}), for large values of $l$ the residue
$Z_{f}(l)$ behaves as
\begin{eqnarray}
Z_{f}(l) \sim l^{-1}. \label{eq:zfl}
\end{eqnarray}
The vanishing of $Z_{f}$ in the lowest energy limit signals the
breakdown of FL theory and the absence of long-lived quasiparticles.

The above solution of $Z_{f}(l)$ can be used to calculate the real
part of the retarded neutron self-energy $\mathrm{Re}\Sigma_R(\omega)$
based on the relation given by Eq.~(\ref{eq:zfdefinition}). The
length scale $l$ and the energy $\omega$ are related via the
following scaling transformation
\begin{eqnarray}
\omega(l) = \omega_{0} e^{-2l},\label{eq:omegalrelation}
\end{eqnarray}
where $\omega_{0}$ stands for some high-energy scale. Making use of
Eq.~(\ref{eq:zfdefinition}), Eq.~(\ref{eq:zfl}), and
Eq.~(\ref{eq:omegalrelation}), we find that
\begin{eqnarray}
\mathrm{Re}\Sigma_R(\omega) \sim \omega
\ln\left(\frac{\omega_0}{\omega}\right).
\end{eqnarray}
By virtue of the Kramers-Kronig relation, one can readily verify that the
neutron damping rate exhibits a linear dependence on $\omega$,
namely
\begin{eqnarray}
\Gamma(\omega) = \mathrm{Im}\Sigma_{R}(\omega) \sim \omega.
\end{eqnarray}
This NFL behavior is well consistent with the perturbative result
Eq.~(\ref{eq:gammaomegaperturbative}). Such an excellent agreement
gives us confidence that the emergence of NFL behavior is a robust
property of superfluid quantum criticality.

Now we remark on the role played by some additional terms. Suppose
that the action contains such a neutron self-coupling term as
$(\psi_{+}^{\ast}\psi_{+})^{1+n}$ with $n$ being a positive integer,
which describes the contact repulsive interaction of neutrons. RG
calculations reveal that its coupling parameter vanishes quickly as
the energy is lowered for any value of $n$. Similar arguments can be
used to prove that all the higher boson self-interactions
$(\phi^{\ast}\phi)^{2+n}$ and all the higher fermion-boson
interactions $\phi^{1+n}(\psi_{+}^{\ast}
\psi_{_{+}}^{\ast})^{1+n^{\prime}}$ are irrelevant perturbations at
low energies and can be neglected.

In realistic neutron matter, the neutrons are coupled to several
types of mesons, including $\pi$, $\sigma$, $\omega$, $\rho$,
etc. The nuclear force mediated by such mesons can be
decomposed into two components: long-range attraction and
short-range repulsion. Previous RG analyses have already verified
that the short-range repulsion between fermions is irrelevant at low
energies \cite{Shankar94}, which is the key ingredient ensuring the
stability of the FL state. This feature is well consistent with the
basic notion that heavy mesons become progressively unimportant as
the energy is lowered. On the contrary, the quantum critical
fluctuations of the superfluid order parameter are gapless. The massless
boson ($r=0$) describing such gapless fluctuations plays an
overwhelming role in the low-energy region and its coupling to
gapless neutrons leads to the breakdown of FL theory. If we ignore
the massless critical boson and consider only the massive mesons,
the neutron damping rate would depend on the energy $\omega$ as
$\Gamma(\omega) \propto \omega^{2}$, as illustrated in
Sec.~\ref{sec:perturbative}. This is a normal FL behavior. If we
consider both massless critical bosons and massive mesons, the
damping rate depends on $\omega$ in the form
\begin{eqnarray}
A \omega + B \omega^{2},
\end{eqnarray}
where $A \omega$ is the NFL term due to superfluid fluctuations and
$B \omega^{2}$ is the FL term induced by mesons. In the limit of
$\omega \rightarrow 0$, the FL term can be discarded since it
vanishes more rapidly than the NFL term. Therefore, the
neutron-meson interactions play a secondary role in the quantum
critical region and their main effect is to change the bare neutron
mass $M_{n}$ to an effective constant mass $M_{n}^{\ast}$.

The influence of long-range attraction is distinct from short-range
repulsion. One can utilize the RG theory to prove that the
attraction between neutrons is a relevant perturbation
\cite{Shankar94}, meaning that the attraction strength parameter
would flow to extremely large values with the decreasing energy.
Such a runaway behavior provides a clear signature of the
instability of the gapless neutron FL state, which is inevitably driven
by the greatly enhanced attraction into the more stable gapped
superfluid state via the formation of Cooper pairs. The presence of
superfluidity is essential to the understanding of NS cooling
and also to the existence of superfluid quantum criticality.
However, since our interest is in the quantum critical phenomena
emerging in the nonsuperfluid critical region, it would be
justified to neglect the long-range attraction, or, equivalently,
the light mesons (like pions), in our calculations of the NFL
behavior.

\section{NFL corrections to specific heat and neutrino emissivity \label{sec:nscooling}}

In this section, we examine the influence of superfluid quantum
criticality on the NS cooling history. As already mentioned, quantum
critical phenomena begin to affect the thermal evolution at a much
earlier time than in the PBF scenario. As a result, the internal
thermal relaxation time could be considerably extended. This would
have observable effects on the cooling history. To address such
effects, we perform a quantitative analysis of the corrections to
the specific heat of neutrons and the neutrino emissivity produced
by the NFL behavior. As shown in Fig.~{\ref{fig:phasediagram}}, the
NFL corrections, driven by $^3P_2$ superfluid quantum criticality,
primarily affect the NS cores, where the thermal properties of the
uniform, asymmetric nuclear matter are predominantly controlled by
the behavior of neutrons. In the NS crust, the thermal evolution is
mainly governed by the ions arranged in a crystalline lattice.
However, in the early stages of NSs lives, the $^3P_2$ superfluid
quantum critical region may extend into the crust, influencing a
small fraction of free neutrons there. Moreover, the NFL behavior,
driven by $^1S_0$ superfluid quantum criticality, also impacts the
neutrons in the inner crust of NSs. Consequently, the NFL region
under consideration also encompasses parts of the crust close to the
core.

The thermal conductivity $\kappa$ may also be affected by the NFL
behavior, as illustrated in Appendix.~\ref{sec:appc}. But the heat
transport of neutrons, compared to that of electrons, makes a minor
contribution to the total thermal conductivity, especially in the
vicinity of the crust \cite{Potekhin15}. For this reason, we neglect
the impact of NFL behavior on the thermal conductivity.

\subsection{Renormalized specific heat \label{sec:specificheat}}

The cooling process of an NS after its birth can be roughly divided
into three main stages \cite{Yakovlev99, Yakovlev01, Yakovlev04,
Page06a, Potekhin15, Tsuruta23}: (a) internal thermal relaxation
stage; (b) neutrino cooling stage; (c) photon cooling stage.

The first stage is characterized by the heat transport caused by the
presence of temperature gradients in the NS interior. This stage
typically lasts for a few hundred years. The thermal relaxation
time $t_{w}$, defined by the time needed for the NS interior to
reach thermal equilibrium, depends sensitively on the heat capacity
$C_{\mathrm{v}}$ and the thermal conductivity $\kappa$ in the crust.
This relationship is inferred from a simplified calculation for a
uniform layer of the star with a crust thickness $\Delta R$
\cite{Yakovlev01, Lattimer94}
\begin{eqnarray}
t_{w} \sim \frac{C_{\mathrm{v}}(\Delta R)^2}{\kappa}.
\label{eq:relaxationtime}
\end{eqnarray}
Once the thermal relaxation stage is terminated, the NS interior
reaches an isothermal state, with the heat-blanketing envelope being
an exception \cite{Yakovlev99, Yakovlev01, Yakovlev04, Page06a,
Potekhin15, Tsuruta23}. Then, the thermal evolution of the interior
can be described by a heat balance equation \cite{Yakovlev99,
Yakovlev01, Yakovlev04, Page06a, Potekhin15, Tsuruta23} of the form
\begin{eqnarray}
C_{\mathrm{v}}\frac{d T}{d t} = -L_{\nu}-L_{\gamma},\label{eq:heatbalance}
\end{eqnarray}
where $L_{\nu}$ denotes its neutrino luminosity, $L_{\gamma}$
represents the surface photon luminosity, and $T$ refers to its
internal temperature. In the neutrino cooling stage, the energy loss
arises predominantly from the emission of neutrinos, implying that
$L_{\nu}$ is significantly larger than $L_{\gamma}$. Within the
period between approximately $10^{5}$ to $10^{6}$ yr, the NS cooling
is controlled mainly by the thermal radiation of photons from the
surface. At this stage, the term $L_{\nu}$ becomes negligible and
can be removed from Eq.(\ref{eq:heatbalance}).
{The NS might undergo a reheating
processes, such as accretion processes \cite{Haensel03} or magnetic
heating processes \cite{Beloborodov16}, during the long history of
thermal evolution. However, the investigation of the reheating
processes is beyond the scope of the present work.}

Heat capacity and neutrino luminosity are two crucial ingredients
throughout all three stages. In this subsection, we analyze how
the heat capacity is modified by the NFL behavior, leaving the
discussion of neutrino luminosity to Sec.~\ref{sec:emissivity}. For
this purpose, we need to generalize the zero-$T$ NFL behavior to
finite temperatures. A convenient way of doing this is to translate
the $l$ dependence of model parameters extracted from the RG
analysis into the $T$ dependence of these parameters. We still
consider the $^{3}P_2(m_J=0)$-wave pairing as an example for
illustration. The other two cases exhibit analogous behaviors.

The total heat capacity $C_{\mathrm{v}}$ is a cumulative quantity of
the heat capacities of various degenerate components that make up
the dense matter of the NS core. The predominant contribution to
$C_{\mathrm{v}}$ arises from neutrons, complemented by a minor
contribution from protons and electrons:
\begin{eqnarray}
C_{\mathrm{v}}=\int_{V} \left(c_{\mathrm{n}} +
c_{\mathrm{p}}+c_{\mathrm{e}}\right) dV.
\label{eq:heatcapcity}
\end{eqnarray}
Here, $c_{\mathrm{p}}$ refers to the specific heat of protons, while
$c_{\mathrm{e}}$ denotes that of electrons. However, the situation
is markedly different in superfluid quantum critical region. In this
region, the effective neutron mass $M_{n}^{\ast}$ receives a
singular contribution from the NFL behavior. The anomalous dimension
of the neutron field does not yield a qualitative change to
$M_{n}^{\ast}$. The qualitative correction to $M_{n}^{\ast}$ arises
mainly from the renormalization of neutron dispersion.
{After including the corrections from
the NFL behavior, Eq.~(\ref{eq:flcn}) can be transformed into
\begin{eqnarray}
c_{n}(T)\sim M^{\ast}_n(l)T.
\label{h0}
\end{eqnarray}
Now consider the following
derivative
\begin{eqnarray}
\frac{d c_{n}(T)}{dT} &\sim& \frac{d\left[M^{\ast}_n(l)T\right]}{dT}\nonumber\\
&=& M^{\ast}_{n}(l) +
\frac{dM^{\ast}_{n}(l)}{dT}T.\label{eq:dCndT}
\end{eqnarray}}

We need to determine the $T$ dependence of $M^{\ast}_{n}(l)$. In
Sec.~\ref{sec:nflbehavior}, we have already obtained the
renormalized neutron mass $M^{\ast}_{n}(l)$, given by
Eq.(\ref{solutionmmm}), from the solutions of RG equations. By
virtue of Eq.~(\ref{eq:omegalrelation}), we converted the
$l$-dependent $M^{\ast}_{n}(l)$ into an energy-dependent function
$M^{\ast}_{n}(\omega)$ at zero temperature. Notice the
correspondence between the thermal energy $k_{\mathrm{B}}T$ and the
single-particle energy $\omega$ given by $k_{\mathrm{B}}T \sim
\omega$. Based on such a correspondence, it is possible to deduce
the following relation \cite{Millis} between an arbitrary
temperature $T$ and an arbitrary length scale $l$:
\begin{eqnarray}
T(l) = T_{0}e^{-2l}.\label{eq:T}
\end{eqnarray}
Here, $T_{0}$ represents the highest temperature of the system and
could be fixed at $\sim 10^{11}~\mathrm{K}$ \cite{Yakovlev05}, the
initial temperature of NSs. {Making
use of the above expression for $T(l)$, we find that the
$T$ dependence of $M_{n}^{\ast}(l)$ is given by
\begin{eqnarray}
\frac{dM_{n}^{\ast}(l)}{dT} &=& -\frac{1}{2T}\frac{d M_{n}^{\ast}(l)}{d l}
\nonumber \\
&=& -\frac{1}{T}\frac{5h^{2}M^{\ast 2}_{n}}{8N\pi^{2}
k^{~}_{\mathrm{F}}}, \label{eq:dmndt}
\end{eqnarray}
where the approximation $k^{~}_{\mathrm{F}}/v^{~}_{\mathrm{F}}
\approx M_{n}^{\ast}$ is utilized. By substituting
Eq.~(\ref{solutionmmm}), Eq.~(\ref{eq:T}), and Eq.~(\ref{eq:dmndt})
into Eq.~(\ref{eq:dCndT}), it is easy to obtain
\begin{eqnarray}
\frac{d c_{n}(T)}{dT} &\sim& M_{n}^{\ast} - \frac{5h^{2} M^{\ast
2}_{n}}{4N\pi^{2}k^{~}_{\mathrm{F}}}l-\frac{5h^{2}M^{\ast
2}_{n}}{8N\pi^{2}k^{~}_{\mathrm{F}}}\nonumber \\
&=& M_{n}^{\ast} + \frac{5h^{2} M^{\ast 2}_{n}}{8N\pi^{2}
k^{~}_{\mathrm{F}}} \left(\ln \frac{T_{0}}{T}-1\right),
\end{eqnarray}}
A straightforward integration of this differential equation leads to
\begin{eqnarray}
\tilde{c}_{n}(T) &=& c_{n} + \frac{5k^{2}_{\mathrm{B}}h^{2} M^{\ast
2}_{n}}{24N\hbar^{3}\pi^{2}}T\ln\frac{T_{0}}{T}\nonumber\\
&=& c_{n} \left[1+ \frac{5h^{2} M^{\ast}_{n}}{8N \pi^{2}
k^{~}_{\mathrm{F}}}\ln\frac{T_{0}}{T}\right]. \label{hcNFL}
\end{eqnarray}
The singular logarithmic correction is the remarkable consequence of
superfluid quantum criticality. This is a unique characteristic of
NFL behavior. A similar $T\ln(1/T)$ correction to the specific heat (of
electrons) has been previously found in the U(1) gauge-interaction
system \cite{Holstein} and also in strange metal systems
\cite{Sachdev23}.

The effect of the logarithmic term $\sim T\ln(T_{0}/T)$ on the NS
cooling is qualitatively different from the linear term $c_{n}$. At
the initial time of the NS cooling history, the temperature is
$T=T_{0}$ and the logarithmic correction is absent since
$\ln(T_{0}/T)=0$. As time goes by and the temperature falls, this
logarithmic correction fades in and eventually dominates over the
linear term $c_{n}$ at low temperatures. It is remarkable that
superfluid quantum criticality exists throughout the whole thermal
evolution of an NS, which can dramatically extend the thermal
relaxation time.

\begin{figure}[htbp]
\centering
\includegraphics[width=3.8in]{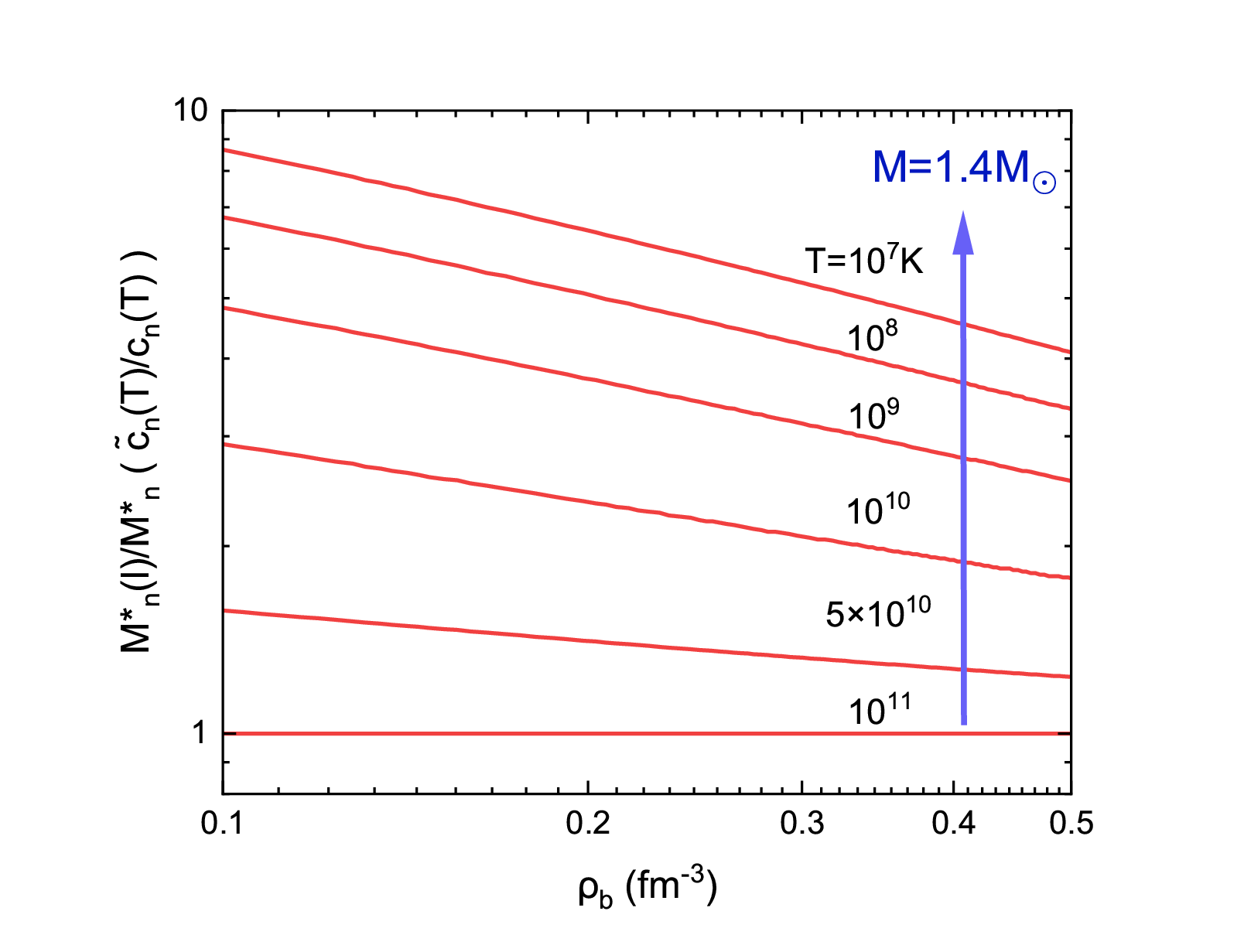}
\caption{{The dependence of the ratio of the NFL-corrected neutron
masses $M_{n}^{\ast}(l)$ ($h=3.0$) to the effective neutron mass
$M^{\ast}_{n}$ ($h=0$) in a $1.4\mathrm{M}_{\odot}$ NS described by
the APR EOS on the baryon density $\rho_b$. The ratio
$\tilde{c}_{n}(T)/c_{n}(T)$ exhibits the same density dependence as
$M_{n}^{\ast}(l)/M^{\ast}_{n}$.} The numbers along the curves
indicate the internal temperatures. Contour levels are set at
$10^{11}$, $5\times10^{10}$, $10^{10}$, $10^{9}$, $10^{8}$, and
$10^{7}$ K. The light purple arrow denotes the direction of
decreasing temperature.} \label{NSrhoM}
\end{figure}

The NFL correction to neutrons' specific heat is mainly attributed
to the modification of the effective neutron mass. Based on the APR
EOS \cite{Akaml98}, we illustrate in Fig.~\ref{NSrhoM} the variation
of the ratio $M_{n}^{\ast}(l)/M^{\ast}_{n}$ as a function of baryon
density $\rho_b$. {We adopt $h=3.0$ as a representative value in the
presence of NFL behavior. For $T < 10^{11}$ K, one can see that
$M_{n}^{\ast}(l)/M_{n}^{\ast}$ decreases with growing $\rho_b$, and
this trend remains unchanged as the temperature is further lowered.
In sharp contrast to the FL state, in which the effective neutron
mass is independent of temperature, the NFL-corrected neutron mass
is strongly temperature dependent. At the initial internal
temperature $T_{0}=10^{11}$ K, the ratio
$M_{n}^{\ast}(l)/M^{\ast}_{n}$ stays at $1$, indicating that the NFL
case coincides with the FL case.} As the temperature decreases, the
impact of NFL behavior on $M_{n}^{\ast}(l)$ is obviously pronounced.
At a fixed $\rho_b$, the magnitude of the NFL-corrected
$M_{n}^{\ast}(l)$ increases progressively in the direction indicated
by the light purple arrow in Fig.~\ref{NSrhoM} and significantly
exceeds $M^{\ast}_{n}$ at sufficiently low temperatures. The marked
increase in the NFL-corrected neutron mass indicates that the NFL
behavior can lead to corrections to other quantities, especially the
pressure-density relation. Consequently, the mass-radius relation of
an NS may also be modified. While a fully consistent treatment would
require incorporating the NFL corrections to the EOS, we defer this
step to future work and, for the present study, adopt the original
APR EOS.

The pronounced NFL correction to the effective neutron mass is
reflected in the variation of the neutron specific heat. As depicted
in Fig.~\ref{NSrhoM}, the ratio of the renormalized neutron specific
heat $\tilde{c}_{n}(T)$ to the neutron specific heat $c_{n}(T)$,
namely $\tilde{c}_{n}(T)/c_{n}(T)$, exhibits the same trend as
$M_{n}^{\ast}(l)/M^{\ast}_{n}$, due to the fact that NFL behavior
contributes the same factor $1+ \frac{5h^{2} M^{\ast}_{n}}{8N
\pi^{2} k^{~}_{\mathrm{F}}}\ln T_{0}/T $ to both the neutron mass
and the neutron specific heat. Now the renormalized total heat
capacity $\tilde{C}_{\mathrm{v}}$ can be determined by
\begin{eqnarray}
\tilde{C}_{\mathrm{v}} = \int_{V} \left(\tilde{c}_{\mathrm{n}} +
c_{\mathrm{p}} + c_{\mathrm{e}}\right) dV.
\label{eq:heatcapcityrenormalized}
\end{eqnarray}

\subsection{Renormalized neutrino emissivity \label{sec:emissivity}}


Then we compute the renormalized neutrino emissivity $Q_{\nu}$
corrected by the NFL behavior. Neutrino emission arises from several
reactions, including DU, MU, NNB, and PBF processes
\cite{Yakovlev99, Yakovlev01, Yakovlev04, Page06a, Potekhin15,
Tsuruta23}. The total value of $Q_{\nu}$ is linked to the luminosity
$L_{\nu}$ by the relationship
\begin{eqnarray}
L_{\nu}=\int_{V} Q_{\nu} dV.\label{eq:lnu}
\end{eqnarray}
It will be shown that the NFL behavior also generates a logarithmic
correction to this quantity.

To proceed with our analysis, we exemplify the MU process of the
neutron branch to estimate the NFL correction to the neutrino
emissivity. For the MU processes of the neutron branch
$n+n\rightarrow p+n+e+\bar{\nu}_e$ and $p+n+e \rightarrow
n+n+\nu_{e}$, the neutrino emissivity has already been calculated
\cite{Friman79, Yakovlev95} and takes the form
\begin{eqnarray}
Q_{\nu}^{\mathrm{MUn}} &=& \frac{11513}{30240}\frac{G^{2}g^{2}_{A}
M^{\ast3}_{n}M^{\ast}_p}{2\pi}\left(\frac{f_{\pi}}{m_{\pi}}
\right)^4\frac{k^{~}_{\mathrm{Fp}}(k_{\mathrm{B}}T)^{8}}{\hbar^{10}c^{8}}
\alpha_{\mathrm{n}}\beta_{\mathrm{n}} \nonumber \\
&\approx&8.55\times 10^{21}\left(\frac{M^{\ast}_{n}}{M_{n}}\right)^3
\left(\frac{M^{\ast}_{p}}{M_{p}}\right) \nonumber\\
&&\times\left(\frac{k^{~}_{\mathrm{Fe}}}{1.68~{\mathrm{fm}}^{-1}}\right)
T^{8}_{9}\alpha_{\mathrm{n}}\beta_{\mathrm{n}}
~\frac{\mathrm{erg}}{{\mathrm{cm}}^{3}\mathrm{s}}.
\label{eq:qnumn}
\end{eqnarray}
Here, $G = G_{F}\cos\theta_{c}=1.436\times
10^{-49}~\mathrm{erg}{\mathrm{cm}}^3$, where $G_F$ is the Fermi weak
interaction constant and $\theta_{c}$ is the Cabibbo angle with
$\sin\theta_{c} = 0.231$. In addition, $g_{A}=1.26$ is the
Gamow-Teller axial-vector coupling constant, $f_{\pi}\approx 1$
denotes the $p$-wave $\pi N$ coupling constant, $m_{\pi}\approx
140~\mathrm{MeV}/\mathrm{c}^2$ represents the mass of the pion,
$\alpha_{\mathrm{n}}=1.13$, $\beta_{\mathrm{n}}=0.68$,
and $c$ is the speed of light. The Fermi momenta of protons
and electrons are denoted by $k^{~}_{\mathrm{Fp}}$ and $k^{~}_{\mathrm{Fe}}$
respectively. $T_{9}$ is defined as
$T_{9}\equiv T/(10^9 K)$. $M_{p}$ and $M^{\ast}_{p}$ are the bare
and effective proton masses, respectively.

Note that superfluid criticality only renormalizes the neutrino
emissivities of the reactions participated by the neutrons. For the
DU process and the proton branch of the MU process, the neutron
contribution is $\sim M^{\ast}_{n}(l)T$. For the NP bremsstrahlung
process, the NNB process, and the proton-proton bremsstrahlung
process, the neutron contributions are $\sim [M^{\ast}_{n}(l)T]^2$,
$\sim [M^{\ast}_{n}(l)T]^4$, and $\sim M^{\ast}_{n}(l)T$,
respectively. For the neutron branch of the MU process, the neutrino
emissivity given by Eq.~(\ref{eq:qnumn}) depends on $T$ as
\begin{eqnarray}
Q_{\nu}^{\mathrm{MUn}}(T) \sim \left[M^{\ast}_n(l)T\right]^3,
\label{h0}
\end{eqnarray}
thereby leading to
\begin{eqnarray}
\frac{dQ_{\nu}^{\mathrm{MUn}}}{dT} \sim 3M^{\ast3}_{n}(l) T^{2} +
3M^{\ast2}_{n}(l) T^{3}\frac{d M^{\ast3}_{n}(l)}{d T}.\label{QMn}
\end{eqnarray}
By substituting Eq.(\ref{eq:dmndt}) into Eq.(\ref{QMn}), we derive the
following result:
\begin{eqnarray}
\frac{dQ_{\nu}^{\mathrm{MUn}}}{dT} &\sim& 3M^{\ast 3}_{n}T^2 +
\frac{375h^{6}M^{\ast 6}_{n}}{512N^{3}\pi^{6}k^{3}_{\mathrm{F}}}
T^{2}\ln^{3}\frac{T_{0}}{T} \nonumber \\
&& +\left(\frac{225 h^{4}M^{\ast 5}_{n}}{64 N^{2}\pi^{4}
k^{2}_{\mathrm{F}}}-\frac{375h^{6} M^{\ast 6}_n}{512
N^{3}\pi^{6}k^{3}_{\mathrm{F}}}\right) T^{2}\ln^{2}
\frac{T_0}{T}\nonumber \\
&& +\left(\frac{45 h^{2}M^{4\ast}_n}{8N \pi^{2}
k^{~}_{\mathrm{F}}}-\frac{75h^{4} M^{\ast 5}_n}{32 N^{2}
\pi^{4}k^{2}_{\mathrm{F}}}\right) T^{2}\ln\frac{T_0}{T}
\nonumber \\
&& -\frac{15h^{2}M^{\ast 4}_n}{8N \pi^{2} k^{~}_{\mathrm{F}}}T^{2}.
\end{eqnarray}
Upon integrating the above differential equation, we get the
corrected expression for the neutrino emissivity of the MU process for
the neutron branch:
\begin{eqnarray}
\tilde{Q}_{\nu}^{\mathrm{MUn}} &=& 8.55\times 10^{21}
\left(\frac{M^{\ast}_{n}}{M_n}\right)^{3}
\left(\frac{M^{\ast}_{p}}{M_p}\right)
\left(\frac{k^{~}_{\mathrm{Fe}}}{1.68~{\mathrm{fm}}^{-1}}
\right)T^{8}_{9}\nonumber \\
&& \times \Bigg[1+ \frac{15h^{2} M^{\ast}_n}{8N\pi^{2}
k^{~}_{\mathrm{F}}}\ln\frac{T_{0}}{T}+\frac{75h^{4} M^{\ast
2}_n}{64N^{2}\pi^{4} k^{2}_{\mathrm{F}}}\ln^{2} \frac{T_0}{T}
\nonumber \\
&& +\frac{125h^{6} M^{\ast3}_n}{512 N^{3}\pi^{6}
k^{3}_{\mathrm{F}}}\ln^{3}\frac{T_0}{T}\Bigg]
\alpha_{\mathrm{n}}\beta_{\mathrm{n}}
~\frac{\mathrm{erg}}{{\mathrm{cm}}^3\mathrm{s}}. \label{MUnNFL}
\end{eqnarray}

The above calculational procedure can be directly applied to
determine the neutrino emissivities arising from other reactions. We
omit the derivational details and simply present the final
results.

For the DU process, the neutrino emissivity of $n\rightarrow
p+e+\bar{\nu}_e$ and $p+e\rightarrow n+\nu_e$ is \cite{Lattimer91}
\begin{eqnarray}
Q^{\mathrm{D}}_{\nu} &=& \frac{457\pi}{10080}G^{2} \left(1 +
3g^{2}_{A}\right)\frac{M^{\ast}_{n}M^{\ast}_{P}M_{e}}{\hbar^{10}
c^{3}} \left(k_{\mathrm{B}}T\right)^{6} \Theta_{\mathrm{npe}}
\nonumber \\
&\approx& 4.24\times 10^{27}\left(\frac{k^{~}_{\mathrm{Fe}}}{1.68
~{\mathrm{fm}}^{-1}}\right)\frac{M^{\ast}_{n}M^{\ast}_{p}}{M^{2}_{n}}
T^{6}_{9}\Theta_{\mathrm{npe}} ~
\frac{\mathrm{erg}}{{\mathrm{cm}}^3\mathrm{s}},
\nonumber \\
\end{eqnarray}
where $M_{e}$ is the electron mass. $\Theta_{\mathrm{npe}}$ is a
step function: $\Theta_{\mathrm{npe}}=1$ if
$k^{~}_{\mathrm{F}}<k^{~}_{\mathrm{Fp}}+k^{~}_{\mathrm{Fe}}$ and
$\Theta_{\mathrm{npe}}=0$ otherwise. After incorporating the
corrections resulting from superfluid quantum criticality, the above
expression becomes
\begin{eqnarray}
\tilde{Q}^{\mathrm{D}}_{\nu} &\approx& 4.24\times 10^{27}
\left(\frac{k^{~}_{\mathrm{Fe}}}{1.68~{\mathrm{fm}}^{-1}}\right)
\frac{M^{\ast}_n M^{\ast}_p}{M^{2}_{n}}T^{6}_{9} \nonumber \\
&&\times \left(1+\frac{5h^{2}M^{\ast}_n}{8N\pi^2
k^{~}_{\mathrm{F}}}\ln \frac{T_{0}}{T}\right)\Theta_{\mathrm{npe}}
~\frac{\mathrm{erg}}{{\mathrm{cm}}^3\mathrm{s}}.
\label{DUNFL}
\end{eqnarray}

For MU processes of the proton branch $n+p \rightarrow
p+p+e+\bar{\nu}_e$ and $p+p+e \rightarrow n+p+\nu_e$, the neutrino
emissivity is \cite{Yakovlev95}
\begin{eqnarray}
Q^{\mathrm{MUp}}_{\nu} &=& \frac{11513}{30240}\frac{G^{2}g^{2}_{A}
M^{\ast}_{n} M^{\ast3}_P}{2\pi}
\left(\frac{f_{\pi}}{m_{\pi}}\right)^{4}\nonumber \\
&&\times \frac{(k^{~}_{\mathrm{Fe}}+3k^{~}_{\mathrm{Fp}}
-k^{~}_{\mathrm{F}})^{2}}{8\hbar^{10}c^{8}k^{~}_{\mathrm{Fe}}}
\left(k_{\mathrm{B}}T\right)^{8}\alpha_{\mathrm{p}}\beta_{\mathrm{p}}
\Theta_{\mathrm{Mp}}
\nonumber \\
&\approx& 8.55 \times 10^{21}\left(\frac{M^{\ast}_n}{M_n}\right)
\left(\frac{M^{\ast}_{p}}{M_{p}}\right)^{3}
\left(\frac{k^{~}_{\mathrm{Fe}}}{1.68~\mathrm{fm}^{-1}}\right)
\nonumber \\
&& \times \frac{\big(k^{~}_{\mathrm{Fe}}+3k^{~}_{\mathrm{Fp}} -
k^{~}_{\mathrm{F}}\big)^2}{8k^{~}_{\mathrm{Fe}}
k^{~}_{\mathrm{Fp}}}T^{8}_{9}\alpha_{\mathrm{p}}
\beta_{\mathrm{p}}\Theta_{\mathrm{Mp}}
~\frac{\mathrm{erg}}{{\mathrm{cm}}^3\mathrm{s}},
\nonumber\\
\end{eqnarray}
where $\alpha_{\mathrm{p}}=\alpha_{\mathrm{n}}$ and
$\beta_{\mathrm{p}}=\beta_{\mathrm{n}}$. $\Theta_{\mathrm{Mp}}$ is
also a step function: $\Theta_{\mathrm{Mp}}=1$ if
$k^{~}_{\mathrm{F}} < 3k^{~}_{\mathrm{Fp}} + k^{~}_{\mathrm{Fe}}$
and $\Theta_{\mathrm{Mp}}=0$ otherwise. It is renormalized by the
NFL behavior to become
\begin{eqnarray}
\tilde{Q}^{\mathrm{MUp}}_{\nu} &\approx& 8.55 \times 10^{21}
\left(\frac{M^{\ast}_n}{M_n}\right)
\left(\frac{M^{\ast}_p}{M_p}\right)^3\left(\frac{k^{~}_{\mathrm{Fe}}}{1.68
~\mathrm{fm}^{-1}}\right)\nonumber \\
&&\times \frac{\big(k^{~}_{\mathrm{Fe}} +
3k^{~}_{\mathrm{Fp}}-k^{~}_{\mathrm{F}}\big)^2}{8
k^{~}_{\mathrm{Fe}}k^{~}_{\mathrm{Fp}}}T^{8}_{9}
\nonumber \\
&& \times \left(1+\frac{5h^{2}M^{\ast}_n}{8N\pi^2
k^{~}_{\mathrm{F}}}\ln \frac{T_0}{T}\right)
\alpha_{\mathrm{p}}\beta_{\mathrm{p}} \Theta_{\mathrm{Mp}}
~\frac{\mathrm{erg}}{{\mathrm{cm}}^3\mathrm{s}}.
\nonumber \\
\label{MUpNFL}
\end{eqnarray}

There are three different NNB processes \cite{Friman79, Yakovlev95}.
The respective neutrino emissivity will be considered below in
order.

1) Process $n+n\rightarrow n+n+\nu_e+\bar{\nu}_e$:
\begin{eqnarray}
Q^{\mathrm{nn}}_{\nu} &=& \frac{41}{14175}\frac{G^{2}_{F} g^{2}_{A}
M^{\ast 4}_n}{2\pi\hbar^{10}
c^8}\left(\frac{f_{\pi}}{m_{\pi}}\right)^4
k^{~}_{\mathrm{F}}\alpha_{\mathrm{nn}}\beta_{\mathrm{nn}}
(k_{\mathrm{B}}T)^{8}N_{\nu}
\nonumber \\
&\approx& 7.4\times 10^{19}\left(\frac{M^{\ast}_n}{M_n}\right)^4
\left(\frac{k^{~}_{\mathrm{F}}}{1.68~\mathrm{fm}^{-1}}\right)
\nonumber \\
&&\times \alpha_{\mathrm{nn}} \beta_{\mathrm{nn}}N_{\nu} T^{8}_{9}
~\frac{\mathrm{erg}}{{\mathrm{cm}}^3\mathrm{s}},
\end{eqnarray}
where $\alpha_{\mathrm{nn}}=0.59$ and $\beta_{\mathrm{nn}}=0.56$.
$N_{\nu}$ is the number of neutrino flavors. Including NFL behavior
turns it into
\begin{small}
\begin{eqnarray}
\tilde{Q}^{\mathrm{nn}}_{\nu} &\approx& 7.4\times 10^{19}
\left(\frac{M^{\ast}_n}{M_n}\right)^{4}
\left(\frac{k^{~}_{\mathrm{F}}}{1.68~\mathrm{fm}^{-1}}\right)
\alpha_{\mathrm{nn}}\beta_{\mathrm{nn}}N_{\nu}T^{8}_{9} \nonumber \\
&&\times \Bigg[1+\frac{625h^{8} M^{\ast4}_n}{4096 N^{4}
\pi^{8}k^{4}_{\mathrm{F}}} \ln^4\frac{T_0}{T} \nonumber \\
&& +\frac{125h^{6} M^{\ast3}_n}{128N^{3}\pi^{6}
k^{3}_{\mathrm{F}}}\ln^{3}\frac{T_0}{T}+\frac{75 h^{4}
M^{\ast2}_n}{32N^{2}\pi^{4}k^{2}_{\mathrm{F}}}\ln^2 \frac{T_0}{T}
\nonumber \\
&& +\frac{5h^{2}M^{\ast}_n}{2N\pi^{2}k^{~}_{\mathrm{F}}}\ln
\frac{T_0}{T}\Bigg]~\frac{\mathrm{erg}}{{\mathrm{cm}}^3\mathrm{s}}.
\label{NNBnNFL}
\end{eqnarray}
\end{small}

2) Process $n+p\rightarrow n+p+\nu_e+\bar{\nu}_e$:
\begin{eqnarray}
Q^{\mathrm{np}}_{\nu} &=& \frac{82}{14175}\frac{G^{2}_{F}g^{2}_{A} M^{\ast
2}_{n} M^{\ast 2}_{p}}{2\pi\hbar^{10}c^{8}}
\left(\frac{f_{\pi}}{m_{\pi}}\right)^4 \nonumber \\
&& \times k^{~}_{\mathrm{Fp}}\alpha_{\mathrm{np}}
\beta_{\mathrm{np}}(k_{\mathrm{B}}T)^{8}N_{\nu} \nonumber \\
&\approx& 1.5\times10^{20}\left(\frac{M^{\ast}_n}{M_n}
\frac{M^{\ast}_p}{M_p}\right)^2\left(\frac{k^{~}_{\mathrm{Fp}}}{1.68
~\mathrm{fm}^{-1}}\right)
\nonumber \\
&&\times\alpha_{\mathrm{np}}\beta_{\mathrm{np}}T^{8}_{9}N_{\nu}
~\frac{\mathrm{erg}}{{\mathrm{cm}}^3\mathrm{s}},
\end{eqnarray}
where $\alpha_{\mathrm{np}}=1.06$ and $\beta_{\mathrm{np}}=0.66$. It
is renormalized to take the form
\begin{eqnarray}
\tilde{Q}^{\mathrm{np}}_{\nu} &\approx& 1.5\times 10^{20}
\left(\frac{M^{\ast}_n}{M_n}\frac{M^{\ast}_p}{M_p}\right)^2
\left(\frac{k^{~}_{\mathrm{Fp}}}{1.68~\mathrm{fm}^{-1}}\right)
\nonumber \\
&& \times \alpha_{\mathrm{np}}\beta_{\mathrm{np}}T^{8}_{9}N_{\nu}
\bigg[1+\frac{25 h^{4} M^{\ast 2}_n}{64N^{2}\pi^{4}
k^{2}_{\mathrm{F}}}\ln^{2}\frac{T_{0}}{T} \nonumber \\
&& +\frac{5h^{2}M^{\ast}_n}{4N\pi^{2}k^{~}_{\mathrm{F}}} \ln
\frac{T_0}{T}\bigg] ~
\frac{\mathrm{erg}}{{\mathrm{cm}}^3\mathrm{s}}. \label{NNBnpNFL}
\end{eqnarray}

3) Process $p+p\rightarrow p+p+\nu_e+\bar{\nu}_e$:
\begin{eqnarray}
Q^{\mathrm{pp}}_{\nu} &=& \frac{41}{14175} \frac{G^{2}_{F}g^{2}_{A}
M^{\ast 4}_{p}}{2\pi\hbar^{10}c^{8}} \Big(\frac{f_{\pi}}{m_{\pi}}
\Big)^{4} k^{~}_{\mathrm{Fp}}\alpha_{\mathrm{pp}}
\beta_{\mathrm{pp}} (k_{\mathrm{B}}T)^{8}N_{\nu}
\nonumber \\
&\approx& 7.4\times 10^{19}\left(\frac{M^{\ast}_{p}}{M_{p}}
\right)^{4} \nonumber \\
&& \times \left(\frac{k^{~}_{\mathrm{Fp}}}{1.68~\mathrm{fm}^{-1}}
\right)\alpha_{\mathrm{pp}}\beta_{\mathrm{pp}}N_{\nu}T^{8}_{9}
~\frac{\mathrm{erg}}{{\mathrm{cm}}^3\mathrm{s}},
\end{eqnarray}
where $\alpha_{\mathrm{pp}}=0.11$ and $\beta_{\mathrm{pp}}=0.7$.
This quantity is not altered by the NFL behavior, since the neutron
mass parameter does not appear.

For the PBF process $\tilde{n}+\tilde{n}\rightarrow
\nu_e+\bar{\nu}_e$, where $\tilde{n}$ represents neutral Bogoliubov
quasiparticles, the corresponding neutrino emissivity is
\cite{Flowers76, Voskresensky87, Yakovlev99b}
\begin{eqnarray}
Q^{\mathrm{PBF}}_{\nu} &=& \frac{4G^{2}_{F}M^{\ast}_{n}
k^{~}_{\mathrm{F}}}{15\pi^{5}\hbar^{10}c^6}(k_{\mathrm{B}}
T)^{7}N_{\nu} \tilde{a}
F\Big[\frac{\Delta_{\mathrm{n}}(T)}{k_{\mathrm{B}}T}\Big]
\nonumber \\
&=& 1.170 \times 10^{21}\left(\frac{M^{\ast}_n}{M_n}\right)
\left(\frac{k^{~}_{\mathrm{F}}}{M_{n}c}\right)T^{7}_{9} \nonumber
\\
&& \times N_{\nu}\tilde{a} F\Big[\frac{\Delta_{\mathrm{n}}(T)}{k_{\mathrm{B}}T}
\Big] ~\frac{\mathrm{erg}}{{\mathrm{cm}}^3\mathrm{s}},
\label{PBFNFL}
\end{eqnarray}
where $\tilde{a} = 2g^{2}_{A}$ and the specific form of the function
$F\Big[\frac{\Delta_{\mathrm{n}}(T)}{k_{\mathrm{B}}T}\Big]$ depends on the
symmetry of the considered superfluid gap. As explained previously,
the NFL behavior and PBF processes exist in different layers, and thus
$Q^{\mathrm{PBF}}_{\nu}$ is not changed by the NFL behavior.


The renormalized total neutrino luminosity $\tilde{L}_{\nu}$ can be
computed as follows
\begin{eqnarray}
\tilde{L}_{\nu}=\int_{V} \tilde{Q}_{\nu} dV,
\label{eq:lnurenormalized}
\end{eqnarray}
where $\tilde{Q}_{\nu}$, the renormalized total neutrino emissivity,
is given by the sum
\begin{eqnarray}
\tilde{Q}_{\nu} &=& \tilde{Q}_{\nu}^{\mathrm{MUn}} +
\tilde{Q}^{\mathrm{D}}_{\nu} + \tilde{Q}^{\mathrm{MUp}}_{\nu}
\nonumber \\
&& +\tilde{Q}^{\mathrm{nn}}_{\nu} + \tilde{Q}^{\mathrm{np}}_{\nu} +
Q^{\mathrm{pp}}_{\nu} + Q^{\mathrm{PBF}}_{\nu}.
\end{eqnarray}
Then the heat balance equation becomes
\begin{eqnarray}
\tilde{C}_{\mathrm{v}}\frac{d T}{d t} = -\tilde{L}_{\nu} -
L_{\gamma}.\label{eq:heatbalancerenormalized}
\end{eqnarray}
This equation will be used to qualitatively analyze the thermal
evolution of NSs.

The above results are obtained for the $^{3}P_{2}(m_J=0)$-wave
superfluid pairing, which is deemed more likely to occur in the NS
core than $^{3}P_{2}(m_J=\pm2)$-wave pairing \cite{Hoffberg70,
Tamagaki70}. Analogous conclusions will be reached if the latter
pairing is considered, with a minor change of the constant
coefficient. In addition to the $^3P_2$-wave superfluid state in the
core, there might be a $^{1}S_0$-wave superfluid state in the crust
region and even a $^{1}S_0$-wave proton superconducting state in the
ultradense core center of NSs. Accordingly, there could be their
respective quantum critical phenomena as well. These phenomena can
be analyzed in an analogous manner.

\section{NS cooling history: theory vs observations \label{sec:numericalresults}}

After specifying the corrections from the NFL-type quantum critical
behavior to the specific heat and the total neutrino emissivity, we
are now ready to examine their effects on the thermal evolution of
NSs.

As shown by Fig.~\ref{fig:phasediagram}, the NS interior is composed
of several layers that are categorized into three distinct classes,
namely the NFL layer, the FL layer, and the superfluid layer. The
NFL layer occupies a large portion of the core at high temperatures,
probably extending its influence to the crust. As the NS is cooling
down, the temperature $T$ decreases and the NFL layer is gradually
narrowed. Meanwhile, the FL layer gets thicker. When $T$ becomes
sufficiently low, the superfluid layer emerges and occupies a
progressively larger portion of the core with decreasing $T$, which
in turn reduces the thickness of the NFL and FL layers. Nevertheless, it
is necessary to emphasize that the NFL and FL layers are always
present.

According to extensive theoretical studies on quantum criticality in
condensed-matter physics, the width of the quantum critical region
is proportional to the thermal energy scale, which is on the order
of $\sim k_{\mathrm{B}} T$ \cite{Sachdevbook}. Thus, the volume of
NFL layers, denoted by $V_{\mathrm{NFL}}$, is reduced with the
decrease of $T$. However, the NFL-corrected specific heat and
neutrino emissivity are enhanced, because the logarithmic correction
factor $\sim \ln(T_{0}/T)$ is amplified. Therefore, the overall
impact of NFL behavior is determined by the net balance between
these two opposite trends. It is worth emphasizing that the NFL
layer and the superfluid layer are mutually independent and do not
influence each other. In the gapped superfluid phase, the quantum
fluctuation of the superfluid order parameter $\phi$ is
significantly suppressed due to the nonzero value of $\langle \Phi
\rangle$. Consequently, NFL behavior does not emerge in the
superfluid layer. On the other hand, since the superfluid transition
temperature $T_{\mathrm{cn}}=0$ at the superfluid QCP, there is no
superfluid Cooper pairing in the NFL layer. It is also obvious that
the thickness of the NFL layer is independent of the value of
$T_{\mathrm{cn}}$.

As aforementioned, there exist two types of Cooper pairing
($^3P_2$-wave and $^{1}S_0$-wave) and totally four QCPs within the
NS interior. One could define four different coupling parameters,
denoted as $h_{1}$, $h_{2}$, $h_{3}$, and $h_{4}$, for these QCPs.
However, the exact values of four critical densities have not yet
been determined. Therefore, the four coupling parameters remain
unknown at present. Furthermore, performing calculations with four
different coupling parameters would be extraordinarily intricate.
Given the difficulty caused by the large uncertainties of the above
parameters and with the intention of simplifying calculations, we
utilize one single parameter $h$ to encapsulate the combined effects
of all four $h$-parameters. This simplification is justified since
the NFL behaviors emerging near all four QCPs are qualitatively the
same. At present, the value of $h$ cannot be computed at a
microscopic level. We consider $h$ as a tuning parameter and
determine its value through fitting to the observational data of NS
cooling.

In order to quantitatively assess the impact of the NFL layer, it is
imperative to incorporate the NFL behavior into the theoretical
analysis of the internal thermal evolution. This incorporation is
beset by the absence of a detailed knowledge of the proportion of
each layer, which is strongly $T$ dependent and hard to ascertain.
Here, we introduce an approximation to handle this difficulty.
Consider, for instance, the specific heat $c_{\mathrm{n}}$. At a
certain temperature above the maximum of superfluid
$T_{\mathrm{cn}}$, we assume that the FL layers and NFL layers
occupy $60\%$ and $40\%$ of the interior, respectively. In this
case, the total specific heat of neutrons would be
\begin{eqnarray}
c_{\mathrm{n},\mathrm{total}} &=& 0.6 c_{\mathrm{n}} + 0.4
\tilde{c}_{\mathrm{n}} \nonumber \\
&=& 0.6 c_{\mathrm{n}} + 0.4\left(c_{\mathrm{n}} +
\frac{5k^{2}_{\mathrm{B}} h^{2} M^{\ast
2}_{n}}{24N\hbar^{3}\pi^{2}}T\ln\frac{T_{0}}{T}\right)
\nonumber \\
&=& c_{\mathrm{n}}+ 0.4\times\frac{5k^{2}_{\mathrm{B}}h^{2} M^{\ast
2}_{n}}{24N \hbar^{3}\pi^{2}}T \ln\frac{T_0}{T},\nonumber
\end{eqnarray}
which can be rewritten as
\begin{eqnarray}
c_{\mathrm{n},\mathrm{total}} = c_{\mathrm{n}} +
\frac{5k^{2}_{\mathrm{B}} \tilde{h}^{2} M^{\ast
2}_{n}}{24N\hbar^{3}\pi^{2}}T \ln\frac{T_0}{T},
\end{eqnarray}
where a new coupling parameter $\tilde{h}^{2} = 0.4 {h}^{2}$ is
defined. For simplicity, we persist in utilizing the symbol $h$ to
denote the new coupling parameter. Upon implementing this approach,
we assume that the entire nonsuperfluid region is occupied by the NFL
state, which makes theoretical analysis more tractable. As $T$
becomes low enough to allow for the onset of superfluidity, the NFL
layers coexist with superfluid layers.

We investigate the cooling history of an NS by using the
NSCool code package developed by Page \cite{Page16}. This
code can be adopted to simulate the thermal evolution of NSs that
respect the spherical symmetry based on the numerical solutions of
the full general-relativistic energy balance equation and the energy
transport equation \cite{Thorne77},{
\begin{eqnarray}
\frac{e^{-2\varphi}}{4\pi R_r^2\left(1+z\right)}\frac{\partial\left(
L_{\gamma} e^{2\varphi}\right)}{\partial R_r}&=&-
Q_{\nu}-\frac{c_{\mathrm{v}}}{e^{\varphi}}\frac{\partial T}{\partial
t}, \label{eq:energybalance} \\
\frac{\kappa}{1+z}\frac{\partial \left(Te^{\varphi}\right)}{\partial
R_r}&=&\frac{- e^{\varphi}L_{\gamma}}{ 4\pi R_r^2}.
\label{eq:energytransport}
\end{eqnarray}
Here, $e^{\varphi}$ is the general-relativistic correction factor
with $\varphi$ being the gravitational potential. In addition, we
use $1 + z = \left(1 - 2M_r/R_r\right)^{-1/2}$, where $M_r$ is the
enclosed mass within the radial distance $R_r$, to denote the
gravitational redshift. In the Newtonian limit,
Eq.~(\ref{eq:energybalance}) is reduced to
Eq.~(\ref{eq:heatbalance}).} We insert the NFL-corrected quantities
given by Eqs.~(\ref{hcNFL}), (\ref{MUnNFL}), (\ref{DUNFL}),
(\ref{MUpNFL}), (\ref{NNBnNFL}), and (\ref{NNBnpNFL}) into the
package. The thermal evolution of an NS is known to be very
complicated, and it could be influenced by a multitude of
ingredients, including the NS EOS, the NS mass, the composition of
the heat-blanketing envelope, the neutron $^1S_0$-wave superfluid
model in the NS crust, the proton $^1S_0$-wave superconductivity
model in the NS core, and the neutron $^3P_2$-wave superfluid model
in the NS core, even though the initial magnetic field is ignored.
Different combinations of the involved parameters may lead to
distinctive cooling trajectories. To make a benchmark analysis, we
choose to utilize the APR EOS \cite{Akaml98} with an iron
heat-blanketing envelope \cite{Potekhin97}. We fix the neutron
\(^1S_0\)-wave superfluid model as the ``SFB'' model \cite{Page04,
Schwenk03} and the proton \(^1S_0\)-wave superconductivity model as
the ``T'' model\cite{Page04, Takatsuka73}, both characterized by a
Gaussian-like curve of transition temperature on the $T$-$\rho$
plane. The former features a maximum transition temperature of
$T^{\mathrm{max}}_{\mathrm{cn}}\sim5.0 \times 10^{9}~\mathrm{K}$,
while the latter has a maximum transition temperature of
$T^{\mathrm{max}}_{\mathrm{cp}}\sim3.3 \times 10^{9}~\mathrm{K}$.
Additionally, we set the initial internal temperature of the NSs in
the cooling simulation to $10^{11}$ K \cite{Yakovlev05}. In the
current analysis, we consider merely nucleons and preclude the
potential existence of exotic particles.

The cooling curves obtained under various conditions are shown in
Fig.~\ref{14feweak}. One can find a generic, condition-independent
tendency that the inclusion of NFL behavior substantially prolongs
the thermal relaxation time, which promotes the cooling process
without invoking DU processes \cite{Lattimer91, Page92, Pethick92}
or the ejection of axions \cite{Iwamoto84}.

Let us first consider an isolated NS having a mass
$1.4\mathrm{M}_{\odot}$ and a weak $^{3}P_2~(m_J=0)$-wave
superfluidity. {The weak superfluid model, denoted as ``a'', is
adapted from Refs.~\cite{Page04, Baldo98}. It exhibits a
Gaussian-like dependence of $T_{\mathrm{cn}}$ on $\rho$, with a
maximum transition temperature of $T^{\mathrm{max}}_{\mathrm{cn}}
\sim 10^{9}~\mathrm{K}$.} The cooling curves, represented by the
solid lines in Fig.~\ref{14feweak}, were generated by varying the
coupling parameter $h$. In the absence of NFL behavior, i.e., when
$h=0$, the red solid line can be divided into three distinct
segments based on its gradient. The most rapid decline occurs
approximately between $10$ and $100$ yr, which characterizes the
stage of internal thermal relaxation. Following this stage is the
neutrino cooling stage, extending from $100$ to $10^5$ yr. The last
stage, the photon cooling stage, starts at the age of $\sim 10^5$
yr.

Next we turn on NFL behavior by choosing a series of values of $h$.
From the blue solid line with $h=1$, we observe that the generic
three-stage shape is maintained. However, the thermal relaxation
stage is postponed from the initial $10$-$100$ yr to a new range of
$40$-$400$ yr. When $h$ increases to $3.0$, as indicated by the dark
green line, the thermal relaxation stage is prolonged to beyond
$1000$ yr. There is a considerable extension of the thermal relaxation
time $t_w$. As $h$ further increases, the starting point of the thermal
relaxation stage is further delayed, and $t_w$ becomes even larger.
This change can be understood by revisiting
Eq.~(\ref{eq:relaxationtime}), which shows that $t_w$ is
proportional to the total heat capacity $C_{\mathrm{v}}$. The NFL
behavior leads to a logarithmic enhancement of the neutrons'
contribution to $\tilde{C}_{\mathrm{v}}$ in the vicinity of the
crust. As $h$ grows, this logarithmic correction is enhanced, which
then increases $t_w$. On the other hand, the solid cooling curves
are immune to the change of $h$ throughout the neutrino cooling
stage, which is attributed to the fact that the numerator and
denominator of the cooling rate $\tilde{L}_\nu/\tilde{C}_{\mathrm{v}}$
are renormalized by NFL behavior to nearly the same extent. As the
crossover to the photon cooling stage occurs, the photon luminosity
surpasses the neutrino luminosity. The heat capacity, while being
reduced in the superfluid state, has nearly the same value as that
during the neutrino cooling stage. Then an increase in $h$ leads to a
gradual decline in the ratio $L_\gamma/\tilde{C}_{\mathrm{v}}$,
which yields a decelerated cooling rate.

\begin{widetext}

\begin{figure}[htbp]
\centering
\includegraphics[width=6.2in]{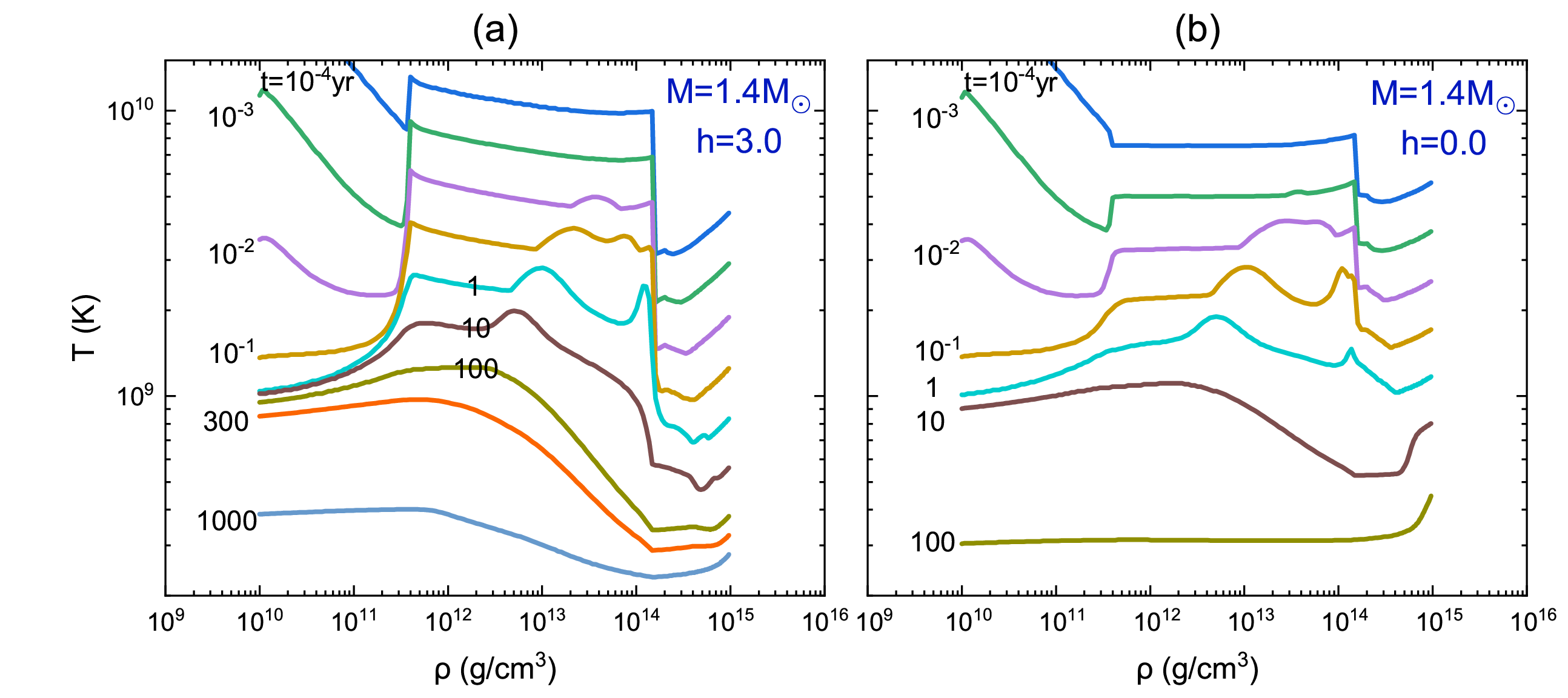}
\caption{{Internal temperature
profiles of a $1.4\mathrm{M}_{\odot}$ NS.} (a): NFL behavior is
present with $h=3.0$. (b): NFL behavior is absent with
$h=0.0$. The numbers adjacent to the curves indicate the stellar
ages. Contour levels are set at $10^{-4}$, $10^{-3}$, $10^{-2}$,
$10^{-1}$, $1$, $10$, $100$, $300$, and $1000$ yr for panel (a), and
at $10^{-4}$, $10^{-3}$, $10^{-2}$, $10^{-1}$, $1$, $10$, and $100$
yr for panel (b).} \label{NSrhoT}
\end{figure}

\end{widetext}

In Fig.~\ref{NSrhoT}, we show the internal temperature profiles of a
$1.4\mathrm{M}_{\odot}$ NS, with and without considering NFL
behavior. In the outer crust with
$\rho<4.0\times10^{11}~\mathrm{g/cm^3}$, the temperature profiles
shown in Figs. 5(a) and 5(b) are indistinguishable before thermal
equilibrium is reached, indicating that the NFL behavior only plays
a minor role. This is because this region is devoid of free
neutrons. In comparison, in the inner crust
($4.0\times10^{11}~\mathrm{g/cm^3}<\rho<2.0\times10^{14}~\mathrm{g/cm^3}$)
and the core region ($\rho>2.0\times10^{14}~\mathrm{g/cm^3}$), the
temperature profiles in Figs. 5(a) and 5(b) manifest obvious
differences. The NFL behavior results in a steeper temperature
gradient between the two regions compared with the case without NFL
behavior, which in turn leads to a longer time scale to reach
thermal equilibrium. As a result, the NS in Figs. 5(b) becomes
isothermal within 100 years, while the one in Figs. 5(a) takes over
1000 years to reach the same state. Notably, within the NFL
scenario, cooling in the inner crust is slower at lower densities
than at higher densities, contrasting with the FL scenario where
cooling is more uniform across all densities. This marked
temperature gradient in the inner crust induced by NFL behavior
yields a cooling curve that displays a period of temperature plateau
during the early stages of cooling, as depicted in
Fig.~\ref{14feweak}.

We compare our theoretical results with the existing observational
data for a number of isolated NSs \cite{Potekhin20}. The data
include the values of the age $t_{\ast}$ and the effective surface
temperature $T^{\infty}_e$ observed by a distant observer. The
selection of the NS age is based on its independence from timing.
From the results shown in Fig.~2, one could find that the variation
of the value of $h$ leads to an excellent agreement between our
theoretical cooling curves and the data set of most isolated NSs,
including all the TINSs and the majority of PSRs.

It is worth mentioning that the observed cooling data of the NS in Cas A
(TINS $6$) are well consistent with the solid cooling curve obtained
by taking $h=1.8$. The rapid decline in its $T^{\infty}_e$ can be
naturally explained if this NS is still in the internal thermal
relaxation stage. This offers an alternative interpretation of the
rapid cooling of the NS in Cas A. We will present a more in-depth
analysis of the cooling history of this NS in a separate work.

Nonetheless, there are two distinct classes of isolated NSs whose
cooling trajectories can hardly be explained by the solid cooling
curves. The first class includes the very young NSs that are found
to be unexpectedly cold, such as the PSRs labeled as $7$ and $11$.
The second class comprises a number of old NSs that are unexpectedly
warm, including the four XINSs labeled as $13-16$. To reproduce
the cooling data of these peculiar NSs, it is necessary to modify
some of the initial conditions. Here, we choose to adjust the NS
mass and/or the maximum value of the neutron $^3P_2$ superfluid
$T_{\mathrm{cn}}$.

For the very young but cold NSs, such as PSRs $7$ and $11$, we
suppose that they have a relatively large mass
$2.0\mathrm{M}_{\odot}$ and then obtain the respective cooling
curves, represented by two dotted lines in Fig.~\ref{14feweak}. We
see that the cooling curve at $h=3.0$ intersects the data bar of
PSR $7$ and the cooling curve at $h=13.0$ intersects the data bar
of PSR $11$. The distinctive fast cooling of these NSs
\cite{Marino24} arises from two ingredients. The first one is the
occurrence of the DU process \cite{Lattimer91, Page92, Pethick92} in the
ultrahigh-density center of massive NSs. The second one is the
logarithmic correction to the DU neutrino emissivity, given by
Eq.~(\ref{DUNFL}), induced by the NFL behavior. However, our results
suggest that the superfluid fluctuations have quite different
strengths in these two massive NSs.

Regarding the old but warm NSs, exemplified by the four XINSs
$13-16$, we choose a mass of $1.4~\mathrm{M}_{\odot}$ and meanwhile
assume the presence of strong $^{3}P_2~(m_J=0)$-wave superfluidity.
The strong superfluid model, labeled as ``c'', is taken from
Refs.~\cite{Page04, Baldo98}. This model exhibits a Gaussian-like
dependence of $T_{\mathrm{cn}}$ on $\rho$, with a
$T^{\mathrm{max}}_{\mathrm{cn}}$ of the order of
$10^{10}~\mathrm{K}$. Since the superfluid $T_{\mathrm{cn}}$ is
relatively high, there is a more significant suppression of the
specific heat and the neutrino emissivity. As depicted in Fig.~2,
the dashed cooling curves pass through the data points for three of
the XINSs, namely $14$, $15$, and $16$, if $h$ takes values within
the range $7-16$. It appears that their thermal evolution can be
explained if they have experienced very strong superfluid quantum
fluctuations and a very large maximum $T_{\mathrm{cn}}$ in their
long cooling histories. The XINS labeled by $13$ cannot be explained
by all the cooling curves with $h \leq 30$ shown in
Fig.~\ref{14feweak}. It is likely that this special NS has gone
through a more complicated thermal evolution than other NSs. In
particular, it may have been reheated by accretion processes
\cite{Haensel03} or magnetic heating processes \cite{Beloborodov16}.

As explained in Appendix \ref{sec:appd}, each of the four coupling
parameters, i.e., $h_{1}$ to $h_{4}$, defined at four QCPs is
related to the central and noncentral components of the two-body
potential $V(\mathbf{r})$. The constant parameter $h$ used in our
calculations serves as the average of these four $h$ parameters.
Apparently, $h$ must reflect the relative magnitudes of the
noncentral potential \(V_{\mathbf{nc}}(\mathbf{r})\) and the central
potential \(V_{\mathbf{c}}(\mathbf{r})\). However, deriving a
stringent constraint on the ratio
\(V_{\mathbf{nc}}(\mathbf{r})/V_{\mathbf{c}}(\mathbf{r})\) is quite
challenging. Currently, the value of $h$ can only be ascertained
through fitting to astrophysical observations of NS cooling. After
performing extensive simulations, we have obtained physically
meaningful cooling curves only when $h\leq 30$. The cooling curves
simulated by the NScool code \cite{Page16} are divergent once
$h>30$. The divergence is encountered whenever
$\tilde{L}_\nu\gg\tilde{C}_{\mathrm{v}}$, which then results in
inconsistent numerical outcomes. It will be interesting to
investigate whether the constraint of $h \leq 30$ leads to any
useful restriction on the two-body potentials in future works.


\section{Summary \label{sec:summary}}

In summary, we have studied the superfluid quantum criticality
emerging in the NS interior and examined its influence on the
thermal evolution. After carrying out an extensive field-theoretic
analysis of the effective model for this quantum criticality, we
have revealed that the quantum critical fluctuations of the
$^3P_{2}$-wave superfluid order parameter produce an unusual NFL
behavior and that this NFL behavior leads to a logarithmic
correction of the $T$ dependence of neutrons' heat capacity and the
neutrino emissivity. Then we incorporated these effects into the
thermal relaxation time and the heat balance equation. Using the
obtained results, we demonstrated that the theoretical cooling
curves can account for the cooling history of a variety of NSs by
adjusting several parameters, including the NS mass, the magnitude
of the superfluid $T_{\mathrm{cn}}$, and the coupling constant $h$.

Our results suggest the existence of an intimate correlation between
superfluid quantum criticality and the thermal evolution of NSs.
Hopefully, the present work will stimulate more applications of
canonical condensed-matter concepts to the theoretical description
of the diverse astrophysical observations of NSs.

Several approximations were employed in our analysis. For instance,
the perturbative and RG calculations were performed at the leading
order of the $1/N$ expansion. Subleading-order corrections may more
or less modify the leading-order results. Additionally, the
parameter $h$ was assumed to be an adjustable constant. In reality,
$h$ is not a constant. Its value depends on several ingredients,
including the two-body nuclear potential (see Appendix
\ref{sec:appd}), the critical densities of all QCPs, the
temperature, and even the cooling history. In order to gain a more
accurate description of the NS cooling trajectory, it is important
to carry out a more elaborate investigation to explore the
dependence of $h$ on these quantities.

Apart from the NS mass, the magnitude of the $^3P_2$-wave superfluid
$T_{\mathrm{cn}}$, and the parameter $h$, the complex NS cooling
history also relies on several other factors, such as the
composition of the heat-blanketing envelope, the proton $^1S_0$-wave
superconducting model, the presence of an initial magnetic field,
and the probable existence of exotic particles \cite{Lattimer04},
e.g., hyperons, Bose-condensed mesons, or deconfined quarks. A
comprehensive analysis of the many possible combinations of these
effects is beyond the scope of this paper, and will be reported in
future works.

Realistic NSs are not static, but are rapidly rotating. This implies
that quantum vortices may be present in the inner crust. The
vortices can considerably increase the neutron specific heat due to
the breaking of Cooper pairs in vortex cores and the superflow
around these vortices \cite{Pecak21}. Allard and Chamel argued
\cite{Chamel23} that the superflow induced by the pinning of
vortices might lead to a gapless neutron superfluidity in the crust,
which provides a possible explanation of the late-time cooling of
some transiently accreting NSs \cite{Chamel24}. We emphasize that
this scenario is entirely different from the superfluid quantum
criticality studied in our work. Such a gapless superfluidity does
not induce NFL behavior, nor does it alter the neutrino emissivity.
Its main impact on NS cooling is to convert the exponential $T$
dependence of the neutron specific heat in the superfluid state of
the inner crust into an FL-like linear $T$ dependence. In
comparison, superfluid quantum criticality is characterized by the
emergence of the NFL behavior and the resulting $\sim T\ln T$
corrections to the neutron specific heat and total neutrino
emissivity. It is possible that superfluid quantum criticality and
gapless superfluidity coexist in one specific NS. In that case, it
would be necessary to combine their effects.

In addition to neutron superfluidity, the $^{1}S_{0}$-wave proton
superconductivity may be formed in the NS core due to Cooper pairing
of protons \cite{Dean03, Lombardoreview, Pagereview, Sedrakian19}.
Its presence is supposed to play a crucial role in the minimal
cooling paradigm \cite{Page11, Kaminker02, Gusakov04}. The proton
superconducting transition temperature $T_{\mathrm{cp}}$ also
exhibits a Gaussian-like density dependence, which is associated
with two superconducting QCPs. Similar to superfluid quantum
criticality, NFL behavior is expected to emerge near these two QCPs.
The impact of NFL behavior on the NS cooling history can be examined
using the approach developed in the present work.

\section*{Acknowledgements}

We would like to thank the anonymous referees for their insightful
and constructive comments, which have contributed to the improvement
of this paper. This work is supported by the National Natural
Science Foundation of China under Grants No.~12433002, No.~12073026,
and No.~12274414, and also by the Anhui Natural Science Foundation
under Grant No.~2208085MA11. H.-F.Z and X.W. acknowledge the support
by the Cyrus Chun Ying Tang Foundations and the 111 Project for
Observational and Theoretical Research on Dark Matter and Dark
Energy (B23042). The numerical calculations in this paper have been
done on the supercomputing system in the Supercomputing Center of
University of Science and Technology of China.

\appendix

\begin{widetext}

\section{Perturbative calculations \label{sec:appa}}

In this appendix, we provide the calculational details that lead to
the polarization function and the neutron self-energy function that
are used in Sec.~\ref{sec:nflbehavior}.

The one-loop polarization function is calculated as follows:
\begin{eqnarray}
\Pi(i\Omega,\mathbf{q})&=& N h^{2}\int\frac{R(\theta)d\omega
d^{3}\mathbf{k}}{(2\pi)^{4}}G_{+}(\omega,\mathbf{k})
G_{+}(\omega+\Omega,\mathbf{k}+\mathbf{q})\nonumber \\
&=& Nh^{2}\int\frac{R(\theta)d\omega d^{2}\mathbf{k}_{\perp} dk_{z}}{(2
\pi)^{4}}\frac{1}{-i\omega+\frac{c^{2}_{f}}{2\mu} \mathbf{k}^{2}_{\perp} +
\frac{c^{2}_{f}k^{~}_{\mathrm{F}}}{\mu}k_{z}}\frac{1}{-i(\omega+\Omega)
+ \frac{c^{2}_{f}}{2\mu}(\mathbf{k}_{\perp}+\mathbf{q}_{\perp})^2 +
\frac{c^{2}_{f}k^{~}_{\mathrm{F}}}{\mu}(k_{z}+q_{z})},
\label{eq:appendixpi}
\end{eqnarray}
where $\mathbf{k}^{2}_{\perp}=k^{2}_{x}+k^{2}_{y}$. We are mainly interested
in the singular contribution of $\Pi(i\Omega,\mathbf{q})$. Such a
contribution is insensitive to which integration variable is
integrated first. We find it convenient to integrate over $k_{z}$
ahead of $\omega$. Defining $k_{z} = \frac{\mu}{c^{2}_{f}
k^{~}_{\mathrm{F}}}\zeta$, we find that the integration over $\zeta$
leads to
\begin{eqnarray}
\Pi(i\Omega,\mathbf{q}) &=& \frac{Nh^{2}\mu}{k^{~}_{\mathrm{F}}
c^{2}_{f}}\int\frac{R(\theta) d\omega d^{2}\mathbf{k}_{\perp}}{(2
\pi)^{4}}\oint d\zeta \frac{1}{-i\omega+\zeta +
\frac{c^{2}_{f}}{2\mu} \mathbf{k}^{2}_{\perp}} \frac{1}{-i(\omega+\Omega) +
\zeta +\frac{c^{2}_{f} k^{~}_{\mathrm{F}}}{\mu}q_{z} +
\frac{c^{2}_{f}}{2\mu}(\mathbf{k}_{\perp}+\mathbf{q}_{\perp})^2}
\nonumber \\
&=& \frac{Nh^{2}\mu}{k^{~}_{\mathrm{F}} c^{2}_{f}}\int
\frac{R(\theta) d\omega d^{2} \mathbf{k}_{\perp}}{(2\pi)^4}2\pi i
\frac{\vartheta(\omega+\Omega)\vartheta(-\omega)-\vartheta(-\omega-\Omega)
\vartheta(\omega)}{i\Omega-\frac{c^{2}_{f}k^{~}_{\mathrm{F}}}{\mu}
q_{z}-\frac{2c^{2}_{f}\mathbf{q}_{\perp}\cdot\mathbf{k}_{\perp} +
c^{2}_{f}\mathbf{q}_{\perp}^2}{2\mu}}
\nonumber \\
&=& \frac{iN h^{2}\mu}{k^{~}_{\mathrm{F}}c^{2}_{f}}\int
\frac{R(\theta)d^{2}\mathbf{k}_{\perp}}{(2\pi)^2}\int
\frac{d\omega}{2 \pi}\frac{\mathrm{sign}(\omega+\Omega) -
\mathrm{sign}(\omega)}{i\Omega-\frac{c^{2}_{f}
k^{~}_{\mathrm{F}}}{\mu}q_{z}-\frac{2c^{2}_{f}\mathbf{q}_{\perp}
\cdot\mathbf{k}_{\perp}+c^{2}_{f}\mathbf{q}_{\perp}^2}{2\mu}}
\nonumber \\
&=& \frac{iNh^{2}\mu\Omega}{k^{~}_{\mathrm{F}}c^{2}_{f}\pi}
\int\frac{R(\theta) d^{2} \mathbf{k}_{\perp}}{(2\pi)^{2}}
\frac{1}{i\Omega-\frac{c^{2}_{f} k^{~}_{\mathrm{F}}}{\mu}q_{z} -
\frac{2c^{2}_{f}\mathbf{q}_{\perp}\cdot\mathbf{k}_{\perp} + c^{2}_{f}
\mathbf{q}_{\perp}^2}{2\mu}} \nonumber \\
&=& \frac{iNh^{2}\mu\Omega}{k^{~}_{\mathrm{F}}c^{2}_{f}\pi}
\int^{\infty}_{0}
\frac{|\mathbf{k}_{\perp}|d|\mathbf{k}_{\perp}|}{(2\pi)^2}
\int^{2\pi}_0 d\theta \frac{R(\theta)}{i\Omega -
\frac{c^{2}_{f}k^{~}_{\mathrm{F}}}{\mu}q_{z}-\frac{2c^{2}_{f}
|\mathbf{q}_{\perp}||\mathbf{k}_{\perp}|\cos\theta + c^{2}_{f}
\mathbf{q}_{\perp}^2}{2\mu}}
\nonumber \\
&=& \frac{N h^{2}\mu^2\Omega}{k^{~}_{\mathrm{F}}c^4_{f}\pi|
\mathbf{q}_{\perp}|}\int^{\infty}_{0}\frac{d|\mathbf{k}_{\perp}|}{(2\pi)^2}
\int^{2\pi}_0 d\theta \frac{R(\theta)}{\frac{\mu\Omega}{c^{2}_{f}
|\mathbf{q}_{\perp}||\mathbf{k}_{\perp}|}+i\frac{k^{~}_{\mathrm{F}}
q_{z}}{|\mathbf{q}_{\perp}||\mathbf{k}_{\perp}|} +
i\frac{|\mathbf{q}_{\perp}|}{2|\mathbf{k}_{\perp}|}+i\cos\theta}.
\label{eq:}
\end{eqnarray}
The integration over $\theta$ leads to
\begin{eqnarray}
\int^{2\pi}_{0} d\theta \frac{R(\theta)}{z+i\cos\theta} =
\begin{cases}
2\pi \mathrm{sign}[\mathrm{Re}(z)]
\left(\frac{1}{\sqrt{z^2+1}}\right) &
^{1}S_0, \\
2\pi \mathrm{sign}[\mathrm{Re}(z)]\left(\sqrt{z^{2}+1}-z\right) &
^{3}P_{2,\pm 2}, \\
2\pi \mathrm{sign}[\mathrm{Re}(z)]\left(\frac{4}{\sqrt{z^{2}+1}} -
3\sqrt{z^{2} + 1}+3z\right) & ^{3}P_{2,0}.\\
\end{cases}
\label{eq:}
\end{eqnarray}
Here,
\begin{eqnarray}
z=\frac{\mu\Omega}{c^{2}_{f}|\mathbf{q}_{\perp}|
|\mathbf{k}_{\perp}|}+i\frac{k^{~}_{\mathrm{F}}
q_{z}}{|\mathbf{q}_{\perp}| |\mathbf{k}_{\perp}|} +
i\frac{|\mathbf{q}_{\perp}|}{2|\mathbf{k}_{\perp}|}.
\end{eqnarray}
We introduce another variable $z^{\prime}=\frac{\mu\Omega}{c^{2}_{f}
|\mathbf{q}_{\perp}|} + i\frac{k^{~}_{\mathrm{F}}
q_{z}}{|\mathbf{q}_{\perp}|} + i\frac{|\mathbf{q}_{\perp}|}{2}$ to
complete the calculation. In the case of the $^{3}P_2~(m_J=0)$-wave gap,
the calculation is performed as follows
\begin{eqnarray}
\Pi(i\Omega,\mathbf{q}) &=& \frac{N h^{2}\mu^{2} \Omega}{
k^{~}_{\mathrm{F}}c^{4}_{f}\pi|\mathbf{q}_{\perp}|}\int^{\infty}_{0}
\frac{d|\mathbf{k}_{\perp}|}{2\pi}\mathrm{sign}\left(\frac{\mu\Omega}{c^{2}_{f}
|\mathbf{q}_{\perp}||\mathbf{k}_{\perp}|}\right)\left[\frac{4}{\sqrt{z^{2}+1}} -
3\sqrt{z^{2}+1}+3z\right] \nonumber \\
&\approx& \frac{N h^{2}\mu^2|\Omega|}{2k^{~}_{\mathrm{F}}c^{4}_{f}
\pi^{2}|\mathbf{q}_{\perp}|}\int^{\Lambda}_{0}d|\mathbf{k}_{\perp}|\left[\frac{4
|\mathbf{k}_{\perp}|}{\sqrt{z^{\prime2}+\mathbf{k}^{2}_{\perp}}} -
3\frac{\sqrt{z^{\prime2} +
\mathbf{k}^{2}_{\perp}}}{|\mathbf{k}_{\perp}|} +
3\frac{z^{\prime}}{|\mathbf{k}_{\perp}|}\right]
\nonumber \\
&\approx& \frac{N h^{2}\mu^{2}\Lambda}{2k^{~}_{\mathrm{F}}c^{4}_{f}
\pi^{2}}\frac{|\Omega|}{|\mathbf{q}_{\perp}|} \nonumber \\
&=& N\gamma\frac{|\Omega|}{|\mathbf{q}_{\perp}|},
\label{eq:appendixlooppi}
\end{eqnarray}
where $\gamma=\frac{h^{2}\mu^{2}\Lambda}{2k^{~}_{\mathrm{F}}
c^{4}_{f}\pi^2}$. The polarization for the other two pairing gaps is
also given by this expression if $\Lambda$ is large enough.

To facilitate the calculation of the one-loop fermion self-energy at an
arbitrary $r$, we redefine $r$ as $r/|\mathbf{q}_{\perp}|$. This
redefinition enables us to derive the following results:
\begin{eqnarray}
\Sigma(i\omega,\mathbf{k};r) &=& -h^{2}\int\frac{R(\theta)d\Omega
d^{3}\mathbf{q}}{(2\pi)^{4}}\tilde{D}(\Omega,\mathbf{q};r)
G_{+}(\omega-\Omega,\mathbf{k}-\mathbf{q})\nonumber \\
&=& \frac{h^{2}}{N}\int\frac{R(\theta)\theta d\Omega d^{2}
\mathbf{q}_{\perp}}{(2\pi)^{d}}\frac{1}{c^{2}_{b}\mathbf{q}^{2}_{\perp}
+\gamma\frac{|\Omega|}{|\mathbf{q}_{\perp}|}+\frac{r}{|\mathbf{q}_{\perp}|}}
\int\frac{dq_{z}}{2\pi}
\frac{1}{i(\omega-\Omega)-\frac{c^{2}_{f}}{2\mu}(\mathbf{k}_{\perp}
-\mathbf{q}_{\perp})^2-\frac{c^{2}_{f}k^{~}_{\mathrm{F}}}{\mu}(k_{z}-q_{z})}
\nonumber \\
&=& \frac{h^{2}}{Nc^{2}_{b}}\int\frac{R(\theta) d\Omega d^{2}
\mathbf{q}_{\perp}}{(2\pi)^d}\frac{1}{\mathbf{q}^{2}_{\perp} +
\frac{\gamma}{c^{2}_{b}}\frac{|\Omega|}{|\mathbf{q}_{\perp}|} +
\frac{r}{c^{2}_{b}|\mathbf{q}_{\perp}|}}\left(\frac{-i\mu}{k^{~}_{\mathrm{F}}
c^{2}_{f}}\right)\mathrm{sign}(\omega-\Omega)
\nonumber \\
&=& -\frac{i\mu h^{2}}{Nc^{2}_{b}k^{~}_{\mathrm{F}}c^{2}_{f}}
\left(\int^{2\pi}_0\frac{R(\theta)d\theta}{2\pi}\right)\int^{\infty}_0
\frac{d|\mathbf{q}_{\perp}|}{2\pi}\int^{\infty}_{-\infty}
\frac{d\Omega}{2\pi}\frac{\mathbf{q}^{2}_{\perp}}{{
|\mathbf{q}_{\perp}|}^{3} + \frac{\gamma}{c^{2}_{b}} |\Omega| +
\frac{r}{c^{2}_{b}}}\mathrm{sign}(\omega-\Omega).
\label{eq:roneloopfermion}
\end{eqnarray}
We integrate $q_{z}$ and $\Omega$ in order. The variable
$\Omega$ is integrated out by using the formula
\begin{eqnarray}
\int d\Omega\frac{\mathrm{sign}(\omega-\Omega)}{{|\mathbf{q}_{\perp}|}^3 +
\frac{\gamma}{c^{2}_{b}}|\Omega|+\frac{r}{c^{2}_{b}}} &=&
\mathrm{sign}(\omega)\frac{2c^{2}_{b}}{\gamma}\ln
\left(\frac{{|\mathbf{q}_{\perp}|}^3+\frac{\gamma}{c^{2}_{b}}|\omega| +
\frac{r}{c^{2}_{b}}}{{|\mathbf{q}_{\perp}|}^3+\frac{r}{c^{2}_{b}}}\right).
\label{eq:omegaintegral2}
\end{eqnarray}

Now, by inserting (\ref{eq:roneloopfermion}) into
(\ref{eq:omegaintegral2}), we obtain
\begin{eqnarray}
\Sigma(i\omega;r) &=& -\frac{i\mu h^{2}}{N
k^{~}_{\mathrm{F}}c^{2}_{f}\pi \gamma}\left(\int^{2\pi}_0
\frac{R(\theta) d\theta}{2\pi}\right)\mathrm{sign}(\omega)
\int^\infty_{0}\frac{d|\mathbf{q}_{\perp}|}{2\pi} \mathbf{q}_{\perp}^{2}
\ln\left(\frac{{|\mathbf{q}_{\perp}|}^3+\frac{\gamma}{c^{2}_{b}} |\omega| +
\frac{r}{c^{2}_{b}}}{{|\mathbf{q}_{\perp}|}^3+\frac{r}{c^{2}_{b}}}\right)
\nonumber \\
&\approx& -\frac{i\mu h^{2}}{N k^{~}_{\mathrm{F}}c^{2}_{f}\pi
\gamma}\left(\int^{2\pi}_0 \frac{R(\theta)
d\theta}{2\pi}\right)\mathrm{sign}(\omega)
\int^\Lambda_{0}\frac{d|\mathbf{q}_{\perp}|}{2\pi} \mathbf{q}_{\perp}^{2}
\ln\left(\frac{{|\mathbf{q}_{\perp}|}^3+\frac{\gamma}{c^{2}_{b}} |\omega| +
\frac{r}{c^{2}_{b}}}{{|\mathbf{q}_{\perp}|}^3+\frac{r}{c^{2}_{b}}}\right)
\nonumber \\
&=&-\frac{i\mu h^2}{6Nk^{~}_{\mathrm{F}}c^2_{\psi}\pi^2\gamma}
\left(\int^{2\pi}_0 \frac{R(\theta)d\theta}{2\pi}\right)
\mathrm{sign}(\omega) \nonumber
\\
&& \times\left[|\omega|\frac{\gamma}{c^2_{b}}\ln\left(1+
\frac{c^2_{b}\Lambda^3}{r+\gamma|\omega|}\right)+{\Lambda^3}
\ln\left(1+\frac{\gamma|\omega|}{c^2_{b}\Lambda^3+r}\right) +
\frac{r}{c^2_{b}} \ln\left((1+\frac{\gamma|\omega|}{c^2_{b}\Lambda^3+r})
(\frac{r}{\gamma|\omega|+r})\right)\right]\nonumber \\
&\approx& -\frac{i\mu h^{2}}{6N c^{2}_{b}k^{~}_{\mathrm{F}}
c^{2}_{f}\pi^{2}}\left(\int^{2\pi}_0\frac{R(\theta)d\theta}{2\pi}
\right)\mathrm{sign}(\omega)|\omega|\ln\left(\frac{c^{2}_{b}
\Lambda^3}{r+\gamma|\omega|}\right)~~ \left(\omega\rightarrow 0,
\Lambda\rightarrow\infty\right). \label{eq:sigmaomegarappen}
\end{eqnarray}
Here, the integration over angle is
\begin{eqnarray}
\int^{2\pi}_0\frac{R(\theta) d\theta}{2\pi}=
\begin{cases}
1 & ^{1}S_0, \\
\frac{1}{2} & ^{3}P_{2,\pm 2},\\
\frac{5}{2} & ^{3}P_{2,0}.\\
\end{cases}
\label{eq:}
\end{eqnarray}
In the above calculations, we have considered the low-energy region
$\omega\rightarrow 0$.

\section{{One-loop RG calculation} \label{sec:appb}}

Here, we present a detailed RG analysis of the effective field
theory for superfluid quantum criticality and derive the flow
equations of all the parameters appearing in the model.

The energy and/or momentum are initially defined within the range
$[0,\Lambda]$. It is customary to divide this range into
$[0,b\Lambda]$ and $[b\Lambda,\Lambda]$, where $b$ is a constant
satisfying the condition $b<1$. The field operators defined within
$[0,b\Lambda]$ and $[b\Lambda,\Lambda]$ are called \cite{Shankar94}
slow modes and fast modes, respectively. We separate all the field
operators into slow and fast modes \cite{Shankar94} as follows
\begin{eqnarray}
&&\phi=\phi_{s}+\phi_{f}, \quad \phi^{\ast} = \phi^{\ast}_{s} +
\phi^{\ast}_{f}, \\
&&\psi_{+}=\psi_{+s}+\psi_{+f}, \quad \psi^{\ast}_{+} = \psi^{\ast}_{+s} +
\psi^{\ast}_{+f}.
\end{eqnarray}
Here, $\phi_{s}$ and $\psi_{+s}$ are slow modes, and $\phi_{f}$ and
$\psi_{+f}$ are fast modes. Then the action $S$ can be formally
decomposed into three parts
\begin{eqnarray}
S=S[s]+S[f]+S[s,f],
\end{eqnarray}
where $S[s]$ contains only slow modes, $S[f]$ contains only fast
modes, and $S[s,f]$ contains both slow and fast modes. Accordingly,
the partition function $Z$ can be expressed as
\begin{eqnarray}
Z &=& \int D\phi_{s}D\phi_{s}^{\ast} D\psi_{+s} D\psi^{\ast}_{+s}
e^{-S^{s}_{0}} \nonumber \\
&\times& \int D\phi_{f} D\phi_{f}^{\ast} D\psi_{+f} D\psi^{\ast}_{+f}
e^{-S^{f}_0}e^{-S_{\phi^4}-S_{\psi_{+}\phi}}.
\end{eqnarray}
Integrating out all fast modes yields
\begin{eqnarray}
Z \rightarrow Z^{f}_{0}\int D\phi_{s}D\phi^{\ast}_{s}D\psi_{+s}
D\psi^{\ast}_{+s}e^{-S^{s}_{0}}\langle e^{-S_{\phi^4} -
S_{\psi_{+}\phi}} \rangle^{~}_{f},\nonumber
\end{eqnarray}
where
\begin{eqnarray}
Z^{f}_{0} &=& \int D\phi_{f} D\phi_{f}^\ast D\psi_{+f} D\psi^{\ast}_{+f}
e^{-S^{f}_0}, \\
\langle e^{-S_{I}}\rangle_f &=& \frac{1}{Z^{f}_{0}}\int D\phi_{f}
D\phi_{f}^{\ast} D\psi_{+f} D\psi^{\ast}_{+f}e^{-S^f_0}e^{-S_{I}}.
\label{eq:meanvalueesi}
\end{eqnarray}
Here, $S_I = S_{\phi^4} + S_{\psi_{+}\phi}$. The calculation of the
expectation value $\langle e^{-S_{I}}\rangle_f$ is the essential
part of RG analysis. Making use of the cumulant expansion method
\cite{Shankar94}, this expectation value can be expanded, up to the
order of $O\left[\left(1/N\right)\right]$, as
\begin{eqnarray}
\langle e^{-S_{I}}\rangle_{f} = e^{-\langle S_I \rangle_{f} -
\frac{1}{2}\langle S^{2}_{I}\rangle_{f}},
\end{eqnarray}
where a $\langle S_{I}\rangle^{2}_{f}$ term is dropped since it only
generates unconnected diagrams and makes zero contribution to the
final results.

The integral measure is decomposed (see Ref.~\cite{Wang17} for a
more comprehensive analysis) into two parts
\begin{eqnarray}
\int^{\Lambda}_{0} d\omega d^3\mathbf{k} &=&
\int^{\Lambda}_{b{\Lambda}} d\omega d^3\mathbf{k} +
\int^{b\Lambda}_{0} d\omega d^3\mathbf{k},
\label{RGscheme}
\end{eqnarray}
where
\begin{eqnarray}
\int^{\Lambda}_{b{\Lambda}} d\omega d^3\mathbf{k} &\equiv&
\int^\infty_{-\infty} d\omega \int^{\infty}_{-\infty}dk_z
\int^{2\pi}_{0} d\theta \int^\Lambda_{b\Lambda} |\mathbf{k}_\perp|
d|\mathbf{k}_{\perp}|, \nonumber \\
\int^{b\Lambda}_{0} d\omega d^3\mathbf{k} &\equiv&
\int^{\infty}_{-\infty} d\omega \int^{\infty}_{-\infty}dk_z
\int^{2\pi}_{0}d\theta\int^{b\Lambda}_{0}|\mathbf{k}_{\perp}|
d|\mathbf{k}_{\perp}|.
\nonumber
\end{eqnarray}

Let us first consider the noninteracting limit. The influence of
interactions will be analyzed later. In this limit, $Z$ is
simplified to
\begin{eqnarray}
Z &=& Z^{f}_{0}\int D\phi_{s}D\phi^{\ast}_{s}D\psi_{+s}
D\psi^{\ast}_{+s}e^{-S^{s}_{0}}\nonumber\\
&\propto&\int D\phi_{s}D\phi^{\ast}_{s}D\psi_{+s}
D\psi^{\ast}_{+s}e^{-S^{s}_{\psi_{+}}-S^{s}_{\phi}}.
\end{eqnarray}
We select the free neutron action $S_{0}$ as the free fixed point
and require that $S_{\psi_{+}}$ remains invariant after performing
the following scaling transformations:
\begin{eqnarray}
\omega &=& \omega^{\prime}b^{2},\label{omega0}\\
k_z &=& k^{\prime}_{z}b^{2},\label{kz0}\\
k_x &=& k^{\prime}_{x}b,\label{kx0}\\
k_y &=& k^{\prime}_{y}b,\label{ky0}\\
v^{~}_{\mathrm{F}} &=& v^{\prime}_{\mathrm{F}}b^{0}.
\label{vf0}
\end{eqnarray}
Applying these transformations to $S^{s}_{\psi_{+}}$ leads to
\begin{eqnarray}
S^{s}_{\psi_{+}}&=& \int^{ b{\Lambda}}_0 \frac{d\omega}{2
\pi}\frac{d^{3}\mathbf{k}}{(2\pi)^3}\psi^{\ast}_{+s} \left[-i\omega
+ \frac{v^{~}_\mathrm{F}}{2 k^{~}_\mathrm{F}}\left(k^{2}_{x} +
k^{2}_{y}\right) +
v^{~}_\mathrm{F} k_{z}\right]\psi_{+s} \nonumber \\
&=& b^8\int^{\Lambda}_0 \frac{d
\omega^{\prime}}{2\pi}\frac{d^{3}\mathbf{k^{\prime}}}{(2 \pi)^3}
\psi^{\ast}_{+s}\left[-i\omega^{\prime} +
\frac{v^{\prime}_\mathrm{F}}{2 k^{~}_\mathrm{F}} \left(k^{\prime
2}_{x} + k^{\prime2}_{y}\right)+v^{\prime}_\mathrm{F}
k^{\prime}_{z}\right]\psi_{+s}.\nonumber \label{freeaction}
\end{eqnarray}
Obviously, this form of $S^{s}_{\psi_{+}}$ is different from the
original $S_{\psi_{+}}$. To eliminate the extra factor $b^{8}$, we
let $\psi_{+s}$ transform as
\begin{eqnarray}
\psi_{+s} = \psi^{\prime}_{+}b^{-4}.\label{eq:scalingspinorfield}
\end{eqnarray}
Then $S^{s}_{\psi_{+}}$ is converted into
\begin{eqnarray}
S^{\prime}_{\psi_{+}} &=&\int^{{\Lambda}}_0
\frac{d\omega^{\prime}}{2\pi}\frac{d^{3}\mathbf{k^{\prime}}}{(2\pi)^3}
\psi^{\prime \ast}_{+}\left[-i\omega^{\prime} +
\frac{v^{\prime}_\mathrm{F}}{2 k^{~}_\mathrm{F}}
\left(k^{\prime2}_{x} + k^{\prime2}_{y}\right) +
v^{\prime}_\mathrm{F} k^{\prime}_{z}\right]\psi^{\prime}_{+},
\nonumber \label{NEWfermionfreeaction}
\end{eqnarray}
which has the same form as the original free action $S_{\psi_{+}}$.

The scaling transformations defined by
Eqs.~(\ref{omega0})$-$(\ref{ky0}) and
Eq.~(\ref{eq:scalingspinorfield}) will be used to handle the free
boson action $S_{\phi}$ and the interaction action $S_{I}$. In the
case of $S_{\phi}$, we consider the following scaling
transformations
\begin{eqnarray}
\Omega &=& \Omega^{\prime}b^{2},\label{Omega0}\\
q_z &=& q^{\prime}_{z}b^{2},\label{qz0}\\
q_x &=& q^{\prime}_{x}b,\label{qx0}\\
q_y &=& q^{\prime}_{y}b.\label{qy0}
\end{eqnarray}
Applying Eqs.~(\ref{Omega0}-\ref{qy0}) to $S^{s}_\phi$ leads to
\begin{eqnarray}
S^{s}_{\phi} = Nb^{8}\int^{{\Lambda}}_0
\frac{d\Omega^{\prime}}{2\pi} \frac{d^{3}\mathbf{q}^{\prime}}{(2
\pi)^{3}}\phi^{\ast}_{s}\left[\mathbf{q}_{\perp}^{\prime2} +
b^{-1}\gamma \frac{|\Omega^{\prime}
|}{|\mathbf{q}^{\prime}_{\perp}|}\right]\phi_{s}.
\label{eq:scaledbosonfreeaction}
\end{eqnarray}
To compare $S^{s}_{\phi}$ to the original form of $S_\phi$, the
boson field $\phi$ and the parameter $\gamma$ should be rescaled as
\begin{eqnarray}
\phi_{s} &=& \phi^{\prime}b^{-4},\label{Zb0}\\
\gamma &=& \gamma^{\prime}b^{1}.\label{gamma0}
\end{eqnarray}
Then $S^{s}_{\phi}$ given by Eq.~(\ref{eq:scaledbosonfreeaction})
becomes
\begin{eqnarray}
S^{\prime}_{\phi} = N\int^{\Lambda}_0 \frac{d
\Omega^{\prime}}{2\pi}\frac{d^{3}\mathbf{q}^{\prime}}{(2\pi)^{3}}
\phi^{\prime \ast}\left[\mathbf{q}_{\perp}^{\prime2} + \gamma^{\prime}
\frac{|\Omega^{\prime}|}{|\mathbf{q}^{\prime}_{\perp}|}\right]\phi^{\prime},
\label{Newbosonfreeaction}
\end{eqnarray}
which coincides with the original free action $S_\phi$. The same
procedure can be applied to the interaction action $S_{I}$ if $h$
and $\lambda$ are rescaled as follows:
\begin{eqnarray}
h &=& h^{\prime}b^{0}, \label{hf0} \\
\lambda &=& \lambda^{\prime}b^{-2}.\label{lambda0}
\end{eqnarray}

The constant $b$ can be rewritten as an exponential function
\begin{eqnarray}
b = e^{-l}, \label{eq:blrelation}
\end{eqnarray}
where $l$ is a positive length scale. The zero-energy limit $\omega
\rightarrow 0$ is equivalent to the long-wavelength limit $l
\rightarrow \infty$. The importance of a parameter can be measured
by its $l$ dependence. Suppose that some parameter $\xi$ is
transformed as
\begin{eqnarray}
\xi = \xi^{\prime}b^{a}.\label{eq:xiscaling}
\end{eqnarray}
For positive $a > 0$, the renormalized parameter $\xi^{\prime} = \xi
e^{al}$ goes to infinity as $l\rightarrow \infty$. Under these
conditions, $\xi$ is said to be relevant at low energies.
Conversely, for $a<0$, the renormalized parameter $\xi^{\prime}=\xi
e^{al}$ vanishes as $l\rightarrow \infty$. In this case, $\xi$ is
irrelevant and its influence on the system is negligible at low
energies. If $a=0$, the parameter $\xi$ does not change under RG
transformations and thus is classified as a marginal parameter. From
the expressions of Eq.~(\ref{vf0}), Eq.~(\ref{gamma0}),
Eq.~(\ref{hf0}), and Eq.~(\ref{lambda0}), we see that
$v^{~}_{\mathrm{F}}$ and $h$ are marginal parameters, $\gamma$ is a
relevant parameter, and $\lambda$ is an irrelevant parameter.
However, these results are valid only at the classical tree level.
To examine the actual effects of model parameters, we should go
beyond the tree level and extend our analysis to include the
loop-level quantum corrections.

We next wish to compute the fermion self-energy corrections.
To this end, we calculate the
expression $\delta S^{s}_{\psi_{+}}=-\frac{1}{2}\langle
S^{2}_{I}\rangle_{f}$, and find that
\begin{eqnarray}
\delta S^{s}_{\psi_{+}} &=& -h^{2}\int_{0}^{b \Lambda}
\frac{d\omega}{2\pi} \frac{d^3\mathbf{k}}{(2\pi)^3}
\psi^{\ast}_{+s}\psi^{~}_{+s} \int_{b \Lambda}^{\Lambda}
\frac{R(\theta)d\Omega}{2\pi}\frac{d^3\mathbf{q}}{(2\pi)^3}
\tilde{D}(\Omega,\mathbf{q})G_{+}(\omega-\Omega,
\mathbf{k}-\mathbf{q}).
\end{eqnarray}
Within the framework of RG theory \cite{Shankar94}, the loop
correction to the neutron action is calculated as follows:
\begin{eqnarray}
\delta S^{s}_{\psi_{+}} &=& -h^{2}\int^{b{\Lambda}}_{0}
\frac{d\omega}{2\pi} \frac{d^3\mathbf{k}}{(2\pi)^3}
\psi^{\ast}_{+s}\psi^{~}_{+s}\int^{\Lambda}_{b{\Lambda}}
\frac{R(\theta)d\Omega}{2\pi}\frac{d^3\mathbf{q}}{(2\pi)^3}
\tilde{D}(\Omega,\mathbf{q};r=0)G_{+}(\omega-\Omega,
\mathbf{k}-\mathbf{q}) \nonumber \\
&=& h^{2}\int^{b{\Lambda}}_0\frac{d\omega}{2\pi}\frac{d^3
\mathbf{k}}{(2\pi)^3}\psi^{\ast}_{+s}\psi^{~}_{+s}
\int^{\Lambda}_{b{\Lambda}} \frac{R(\theta) d\Omega}{2\pi}
\frac{d^3\mathbf{q}}{(2\pi)^3}\frac{1}{\mathbf{q}^{2}_{\perp}+\gamma
\frac{|\Omega|}{|\mathbf{q}_{\perp}|}}\frac{1}{i(\omega-\Omega) -
\frac{v^{~}_{\mathrm{F}}}{2k^{~}_{\mathrm{F}}}(\mathbf{k}_{\perp} -
\mathbf{q}_{\perp})^2-v^{~}_{\mathrm{F}}(k_{z}-q_{z})}\nonumber \\
&=&\frac{h^{2}}{N}\int^{ b{\Lambda}}_0\frac{d\omega}{2\pi}
\frac{d^3\mathbf{k}}{(2\pi)^3}\psi^{\ast}_{+s}\psi^{~}_{+s}
\int^{\Lambda}_{b{\Lambda}}\frac{R(\theta)d^2\mathbf{q}_{\perp}}{(2
\pi)^2}\int^{\infty}_{-\infty}\frac{d\Omega}{2\pi}
\frac{1}{\mathbf{q}^{2}_{\perp}+\gamma\frac{|\Omega|}{|\mathbf{q}_{\perp}|}}
\int^{\infty}_{-\infty}\frac{dq_{z}}{2\pi}\frac{1}{i(\omega-\Omega)
- \frac{v^{~}_{\mathrm{F}}}{2k^{~}_{\mathrm{F}}}
(\mathbf{k}_{\perp}-\mathbf{q}_{\perp})^2 -
v^{~}_{\mathrm{F}}(k_{z}-q_{z})}\nonumber \\
&=& \frac{h^{2}}{N}\int^{ b{\Lambda}}_0\frac{d\omega}{2\pi}
\frac{d^3\mathbf{k}}{(2\pi)^3}\psi^{\ast}_{+s}\psi^{~}_{+s}
\int^{\Lambda}_{b{\Lambda}}\frac{R(\theta)d^2\mathbf{q}_{\perp}}{(2
\pi)^2}\int^{\infty}_{-\infty}\frac{d\Omega}{2\pi}
\frac{|\mathbf{q}_{\perp}|}{{|\mathbf{q}_{\perp}|}^{3}+\gamma|\Omega|}
\frac{-i}{v^{~}_{\mathrm{F}}}\mathrm{sign}(\omega-\Omega)
\nonumber \\
&=& -\frac{i h^{2}}{Nv^{~}_{\mathrm{F}}}\left(\int^{2\pi}_0
\frac{R(\theta) d\theta}{2\pi}\right)\int^{ b{\Lambda}}_0
\frac{d\omega}{2\pi}\frac{d^3\mathbf{k}}{(2\pi)^3}\psi^{\ast}_{+s}
\psi^{~}_{+s}\int^{\Lambda}_{b{\Lambda}}\frac{d|\mathbf{q}_{\perp}|}{2\pi}
\int^{\infty}_{-\infty}\frac{d\Omega}{2\pi}
\frac{\mathbf{q}^{2}_{\perp}}{{|\mathbf{q}_{\perp}|}^3+\gamma|\Omega|}
\mathrm{sign}(\omega-\Omega) \nonumber \\
&=&-\frac{ih^{2}}{Nv^{~}_{\mathrm{F}}\pi\gamma}\left(\int^{2\pi}_0
\frac{R(\theta) d\theta}{2\pi}\right)\int^{ b{\Lambda}}_0
\frac{d\omega}{2\pi} \frac{d^3\mathbf{k}}{(2\pi)^3} \psi^{\ast}_{+s}
\psi^{~}_{+s}\mathrm{sign}(\omega)\int^\Lambda_{b\Lambda}
\frac{d|\mathbf{q}_{\perp}|}{2\pi} \mathbf{q}^{2}_{\perp}
\ln\frac{{|\mathbf{q}_{\perp}|}^{3} +
\gamma|\omega|}{{|\mathbf{q}_{\perp}|}^3}.
\label{eq:RGloopintegral}
\end{eqnarray}
Using the RG scheme (\ref{RGscheme}) and introducing a new
variable $\chi \equiv\frac{{|\mathbf{q}_{\perp}|}^3}{|\omega|}$, it
is found that
\begin{eqnarray}
\delta S^{s}_{\psi_{+}} &=& -\frac{ih^{2}}{3N v^{~}_{\mathrm{F}}\pi
\gamma}\left(\int^{2\pi}_0\frac{R(\theta)d\theta}{2\pi}\right)
\int^{b{\Lambda}}_0\frac{d\omega}{2\pi}
\frac{d^3\mathbf{k}}{(2\pi)^3}\psi^{\ast}_{+s}\psi^{~}_{+s}
\mathrm{sign}(\omega)|\omega|\int^{{\Lambda^3}
|\omega|^{-1}}_{{b^{3}\Lambda^{3}}|\omega|^{-1}}\frac{d\chi}{2\pi}
\ln\frac{\chi + \gamma}{\chi} \nonumber \\
&\approx& \frac{h^{2}}{6N v^{~}_{\mathrm{F}}\pi^{2}\gamma}
\left(\int^{2\pi}_0\frac{R(\theta)d\theta}{2\pi}\right)
\int^{b{\Lambda}}_0\frac{d\omega}{2\pi}\frac{d^3\mathbf{k}}{(2
\pi)^3}\psi^{\ast}_{+s}\psi^{~}_{+s}(-i\omega)\left[\gamma
\ln(b^{-3})\right] ~~(\Lambda\rightarrow\infty)\nonumber
\\
&=&\int^{b{\Lambda}}_0 \frac{d\omega}{2\pi}\frac{d^3\mathbf{k}}{(2
\pi)^3}\psi^{\ast}_{+s}\psi^{~}_{+s}(-i\omega)\left[\frac{h^{2}}{2 N
v^{~}_{\mathrm{F}}\pi^{2}}\left(\int^{2\pi}_{0}
\frac{R(\theta)d\theta}{2\pi}\right)\ln(b^{-1})\right].\label{eq:}
\end{eqnarray}
After performing straightforward computations, we obtain
\begin{eqnarray}
\delta S^{s}_{\psi_{+}} = \begin{cases} \int_{0}^{b \Lambda}
\frac{d\omega}{2\pi}\frac{d^3\mathbf{k}}{(2\pi)^3} \psi^{\ast}_{+s}
\psi_{+s}(-i\omega)C_{1}\ln(b^{-1}) &
^{1}S_0, \\
\\
\int_{0}^{b \Lambda}\frac{d\omega}{2\pi} \frac{d^3\mathbf{k}}{(2
\pi)^3}\psi^{\ast}_{+s}\psi_{+s} (-i\omega)C_{2}\ln(b^{-1}) &
^{3}P_{2,\pm2}, \\
\\
\int_{0}^{b \Lambda}\frac{d\omega}{2\pi}\frac{d^3
\mathbf{k}}{(2\pi)^3}\psi^{\ast}_{+s} \psi_{+s}(-i\omega)
C_{3}\ln(b^{-1}) & ^{3}P_{2,0},
\end{cases}
\end{eqnarray}
where we have defined three constants:
\begin{eqnarray}
C_{1} = \frac{h^2}{2N v^{~}_{\mathrm{F}}\pi^2}, \\
C_{2} = \frac{h^2}{4Nv^{~}_{\mathrm{F}}\pi^2}, \\
C_{3} = \frac{5h^2}{4Nv^{~}_{\mathrm{F}}\pi^2}.
\end{eqnarray}
In terms of the length parameter $l$, $\delta S^{s}_{\psi_{+}}$ is
expressed as
\begin{eqnarray}
\delta S^{s}_{\psi_{+}} = \begin{cases} \int_{0}^{b \Lambda}
\frac{d\omega}{2\pi}\frac{d^3\mathbf{k}}{(2\pi)^3}
\psi^\ast_{+s}\psi_{+s}(-i\omega)C_{1}l & ^{1}S_0, \\
\\
\int_{0}^{b \Lambda} \frac{d\omega}{2\pi} \frac{d^3
\mathbf{k}}{(2\pi)^3}\psi^\ast_{+s}\psi_{+s}(-i\omega)C_{2}l &
^{3}P_{2,\pm2}, \\
\\
\int_{0}^{b \Lambda}\frac{d\omega}{2\pi} \frac{d^3\mathbf{k}}{(2
\pi)^3}\psi^\ast_{+s}\psi_{+s}(-i\omega)C_{3}l & ^{3}P_{2,0}.
\end{cases}
\end{eqnarray}

The $\delta S^{s}_{\psi_{+}}$ term is bilinear in the spinor field.
Apparently, it represents the neutron self-energy correction.
Incorporating this correction changes the original free neutron
action into
\begin{eqnarray}
S^{s}_{\psi_+} &=& \int_{0}^{b \Lambda}\frac{d\omega}{2\pi}
\frac{d^3\mathbf{k}}{(2\pi)^3} \psi^{\ast}_{+s}\Big[-i\omega
e^{C_{i}l} +\frac{v^{~}_{\mathrm{F}}}{2k^{~}_{\mathrm{F}}}\left(k^{2}_{x} +
k^{2}_{y}\right)+v^{~}_{\mathrm{F}}k_z\Big]\psi_{+s}.
\end{eqnarray}
The appearance of the exponential function $e^{C_{i}l}$ is a
consequence of the fermion-boson interaction. To ensure that
$S^{s}_{\psi_+}$ has the same form as the original free neutron
action, we need to redefine the scaling relation of the spinor
field:
\begin{eqnarray}
\psi_{+s} = \psi^{\prime}_{+} e^{\left(4-\frac{C_3}{2}\right)l}.
\label{Zf}
\end{eqnarray}
Apparently, the inclusion of the self-energy correction alters the
scaling behavior of $\psi_{+s}$. Comparing to the scaling
transformation given by Eq.~(\ref{eq:scalingspinorfield}) determined
in the noninteracting limit, the interaction correction yields an
extra factor $-\frac{C_{3}}{2}$, which is usually called the
anomalous dimension of the neutron field. To convert the above
$S^{s}_{\psi_+}$ back to its original form, it is necessary to
rescale the fermion velocity $v^{~}_{\mathrm{F}}$ as
\begin{eqnarray}
v^{~}_{\mathrm{F}} = v^{\prime}_{\mathrm{F}}e^{C_{3}l}.\label{vf}
\end{eqnarray}
The anomalous dimension of the neutron field modifies the fermion-boson
interaction term. This modification can be eliminated by rescaling
the coupling parameter $h$ in the following manner
\begin{eqnarray}
h = h^{\prime}e^{C_{3}l} \label{hf}.
\end{eqnarray}
Other dimensional quantities are transformed in terms of the length
scale $l$ as follows:
\begin{eqnarray}
\omega &=& \omega^{\prime}e^{-2l},\\
k_z &=& k^{\prime}_ze^{-2l},\\
k_x &=& k^{\prime}_xe^{-l},\\
k_y &=& k^{\prime}_ye^{-l},\\
\Omega &=& \Omega^{\prime}e^{-2l},\\
q_z &=& q^{\prime}_ze^{-2l},\\
q_x &=& q^{\prime}_xe^{-l},\\
q_y &=& q^{\prime}_ye^{-l},\\
\phi_{s}&=&\phi^{\prime}e^{4l},\\
\gamma &=& \gamma^{\prime}e^{-l},\label{gamma}\\
\lambda &=& \lambda^{\prime}e^{2l}.\label{lambda}
\end{eqnarray}
Notice that the boson field $\phi$ does not acquire an anomalous
dimension. This is because the boson self-energy, namely the
one-loop polarization function, is already incorporated into the
free boson action $S_{\phi}$ given by
Eq.~(\ref{eq:scaledbosonfreeaction}). The interaction corrections
are absorbed into the scaling property of the parameter $\gamma$.
Taking $^{3}P_2~(m_J=0)$ wave as an example, the above
transformations change the interaction actions to
\begin{eqnarray}
S^{\prime}_{\phi^{4}} &=& \frac{\lambda^{\prime}}{4}\int
\prod^{4}_{i=1} \int\frac{d\Omega^{\prime}_i}{2\pi} \frac{d^3
\mathbf{q}^{\prime}_i}{(2\pi)^3} \delta \left(\Omega^{\prime}_{1}
+\Omega^{\prime}_{3} - \Omega^{\prime}_{2} -
\Omega^{\prime}_{4}\right)  \delta^{3}(\mathbf{q}^{\prime}_{1} +
\mathbf{q}^{\prime}_{3}-\mathbf{q}^{\prime}_{2}
-\mathbf{q^{\prime}}_{4})|\phi^{\prime\ast}\phi^{\prime}|^2, \\
S^{\prime}_{\psi_+\phi} &=& h^{\prime}\prod^{2}_{i=1}\int
\frac{d\omega^{\prime}_i}{2\pi} \frac{d^{3}
\mathbf{k}^{\prime}_i}{(2\pi)^3} \frac{d\Omega^{\prime}}{2\pi}
\frac{d^3\mathbf{q}^{\prime}}{(2\pi)^3} \delta(\omega^{\prime}_{1} +
\omega^{\prime}_{2} -
\Omega^{\prime})\delta^{3}(\mathbf{k}^{\prime}_{1} +
\mathbf{k}^{\prime}_{2}-\mathbf{q}^{\prime}) \nonumber \\
&& \times \big[\left(\hat{q}^{\prime}_{x} + i\hat{q}^{\prime}_{y}
\right) \phi^{\prime\ast}\psi^{\prime}_{+\uparrow}
\psi^{\prime}_{+\uparrow}
+\left(-\hat{q}^{\prime}_{x}+i\hat{q}^{\prime}_{y}\right)
\phi^{\prime\ast} \psi^{\prime}_{+\downarrow}
\psi^{\prime}_{+\downarrow}  +2\hat{q}^{\prime}_{z}\phi^{\prime\ast}
\left(\psi^{\prime}_{+\uparrow} \psi^{\prime}_{+\downarrow} +
\psi^{\prime}_{+\downarrow} \psi^{\prime}_{+\uparrow}\right)
\nonumber \\
&& +\left(\hat{q}^{\prime}_{x}-i\hat{q}^{\prime}_{y}\right)
\phi^{\prime}\psi^{\prime \ast}_{+\uparrow} \psi^{\prime
\ast}_{+\uparrow}+\left(-\hat{q}^{\prime}_{x}-i\hat{q}^{\prime}_{y}\right)
\phi^{\prime}\psi^{\prime\ast}_{+\downarrow}\psi^{\prime\ast}_{+\downarrow}
+ 2\hat{q}^{\prime}_{z}\phi^{\prime}\left(\psi^{\prime \ast}_{+
\uparrow} \psi^{\prime\ast}_{+\downarrow} + \psi^{\prime \ast}_{+
\downarrow} \psi^{\prime \ast}_{+\uparrow}\right)\big].
\end{eqnarray}
These two actions now have the same forms as their original actions.

The original spinor field $\psi_{+}$ and the renormalized one
$\psi_{+}'$ are related via Eq.~(\ref{Zf}). The quasiparticle
residue $Z_{f}$ is actually the renormalization factor of the spinor
field \cite{Varma}. From Eq.~(\ref{Zf}), we derive the RG flow
equation of $Z_{f}$ and show its expression in Eq.~(\ref{Zf1}).
Making use of Eqs.~(\ref{vf}), (\ref{hf}), (\ref{gamma}), and
(\ref{lambda}), we obtain the flow equations of
$v^{~}_{\mathrm{F}}$, $h$ $\gamma$ and $\lambda$, which are listed
in Eqs.~(\ref{vf1})$-$(\ref{lambda1}).

\section{{NFL correction to thermal conductivity}}\label{sec:appc}

The thermal conductivity $\kappa$ in the NS core can be decomposed
into three parts \cite{Potekhin15}:
\begin{eqnarray}
\kappa= \kappa_{n}+\kappa_{p}+\kappa_{e}, \label{eq:kappa}
\end{eqnarray}
where $\kappa_{n}$, $\kappa_{p}$, and $\kappa_{e}$ denote the
thermal conductivity of neutrons, protons, and electrons
respectively. The NFL correction to the thermal conductivity mainly
originates from the modification of the neutron thermal
conductivity. In the FL state, the neutron thermal conductivity has
the form
\begin{eqnarray}
\kappa_{n}(T) = \frac{\pi^2}{3}\frac{\rho_{n}
\tau_{n}}{M^{\ast}_{n}}k^{2}_{\mathrm{B}}T, \label{eq:flkappa}
\end{eqnarray}
where $\rho^{~}_n$ is the neutron density and $\tau_n$ is the relaxation
time in neutron scattering processes. Upon incorporating the
corrections from the NFL behavior, Eq.~(\ref{eq:flkappa}) can be
modified to
\begin{eqnarray}
\kappa_{\mathrm{n}}(T)\sim \frac{T}{M^{\ast}_{n}(l)}. \label{eq:nflkappa}
\end{eqnarray}
In principle, $\tau_n$ should also be affected by NFL behavior. In
the FL state, $\tau_n$ behaves as $1/T^2$ \cite{Baiko01}. Upon
incorporating the NFL correction, $\tau_n$ would deviate from FL
behavior and exhibit a $1/T$ dependence instead. Similar behaviors
have also been found in degenerate quark plasma \cite{Heiselberg93}
and relativistic degenerate electron plasma \cite{Sarkar13}.

\begin{figure}[htbp]
\centering
\includegraphics[width=3.8in]{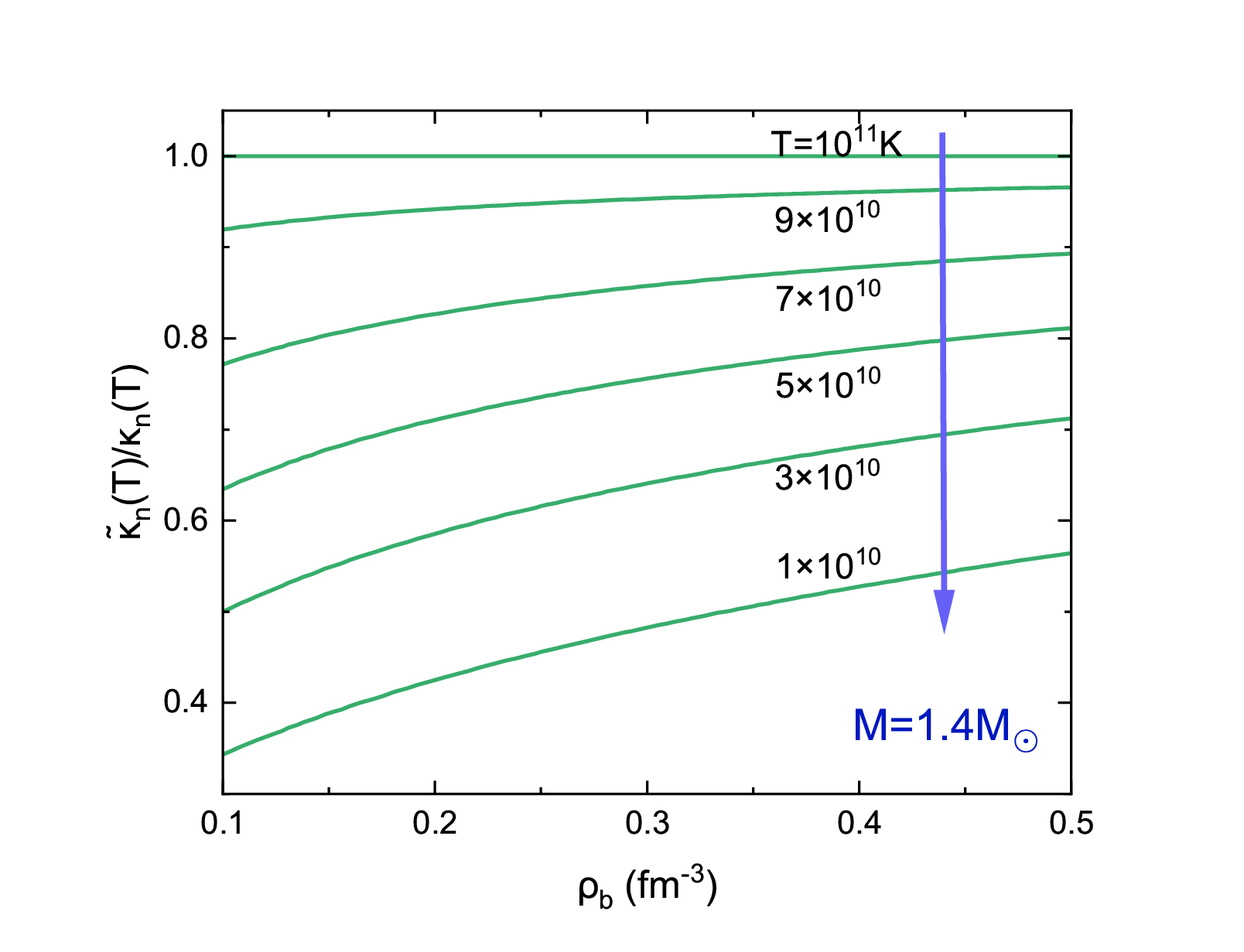}
\caption{{The dependence of the ratio of the renormalized neutron thermal
conductivity} $\tilde{\kappa}_{n}(T)$ to the neutron thermal conductivity
$\kappa_{n}(T)$ in a $1.4\mathrm{M}_{\odot}$ NS described by the APR
EOS on the baryon density $\rho_b$.} \label{NSrhokappa}
\end{figure}

We now turn our attention to the following derivative:
\begin{eqnarray}
\frac{d\kappa_{n}(T)}{dT} &\sim& \frac{d
\left[T/M^{\ast}_{n}(l)\right]}{dT}\nonumber \\
&=& \frac{1}{M^{\ast}_{n}(l)} -\frac{T}{M^{\ast 2}_{n}(l)}
\frac{dM^{\ast}_{n}(l)}{dT}\nonumber \\
&=& \frac{1}{M_{n}^{\ast} + \frac{5h^{2} M^{\ast 2}_{n}}{8N
\pi^{2}k^{~}_{\mathrm{F}}} \ln \frac{T_{0}}{T}} +
\frac{1}{\left(M_{n}^{\ast} + \frac{5h^{2} M^{\ast
2}_{n}}{8N\pi^{2}k^{~}_{\mathrm{F}}}\ln\frac{T_{0}}{T}\right)^2}
\frac{5h^{2}M^{\ast 2}_{n}}{8N\pi^{2}k^{~}_{\mathrm{F}}}.
\label{eq:dkappadT}
\end{eqnarray}
After integrating the above differential equation, we derive the
corrected expression for the neutron thermal conductivity:
\begin{eqnarray}
\tilde{\kappa}_{n}(T) &=&\frac{\pi^2}{3}\frac{\rho_{n}
\tau_{n}}{M^{\ast}_{n}}k^{2}_{\mathrm{B}}T\left[1+\frac{5h^{2}
M^{\ast }_{n}}{8N\pi^{2}k^{~}_{\mathrm{F}}} \ln
\frac{T_{0}}{T}\right]^{-1}\nonumber\\
&=&\kappa_{n}(T)\left[1+\frac{5h^{2} M^{\ast
}_{n}}{8N\pi^{2}k^{~}_{\mathrm{F}}} \ln
\frac{T_{0}}{T}\right]^{-1}.
\label{eq:nflkappad}
\end{eqnarray}

In Fig.~\ref{NSrhokappa}, we show the variation of the ratio
$\tilde{\kappa}_{n}(T)/\kappa_{n}(T)$ as a function of baryon
density $\rho_b$. The neutron thermal conductivity in the NFL state
($h=3.0$) is reduced compared with that in the FL state ($h=0$). The
reduction in $\tilde{\kappa}_{n}(T)$ is more pronounced at low
densities than at high densities, and it becomes increasingly
significant as the NS cools down. However, in the low-density
regions of the NS, thermal conductivity is mainly determined by the
electrons \cite{Potekhin15}. Meanwhile, the NFL correction to
$\kappa_{n}(T)$ in high-density regions of the NS is minor, as
illustrated in Fig.~\ref{NSrhokappa}. Therefore, the overall impact
of NFL behavior on the total thermal conductivity is negligible.

\section{Relation between coupling parameter and two-body potential}\label{sec:appd}

Here, we demonstrate how to estimate the coupling parameter, using
$h_1$ as an example, based on the microscopic details of the
two-body nuclear potential. The consideration is based on an
effective four-fermion-type pairing interaction in the spin-singlet
Cooper channel:
\begin{eqnarray}
&&\exp\left[{-\int dt d^3\mathbf{r}V(\mathbf{r})
\psi^{\ast}_{\uparrow}\psi^{\ast}_{\downarrow}
\psi_{\downarrow}\psi_{\uparrow}}\right] \nonumber \\
&=& \exp\left[{-\int dt d^3\mathbf{r}V(\mathbf{r})\Delta^{\ast}
\Delta}\right].
\end{eqnarray}
The potential function is real: $V(\mathbf{r}) =
V^{\ast}(\mathbf{r})$. It is convenient to define two composite
operators $\Delta=\psi_{\downarrow}\psi_{\uparrow}$ and
$\Delta^{\ast}=\psi^{\ast}_{\uparrow}\psi^{\ast}_{\downarrow}$ to
describe Cooper pairing. We then introduce an auxiliary bosonic field
$\phi$. After performing a Hubbard-Stratonovich transformation
\cite{Stratonovich58, Hubbard59}, the above term becomes
\begin{eqnarray}
&&\exp\left[{-\int dt d^3\mathbf{r}V(\mathbf{r})\Delta^{\ast}
\Delta}\right] \nonumber \\
&=& \int D\phi D\phi^\ast \exp\left[-\int dt d^3\mathbf{r}
\left(-\frac{1}{V(\mathbf{r})}\phi^\ast\phi+\phi^\ast
\Delta+\phi\Delta^{\ast}\right)\right]\nonumber \\
&=&\int D\phi D\phi^\ast \exp\left[-\int dt d^3\mathbf{r}
\left(-\frac{1}{V(\mathbf{r})}\phi^\ast\phi+\phi^\ast
\psi_{\downarrow}\psi_{\uparrow}+\phi\psi^{\ast}_{\uparrow}
\psi^{\ast}_{\downarrow}\right)\right].
\label{partitionfunction}
\end{eqnarray}
The auxiliary boson field $\phi$ denotes a dynamically fluctuating
complex field, which represents the quantum fluctuation of the
superfluid order parameter.

The two-body potential for neutrons is intricate, encompassing both
the central potential and the noncentral components (such as the
spin-orbit and tensor potentials). It can be expressed in the form
\begin{eqnarray}
V(\mathbf{r}) = V_{\mathbf{c}}(\mathbf{r}) +
V_{\mathbf{nc}}(\mathbf{r}),
\end{eqnarray}
where $V_{\mathbf{c}}(\mathbf{r})$ and $V_{\mathbf{nc}}(\mathbf{r})$
denote the central and noncentral potentials, respectively. In
order to find the relation between $h_1$ and $V(\mathbf{r})$, we
decompose the potential function as
\begin{eqnarray}
V(\mathbf{r}) = V_{\mathbf{c}}(\mathbf{r}) \left(1 +
\frac{V_{\mathbf{nc}}(\mathbf{r})}{V_{\mathbf{c}}(\mathbf{r})}\right),
\label{decomposepotential}
\end{eqnarray}
and then redefine the boson field as
\begin{eqnarray}
\phi^{\prime}(t, \mathbf{r}) = \frac{1}{\sqrt{1+\frac{V_{\mathbf{nc}}
(\mathbf{r})}{V_{\mathbf{c}}(\mathbf{r})}}}\phi(t,\mathbf{r}).
\label{rebosefield}
\end{eqnarray}
Equations (\ref{decomposepotential}) and (\ref{rebosefield}) enable us to
include the central potential in the boson free term and absorb
the remaining contributions (noncentral components) into the
fermion-boson coupling term. After substituting these equations into
the partition function given by Eq.~(\ref{partitionfunction}), we
obtain
\begin{eqnarray}
&&\int D\phi^{\prime} D\phi^{\prime\ast} \exp\left[-\int dt
d^{3}\mathbf{r}\left(-\frac{1}{V_{\mathbf{c}}(\mathbf{r})}
\phi^{\prime\ast}\phi^{\prime}+
\left(\sqrt{1+\frac{V_{\mathbf{nc}}(\mathbf{r})}{V_{\mathbf{c}}
(\mathbf{r})}}\right)\left(\phi^{\prime\ast}\psi_{\downarrow}
\psi_{\uparrow}+\phi^{\prime}\psi^{\ast}_{\uparrow}
\psi^{\ast}_{\downarrow}\right)\right)\right]\nonumber \\
&=& \int D\phi^{\prime} D\phi^{\prime\ast} \exp\left[-\int dt
d^{3}\mathbf{r} \left(-\frac{1}{V_{\mathbf{c}}(\mathbf{r})}
\phi^{\prime\ast}\phi^{\prime}+ h_1\left(\phi^{\prime\ast}
\psi_{\downarrow}\psi_{\uparrow}+\phi^{\prime}
\psi^{\ast}_{\uparrow}\psi^{\ast}_{\downarrow}\right)\right)\right].
\end{eqnarray}
We eventually find that $h_1$ is linked to the potential functions
as follows:
\begin{eqnarray}
h_1 = \sqrt{1+\frac{V_{\mathbf{nc}}
(\mathbf{r})}{V_{\mathbf{c}}(\mathbf{r})}}. \label{hpotential}
\end{eqnarray}
Similar relations can be derived for the other three $h$ parameters,
namely $h_{2}$, $h_{3}$, and $h_{4}$.

\end{widetext}

\end{document}